\documentclass[10pt]{article}
\usepackage[top=2cm, bottom=2cm, left=2.5cm, right=2.5cm]{geometry}

\usepackage{natbib}
\usepackage{multirow}
\usepackage{booktabs}
\usepackage{graphicx} 
\usepackage{longtable}
\usepackage{multirow}
\usepackage{adjustbox}
\usepackage{float}
\usepackage{hyperref}
\usepackage{tikz}
\usetikzlibrary{shapes,arrows,positioning,calc}
\usepackage{amsmath}
\usepackage{xcolor}
\usepackage{tikz}
\usetikzlibrary{backgrounds, positioning}
\usetikzlibrary{fit} 
\usetikzlibrary{positioning, arrows.meta, fit, backgrounds}
\usepackage{tikz}
\usetikzlibrary{decorations.pathreplacing,calligraphy,positioning,fit,calc}
\usepackage{makecell}

\usepackage{tikz}
\usetikzlibrary{mindmap, backgrounds, shapes.geometric, arrows, positioning, shadows}

\usepackage{graphicx} 
\usepackage{booktabs} 
\usepackage{amssymb}  
\usepackage{array}    
\usepackage{amssymb} 
\usepackage{tabularx}
\usepackage[utf8]{inputenc}

\title{\textbf{Intelligent Agents with Emotional Intelligence: Current Trends, Challenges, and Future Prospects}}

\author{Raziyeh Zall\textsuperscript{1} \and Alireza Kheyrkhah\textsuperscript{1} \and Erik Cambria\textsuperscript{2}  \and Zahra Naseri\textsuperscript{1} \and M.Reza Kangavari\textsuperscript{1}}
\date{\today}

\begin{document}
	
	\maketitle
	
	\noindent
	\textsuperscript{1}School of Computer Engineering, Iran University of Science and Technology, Tehran, Iran \\
	\textsuperscript{2}College of Computing and Data Science, Nanyang Technological University, Singapore \\
	\textbf{Email:} zall\_razieh@comp.iust.ac.ir, zall.raziyeh@gmail.co

\begin{abstract}
	The development of agents with emotional intelligence is becoming increasingly vital due to their significant role in human-computer interaction and the growing integration of computational systems across various sectors of society. Affective computing aims to design intelligent systems that can recognize, evoke, and express human emotions, thereby emulating human emotional intelligence. While previous reviews have focused on specific aspects of this field, there has been limited comprehensive research that encompasses emotion understanding, elicitation, and expression, along with the related challenges. This survey addresses this gap by providing a holistic overview of core components of artificial emotion intelligence into one cohesive map for researchers. It covers emotion understanding through multimodal data processing, as well as affective cognition, which includes cognitive appraisal, emotion mapping, and adaptive modulation in decision-making, learning, and reasoning. Additionally, it addresses the synthesis of emotional expression across text, speech, and facial modalities to enhance human-agent interaction.
	This paper identifies and analyzes the key challenges and issues encountered in the development of affective systems, covering state-of-the-art methodologies designed to address them. Finally, we highlight promising future directions, with particular emphasis on the potential of generative technologies to advance affective computing.
\end{abstract}

\section{Introduction}\label{sec1}
\begin{figure*}[t]
	\centering
	\includegraphics[width=1\textwidth]{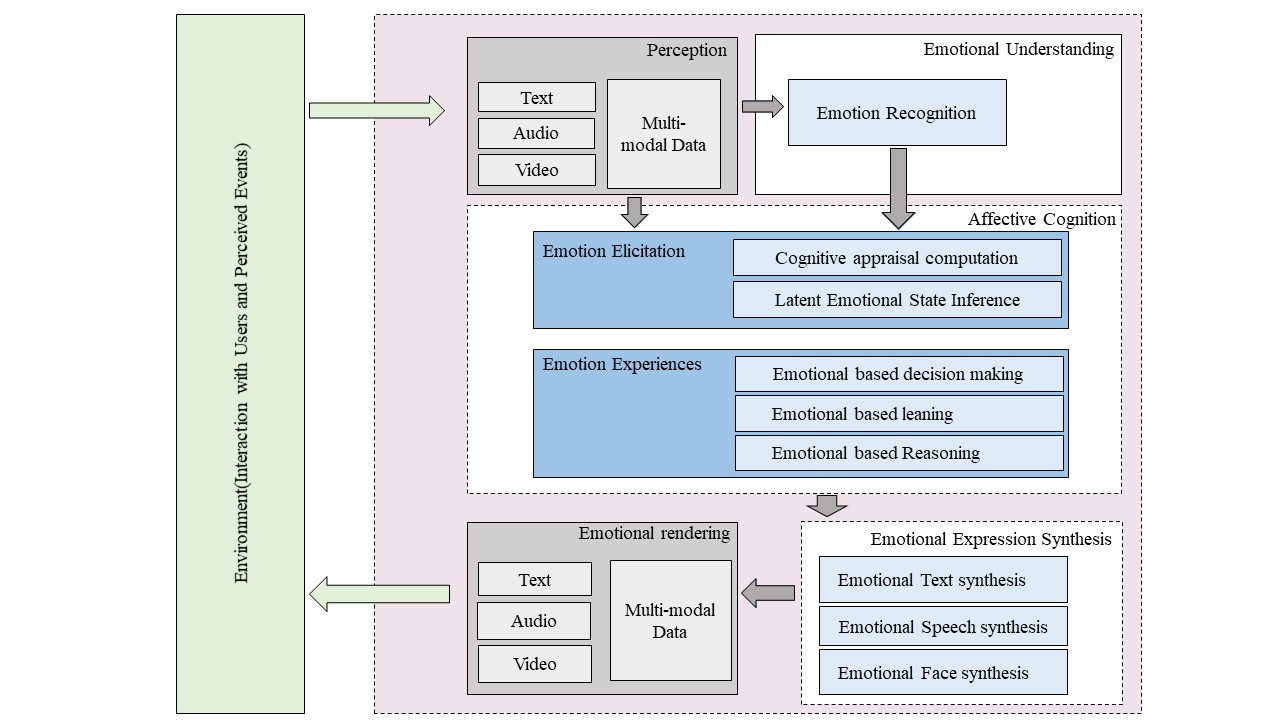}
	\caption{Overview of intelligent agent with emotional intelligence}
	\label{fig:overview}
\end{figure*}

Developing intelligent agents that possess human-level intelligence is a key goal in the field of Human-Computer Interaction (HCI) and general artificial intelligence \cite{Jeon2017}. A crucial aspect of achieving this goal is the incorporation of emotional intelligence, which is essential for human cognition and social interaction, into these intelligent agents. Emotional intelligence encompasses three interrelated capabilities.  First, emotion understanding involves accurately detecting and interpreting affective signals; for example, recognizing when a user is feeling frustrated during an interaction by analyzing their tone of voice or facial expressions. Second, emotion elicitation and experiences refer to interpreting the causes, context, and implications of emotions for both the individual and the interaction. For instance, an agent can perform a cognitive appraisal of the surrounding environment and contextual factors to infer its internal emotional state, which subsequently guides appropriate decision-making in the given situation. Third, emotion expression encompasses the capacity to generate, modulate, and convey appropriate emotional responses in a socially meaningful way, such as responding with a reassuring message or a sympathetic tone when the user is upset.\\
Affective Computing, coined by Rosalind Picard \cite{Picard2000}, emerged as a discipline dedicated to equipping machines with emotional intelligence, enabling them to recognize, interpret, and respond to human emotions. By embedding emotional intelligence into intelligent agents, affective computing facilitates more naturalistic, adaptive, and socially competent interactions, which in turn enhance user trust, engagement, and satisfaction \cite{zall2024towards}. Such emotionally intelligent systems not only improve usability but also enable advanced functionalities, including personalized assistance, empathetic dialogue, and context-aware decision-making.\\
Figure~\ref{fig:overview} presents the conceptual framework for intelligent agents with emotional intelligence. Building upon prior research, three fundamental components are identified as the core elements underpinning the development of emotional intelligence, described as follows:

\begin{enumerate}
	\setlength{\itemindent}{1.5em}
	\setlength{\leftmargin}{2.5em}
	
	\item \textbf{Emotion Understanding:}  
	This stage involves analyzing the affective features embedded in user input, enabling the agent to accurately detect and interpret the user’s emotional state during interaction \cite{afzal2024comprehensive, zhao2025leveraging}.
	
	\item \textbf{Affective Cognition:}  
	In this phase, the agent assesses emotional events through cognitive reasoning processes to ensure accurate and context-sensitive interpretation. Subsequently, it performs emotional elicitation modeling to generate an internal affective state, which modulates higher-order cognitive functions such as learning, inference, and decision-making \cite{liu2024emotion, Raggioli2025}. This internal affective regulation drives the agent’s adaptive behavior, resulting in responses aligned with the user’s emotional context.
	
	\item \textbf{Emotional Expression Synthesis:}  
	Finally, the agent externalizes its emotional states through multimodal communication channels, including text, speech, and visual expressions, where cross-modal synchronization is essential for fostering coherence, authenticity, and naturalness in emotional interactions \cite{chen2023, abilbekov2024kazemotts21}.
	
\end{enumerate}
Emotion understanding in intelligent agents encounters substantial challenges in accurately detecting subtle or ambiguous affective signals, especially within diverse cultural contexts or in the presence of environmental noise \cite{rahmani2023transfer} \cite{kamran2023emodnn}.
These challenges arise from practical limitations, including datasets with restricted size, noise, imbalance, and suboptimal quality \cite{ye2024semi, oh2024noise}. Further obstacles are presented by learning model deficiencies, such as low accuracy, limited interpretability, poor generalizability, and the lack of standardized evaluation metrics \cite{umair2024emotion} \cite{cambria2024senticnet}. Multimodal integration introduces additional complexity, particularly in the combination of heterogeneous data types and the management of missing modalities \cite{kumar2024multimodal, geetha2024multimodal, alsaadawi2024tac, khan2024exploring}. The inherent complexity of emotions, including their diversity, overlap, and the challenges associated with experimental design, further complicates emotion understanding \cite{ye2024semi, sharma2024emotion}. Moreover, contemporary large-scale models are affected by issues such as hallucinations, high annotation costs, and limited contextual comprehension \cite{farquhar2024detecting, schuller2024affective}. Recent advancements, such as data augmentation, synthetic data generation, transfer, and semi-supervised learning, model interpretability methods, enhanced evaluation standards, and multimodal fusion, present promising avenues for addressing these challenges.
Affective cognition, a core component of emotional intelligence in intelligent agents, extends beyond emotion recognition to encompass reasoning about emotions and the generation of contextually appropriate responses through the integration of cognitive and affective theories of mind. Modeling affective cognition is inherently challenging due to the complexity of capturing contextual and causal factors, often necessitating sophisticated cognitive frameworks enriched with domain-specific knowledge.

This process involves identifying events and mental states that elicit emotions (emotion elicitation) and interpreting the resulting behavioral and cognitive outcomes (emotional experiences), as described in cognitive appraisal theories and data-driven approaches \cite{zall2024towards, liu2024emotion, jokinen2025introduction}. However, the development of affective cognition models faces several significant challenges, including limited and ambiguous datasets, difficulties in computing cognitive appraisal variables, and scalability constraints in computational cognitive models \cite{gandhi2024human, somarathna2022virtual, bayro2025systematic}.
Additional challenges arise in large language model (LLM)-based systems, such as contextual misinterpretation and limitations in emotional reasoning capabilities \cite{tak2024journal, khan2025large, Raggioli2025}. To address these issues, recent studies have proposed solutions including advanced cognitive architectures, virtual reality–based data collection, reinforcement learning integration, and explainable emotion-alignment frameworks. This study investigates these challenges and solutions to advance the development of emotionally intelligent agents capable of naturalistic and socially appropriate interactions.\\
Emotional text, speech, and face synthesis have become essential components of affective computing. They allow agents to generate responses that are contextually appropriate and emotionally resonant. In the realm of text synthesis, advancements like style transfer, conditional generation, and fine-tuned LLMs enhance expressiveness. However, challenges such as data sufficiency, consistency, and multimodal alignment still exist. Frameworks like MOPO and EmoBench aim to address these issues \cite{firdaus2023multi65, resendiz2024mopo43, sabour2024emobench}. Emotional Speech Synthesis (ESS) and voice conversion increasingly utilize non-parallel and diffusion-based methods, such as EmoConv-Diff, to improve scalability and emotional fidelity \cite{zhou2020transforming, prabhu2024}. Simultaneously, emotional face synthesis utilizes GANs, diffusion models, and multimodal systems such as UniPortrait and EmotiveTalk to address challenges related to data imbalance and synchronization. These advancements enable the creation of lifelike and coherent multimodal affective expressions.
\subsection{Research questions}
To guide our investigation into the integration of emotional intelligence in intelligent agents, we pose the following research questions:

\begin{itemize}
	\item \textit{RQ1}: How can intelligent agents robustly detect and interpret affective signals across diverse environments despite data and model limitations?
	\item \textit{RQ2}: What are effective approaches for modeling emotion elicitation and experience in affective cognition, given challenges in ambiguity, computation, and scalability?
	\item \textit{RQ3}: How can intelligent agents generate contextually appropriate and emotionally coherent responses across multiple modalities?
\end{itemize}
\subsection{Contribution}
This study investigates the fundamental components of emotional intelligence in intelligent agents, with an emphasis on their functional roles, the specific challenges that impede effective implementation, and the solutions proposed in existing literature. Through a systematic examination of these dimensions, the study aims to elucidate how emotional intelligence can be effectively integrated into artificial agents to enable more natural, adaptive, and human-like interactions. By synthesizing current research findings, this work highlights both the progress achieved and the critical gaps that remain unaddressed in the pursuit of genuine emotional intelligence in intelligent agents. Despite the substantial potential of emotionally intelligent agents, persistent challenges across core capabilities continue to constrain their performance and applicability in real-world environments.  \\
\begin{table*}[htbp]
	\centering
	\caption{Comparison of the present survey with recent studies in affective computing (2024--2025) based on covered emotional aspects and modalities}
	\resizebox{\textwidth}{!}{%
		\begin{tabular}{>{\raggedright\arraybackslash}p{8cm}cccccc}
			& \multicolumn{3}{c}{\textbf{Emotion}} & \multicolumn{3}{c}{\textbf{Modality}} \\
			\cmidrule(lr){2-4} \cmidrule(lr){5-7}
			& Recognition & Elicitation & Expression & Text & Speech & Vision/Facial \\
			\midrule
			Recent Trends of Multimodal Affective Computing: A Survey from an NLP Perspective \cite{hu2024recent} & $\checkmark$ &-- & -- & $\checkmark$ & $\checkmark$ & $\checkmark$ \\ \hline
			A Review of Human Emotion Synthesis Based on Generative Technology \cite{ma2025review} & $\checkmark$ & -- & $\checkmark$ & $\checkmark$ & $\checkmark$ & -- \\ \hline
			Affective Computing in the Era of Large Language Models: A Survey from the NLP Perspective \cite{zhang2024affective}	 & $\checkmark$ & -- & $\checkmark$ & $\checkmark$ & -- & -- \\ \hline
			Artificial Emotion: A Survey of Theories and Debates on Realising Emotion in Artificial Intelligence \cite{li2025artificial}	 & -- & $\checkmark$ & -- & -- & -- & -- \\ \hline
			Emotion recognition and generation: a comprehensive review of face, speech, and text modalities \cite{mobbs2025emotion}	
			& $\checkmark$ & -- & $\checkmark$ & $\checkmark$ & $\checkmark$ & $\checkmark$ \\ \hline
			Intelligent Agents with Emotional Intelligence: Current Trends, Challenges, and Future Prospects & $\checkmark$ & $\checkmark$ & $\checkmark$ & $\checkmark$ & $\checkmark$ & $\checkmark$ \\
			\bottomrule
		\end{tabular}%
	}
	\label{tab:comparison}
\end{table*}
\noindent
Table~\ref{tab:comparison} provides a comprehensive comparison between this survey and recent studies in the field of affective computing published from 2024 to 2025. The comparison is organized along two principal dimensions: (1) emotional aspects, encompassing emotion recognition, elicitation, and expression, and (2) modalities, including text, speech, and visual or facial cues. This structure underscores the distinctive contribution of the present study, which offers an integrative perspective by addressing all relevant emotional dimensions across multiple modalities—thereby delivering a more holistic and unified understanding compared to previous research.
The primary contributions of this study can be summarized as follows:
\begin{itemize}
	\item To the best of our knowledge, this work represents the first comprehensive review of intelligent agents equipped with the full spectrum of emotional intelligence capabilities—namely, emotion understanding, affective cognition, and emotional expression, providing a cohesive framework for advancing naturalistic and empathetic human-computer interaction (HCI).
	\item It offers an in-depth analysis and categorization of the key challenges that hinder the effective realization of these three core capabilities within intelligent agents.
	\item It systematically evaluates recent methodological advancements proposed to address these challenges and delineates promising future research directions for the development of emotionally intelligent systems.
\end{itemize}

\subsection{Paper organization}
The remainder of this paper is organized as follows. Section 2 describes the systematic literature review methodology. Section 3 examines challenges and emerging solutions in emotion understanding. Section 4 explores challenges and emerging solutions in affective cognition, focusing on emotion elicitation in interactive systems and emotional experiences. Sections 5 to 7 address challenges and emerging solutions in emotional expression synthesis across multiple modalities: Section 5 focuses on Emotional Text Synthesis (ETS), Section 6 on ESS, and Section 7 on emotional face synthesis. Section 8 offers a comprehensive discussion of challenges and future research directions. Finally, Section 9 concludes with a summary of contributions and a vision for developing emotionally intelligent agents.
\section{Methodology: Systematic Literature Review Process}
This paper adheres to the Preferred Reporting Items for Systematic Reviews and Meta-Analyses (PRISMA) guidelines to ensure transparency and reproducibility in the study selection process. We adopted the PRISMA framework to systematically identify, screen, and include relevant literature on emotional intelligence in intelligent agents. Our focus is on three core capabilities: emotion understanding (emotion recognition), affective cognition (emotional elicitation and experiences), and emotional expression synthesis.
\subsection{Identification}
We conducted a comprehensive literature search across several electronic databases, including Google Scholar, IEEE Xplore, ACM Digital Library, Scopus, Web of Science, and arXiv. The search terms were crafted to encompass key themes and included combinations such as:
("affective computing" OR "emotional AI" OR "emotion recognition" OR "emotion synthesis") AND ("textual emotion" OR "speech emotion" OR "facial expression" OR "multimodal emotion") AND ("cognitive architectures" OR "computational models of emotion" OR "appraisal theories" OR "reinforcement learning").
Our search focused on publications from 2017 to 2025 to capture the latest advancements in deep learning-based approaches. Initially, no date restrictions were applied to include foundational works; however, we prioritized publications from 1990 to 2025 to emphasize modern developments. 
We also identified additional records through backward citation searching (reviewing the references of key papers) and forward citation searching (using tools like Google Scholar's "Cited by" feature). Grey literature, such as preprints and conference proceedings, was included if it was peer-reviewed or highly cited.
\subsection{Screening}
The titles and abstracts of all 2,500 records were screened for relevance. Studies were excluded at this stage if they were clearly outside the scope of affective computing (e.g., psychological studies on human emotion, medical studies on emotional disorders) or focused solely on human subjects without application to intelligent agents. This screening process reduced the number of records to 500.
\begin{figure}[h]
	\centering
	\includegraphics[width=0.6\textwidth]{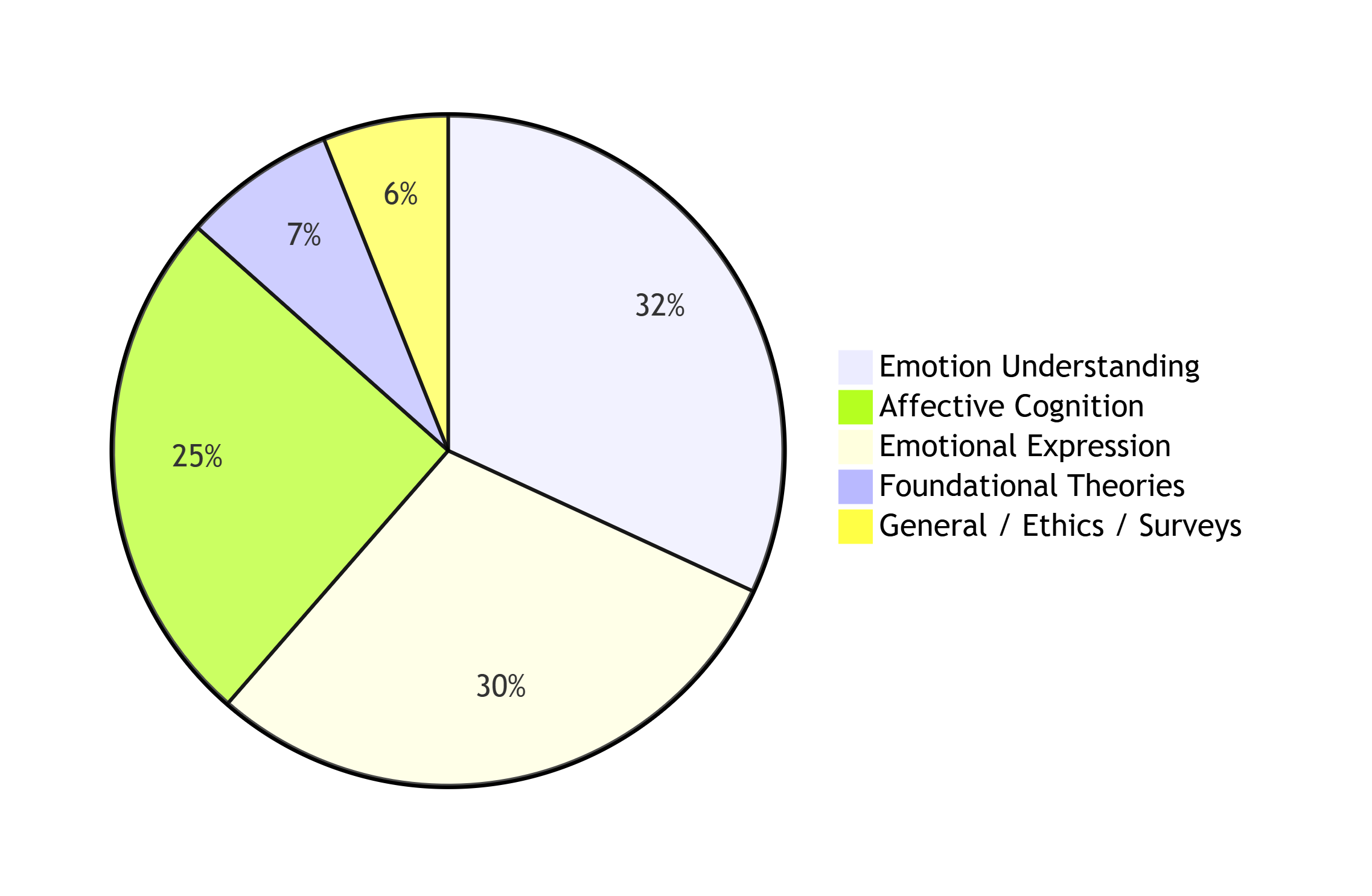}
	\caption{Distribution of the 298 included studies across the three core capabilities of emotional intelligence in intelligent agents.}
	\label{fig:pie_capabilities}
\end{figure}
\begin{figure}[h]
	\centering
	\includegraphics[width=0.9\textwidth]{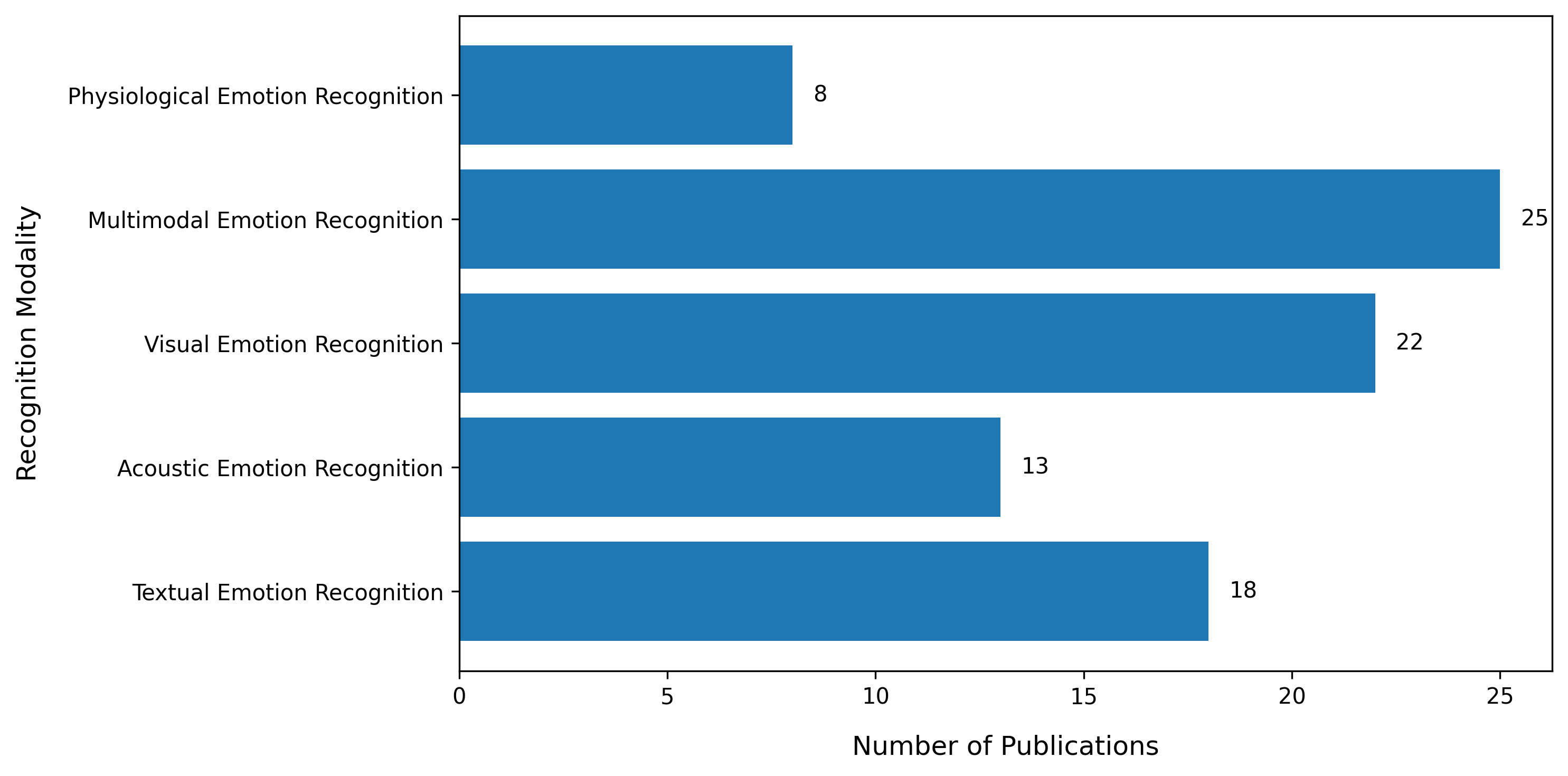}
	\caption{Breakdown of emotion understanding research by modality}
	\label{fig:bar_understanding}
\end{figure}
\subsection{Eligibility}
The full text of the 500 remaining reports was retrieved and assessed for eligibility based on the following criteria:
\begin{itemize}
	\item{Inclusion criteria:}\\
	The review incorporated studies published in English, including peer-reviewed journal articles, conference proceedings, books, and preprints from reputable repositories. To be included, a publication was required to explicitly address emotional intelligence in intelligent agents and demonstrate relevance to at least one of the three core capabilities examined in this study: (i) emotion understanding, such as emotion recognition from textual, acoustic, visual, or physiological data; (ii) affective cognition, including computational models of emotion elicitation, internal emotional state representation, or cognitive–affective architectures; and (iii) emotional expression synthesis, encompassing the generation of affective responses through textual, acoustic, or visual modalities. Both empirical studies presenting quantitative evaluations and theoretical or review articles offering insights into AI architectures, models, datasets, or applications were considered. Duplicate records identified across multiple databases were systematically removed to ensure a unique set of studies for analysis.
	
	\item {Exclusion criteria:}\\
	Non-English publications were excluded to maintain consistency in analysis and interpretation. Studies that did not explicitly focus on intelligent agents—such as works limited to general human psychology without a clear artificial intelligence context—were omitted. Additionally, research addressing unrelated topics, including non-affective AI tasks (e.g., conventional pathfinding or image classification without emotional components), was excluded. Low-quality sources, such as non-peer-reviewed blogs, opinion pieces, and outdated technical reports lacking empirical validation or academic citations, were also filtered out. Finally, all duplicate publications across the searched databases were removed to maintain a non-redundant corpus for the systematic review.
	
\end{itemize}
\begin{figure}[h]
	\centering
	\includegraphics[width=0.9\textwidth]{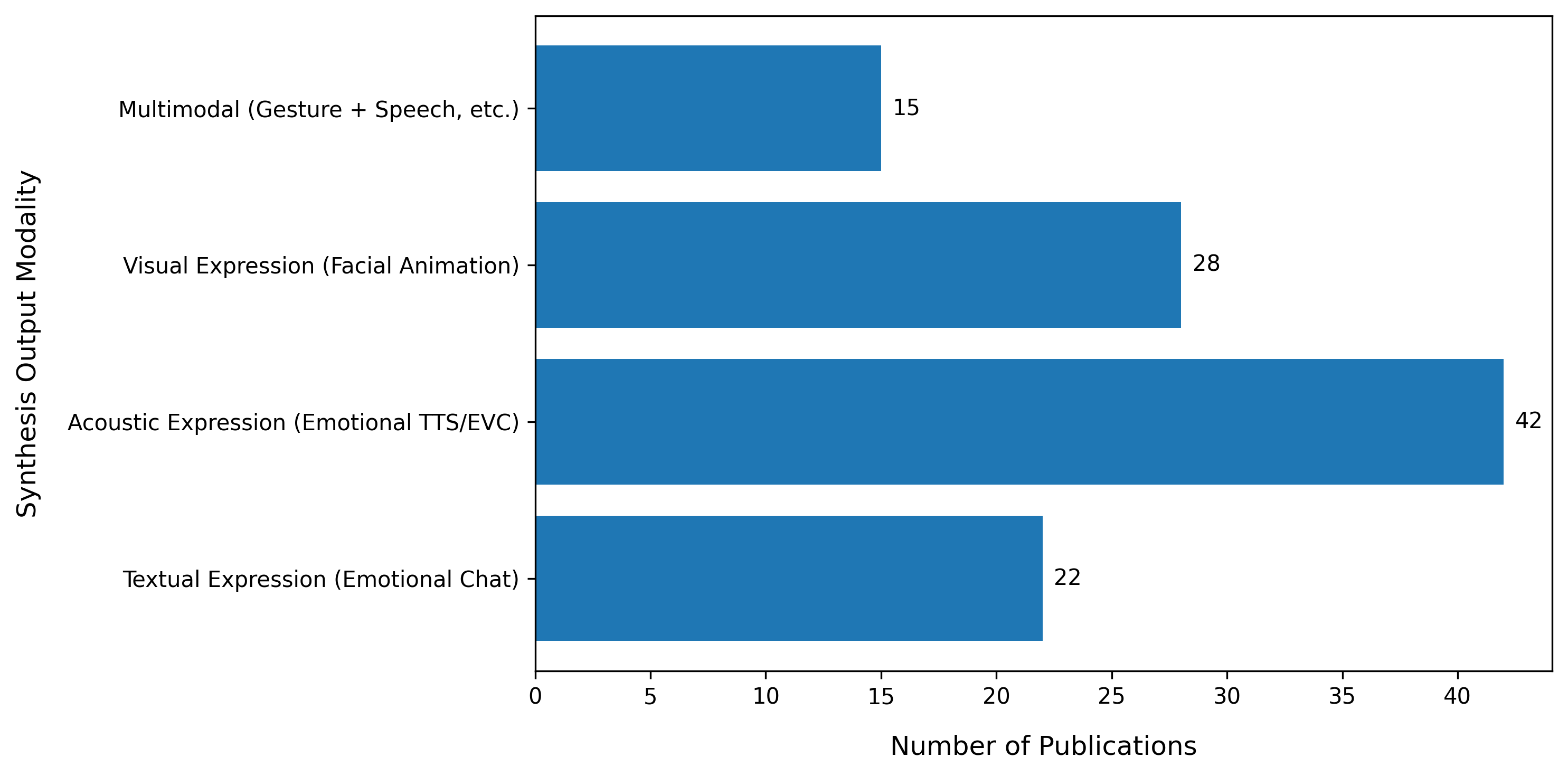}
	\caption{Breakdown of emotional expression synthesis research by output modality.}
	\label{fig:bar_synthesis}
\end{figure}
After this rigorous assessment, 202 reports were excluded, leaving 298 studies that formed the core basis of our survey.
\subsection{Inclusion}
\begin{figure}[h]
	\centering
	\includegraphics[width=0.9\textwidth]{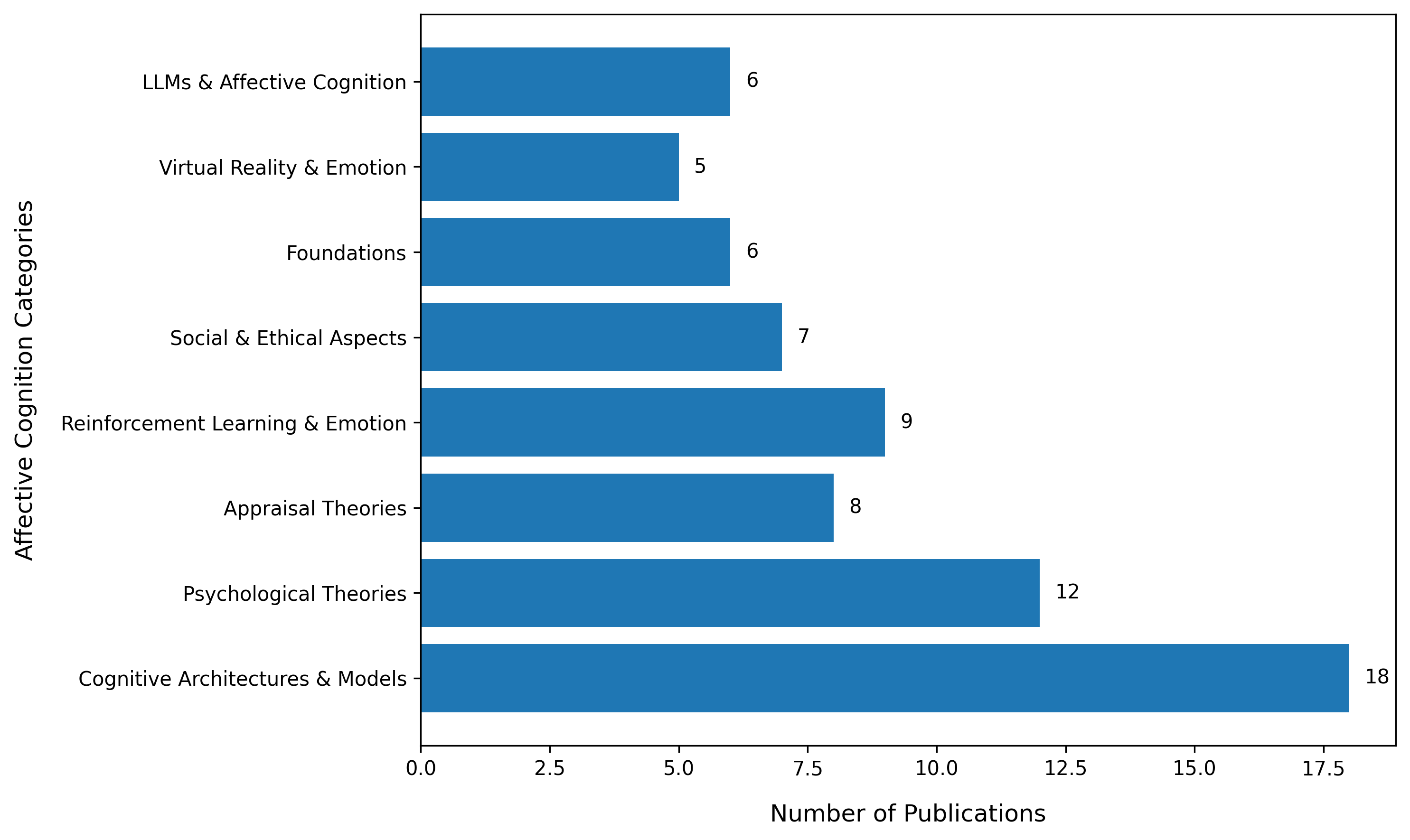}
	\caption{Categorization of affective cognition research by theoretical and computational focus.}
	\label{fig:bar_cognition}
\end{figure}

The final 298 studies were subjected to an in-depth analysis to synthesize key findings, identify emerging trends, compare methodologies, and explore challenges and future directions outlined in this paper. To offer a comprehensive view of the research landscape, we categorized these studies based on their primary focus across the three core capabilities of emotional intelligence in intelligent agents.
As depicted in Fig. \ref{fig:pie_capabilities}, the distribution of the 298 studies highlights that 32\% of the research effort is devoted to emotion understanding, which includes techniques for recognizing emotions from textual, acoustic, visual, and physiological data. Emotional expression synthesis, focusing on generating emotional responses in text, speech, or visual forms, accounts for 30\% of the corpus. The remaining 25\% addresses affective cognition, exploring mechanisms for eliciting, representing, and integrating emotions into cognitive architectures. The residual 13\% comprises foundational works and general surveys, providing a broader context for the field.
A more detailed breakdown of each capability is presented in Figs. \ref{fig:bar_understanding}, \ref{fig:bar_cognition}, and \ref{fig:bar_synthesis}, which highlight specific sub-categories within these domains. Fig. \ref{fig:bar_understanding} illustrates the modalities within emotion understanding, revealing a strong emphasis on multimodal recognition frameworks. Fig. \ref{fig:bar_synthesis} delineates the output modalities for emotional expression synthesis. Lastly, Fig. \ref{fig:bar_cognition} categorizes the diverse theoretical and computational models in affective cognition, ranging from cognitive architectures to appraisal theories. This structured analysis serves as the foundation for the detailed state-of-the-art review presented in the subsequent sections of this paper.

\section{Emotion Understanding}
\label{elicitation}
Emotion recognition is a fundamental task in artificial intelligence (AI) that aims to understand and interpret human emotions expressed through various forms of data \cite{afzal2024comprehensive}. This process identifies specific emotional states using facial expressions \cite{canal2022survey}, vocal tones \cite{wani2021comprehensive}, language \cite{deng2021survey}, and physiological signals \cite{dadebayev2022eeg}. Emotion recognition is crucial in enhancing HCI, mental health monitoring, and other applications where understanding human emotional responses is essential \cite{nayak2021human}.
The inputs for emotion recognition are diverse and multimodal, making the task challenging and an intriguing opportunity. Text-based inputs, such as social media posts, reviews, or dialogues, are processed using Natural Language Processing (NLP) techniques to identify emotional nuances in the language. Machine learning models analyze words, phrases, sentence structure, and context to predict the underlying emotional tone of a text. In contrast, vision-based inputs leverage computer vision techniques to analyze facial expressions, body movements, and gestures, often using deep learning models for recognition. These visual cues offer valuable data to help AI systems accurately interpret emotions. This is especially important in real-time applications such as video conferencing or emotion-aware robotics. In addition to text and vision, sound or speech signals are critical for emotion recognition. Audio inputs, such as tone of voice, pitch, volume, and rhythm, are key indicators of emotional states. Finally, physiological signals, including heart rate, skin conductance, and facial muscle activity, provide additional insights into emotional responses, obtained through wearable devices or biosensors. These physiological markers reflect the autonomic nervous system’s reactions to emotional stimuli. They are instrumental in applications like mental health monitoring and affective computing, where a deeper understanding of emotional states is necessary for more personalized interactions.
\subsection{Approaches}
\begin{figure}[H]
	\centering
	\includegraphics[scale=0.8]{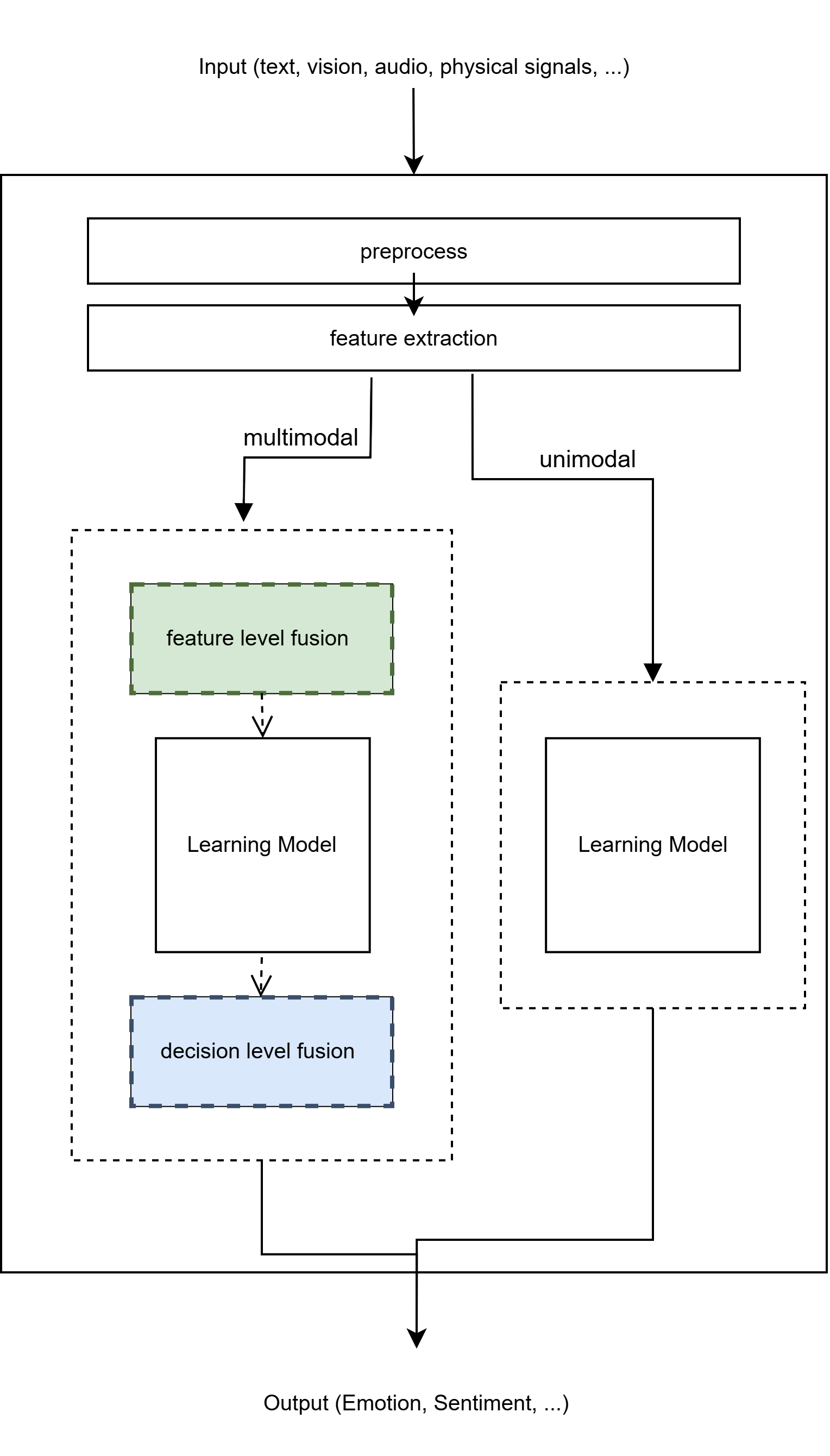}
	\caption{The overall framework of emotion recognition.}
	\label{fig:emotion_recognition_framework}
\end{figure}

Figure~\ref{fig:emotion_recognition_framework} illustrates the overall emotion recognition framework, highlighting the flow of data through successive processing stages and model learning components. 
The process begins with input data acquisition, which may involve multiple modalities, including textual data (e.g., written language), visual data (images or video), acoustic data (speech or audio signals), and physiological signals (e.g., heart rate or electroencephalography (EEG)). Each modality offers complementary cues for inferring emotional states. For example, textual data can convey sentiment through lexical and semantic patterns, while facial expressions captured in visual data provide observable emotional indicators. Similarly, physiological signals such as heart rate variability can reflect stress or relaxation levels. The diversity of input modalities reflects the multifaceted nature of human emotions and motivates the integration of multimodal data for robust emotion recognition. Following data acquisition, preprocessing is applied to clean and standardize the raw inputs, which often contain noise, artifacts, or inconsistencies. Preprocessing techniques vary by modality and may include background noise removal in audio signals, normalization and tokenization in textual data, or resizing and cropping in visual data. This stage is important for improving data quality and ensuring that learning models are not adversely affected by irrelevant variations or noise. Feature extraction aims to derive informative representations from the preprocessed data. In textual modalities, this may involve extracting emotion-related lexical features or syntactic patterns. Visual data can yield features related to facial expressions or motion dynamics, while acoustic features such as pitch, intensity, and spectral characteristics are commonly used to infer emotional states. For physiological signals, features such as heart rate variability or EEG frequency bands are extracted to capture underlying affective responses. These features constitute the foundational representations used for emotion classification. Modality processing determines how information from different data sources is handled. In unimodal processing, each modality is analyzed independently using a dedicated learning model. In contrast, multimodal processing leverages complementary information across modalities to improve recognition performance through data fusion. Two primary fusion strategies are commonly employed:
\begin{itemize}
	\item \textbf{Feature-level fusion:} Features from multiple modalities are combined at an early stage to form a unified representation that is subsequently input to a learning model.
	\item \textbf{Decision-level fusion:} Each modality is processed independently, and the resulting predictions are aggregated at a later stage to produce a final emotion classification.
\end{itemize}
Finally, learning models map the extracted and fused features to emotional outcomes. These models are trained to identify patterns within the feature space and to classify emotional states based on learned representations.
\subsection{Challenges}
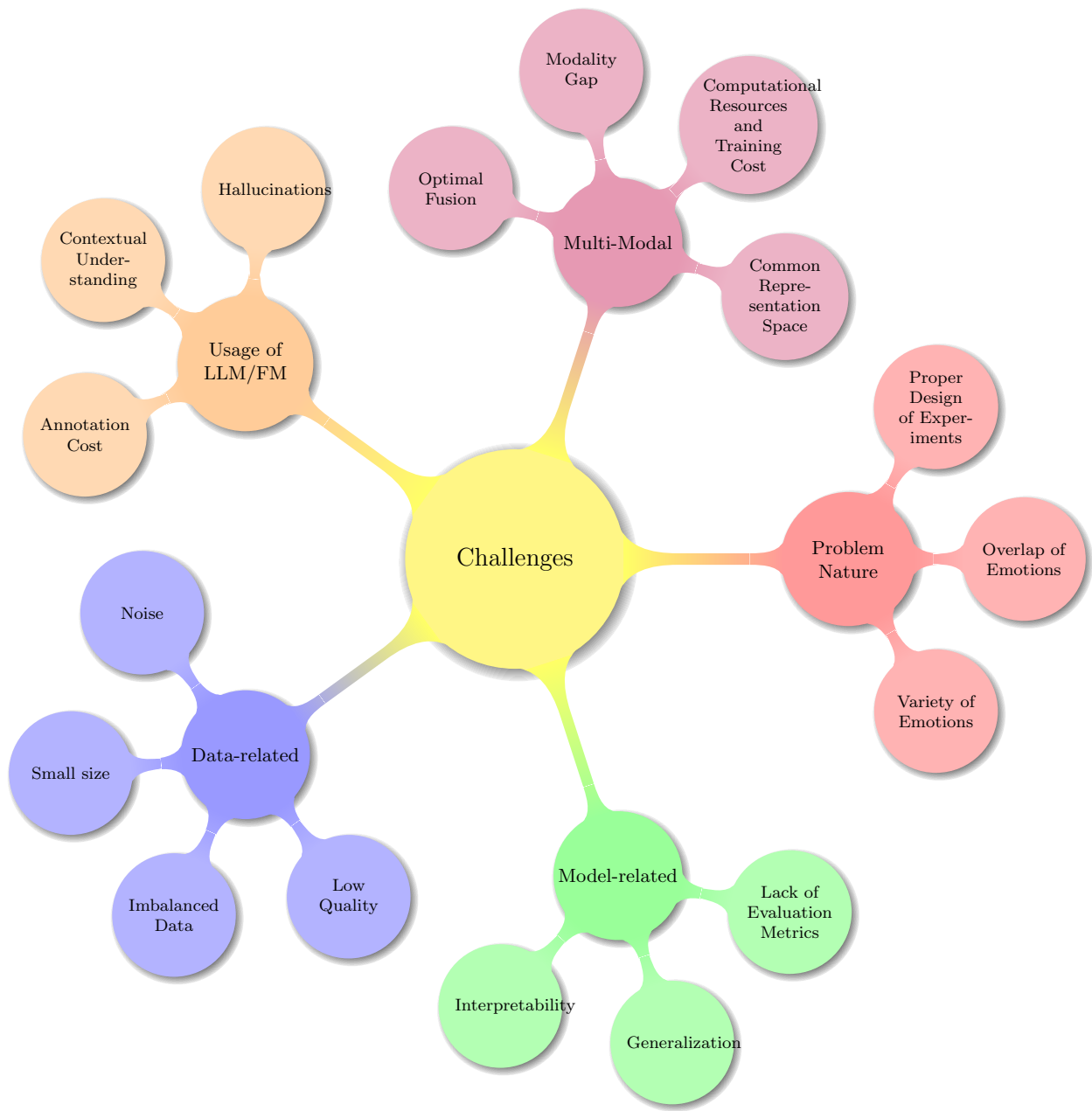
\begin{figure*}[t]
	\centering
	\resizebox{1\textwidth}{!}{
		\begin{tikzpicture}[mindmap, grow cyclic, every node/.style={concept, circular drop shadow, minimum size=2.0cm}, 
			concept color=yellow!60, text=black,level 1/.append style={level distance=5.5cm,sibling angle=72}]
			\node[concept] {Challenges}
			child[concept color=blue!40] { 
				node {Data-related}
				child[concept color=blue!30] { 
					node {Noise}
				}
				child[concept color=blue!30] { 
					node {Small size}
				}
				child[concept color=blue!30] { 
					node {Imbalanced Data}
				}
				child[concept color=blue!30] { 
					node {Low Quality}
				}
			}
			child[concept color=green!40] { 
				node {Model-related}
				child[concept color=green!30] { 
					node {Interpretability}
				}
				child[concept color=green!30] { 
					node {Generalization}
				}
				child[concept color=green!30] { 
					node {Lack of Evaluation Metrics}
				}
			}
			child[concept color=red!40] { 
				node {Problem Nature}
				child[concept color=red!30] { 
					node {Variety of Emotions}
				}
				child[concept color=red!30] { 
					node {Overlap of Emotions}
				}
				child[concept color=red!30] { 
					node {Proper Design of Experiments}
				}
			}
			child[concept color=purple!40] { 
				node {Multi-Modal}
				child[concept color=purple!30] { 
					node {Common Representation Space}
				}
				child[concept color=purple!30] { 
					node {Computational Resources and Training Cost}
				}
				child[concept color=purple!30] { 
					node {Modality Gap}
				}
				child[concept color=purple!30] { 
					node {Optimal Fusion}
				}
			}
			child[concept color=orange!40] { 
				node {Usage of LLM/FM}
				child[concept color=orange!30] { 
					node {Hallucinations}
				}
				child[concept color=orange!30] { 
					node {Contextual Understanding}
				}
				child[concept color=orange!30] { 
					node {Annotation Cost}
				}
			};
		\end{tikzpicture}
	}
	\caption{Challenges in emotion understanding }
	\label{fig:emotion_recognition_challenges}
\end{figure*}

In this section, we aim to address the existing challenges in emotion recognition and explore the solutions that have been proposed for each challenge. We will review relevant studies that address these issues and their suggested remedies. These challenges arise from the complex nature of human emotions and the limitations of current technologies used to process and interpret them. As depicted in Figure~\ref{fig:emotion_recognition_challenges}, we classify these challenges into three major domains: data, learning models, and problem nature. Table~\ref{tab:emotional_understanding} summarizes these categories, outlining their respective sub-issues and highlighting studies that address these challenges. Next, we will discuss the challenges and the methods proposed to overcome them.

\clearpage
\begin{longtable}{@{}p{2.5cm}p{2.8cm}p{9.5cm}@{}}
	\caption{Summary of challenges and solutions in emotional understanding}
	\label{tab:emotional_understanding} \\
	\toprule
	\textbf{Challenges} & \textbf{Sub-challenge} & \multicolumn{1}{c}{\textbf{Solutions}} \\
	\midrule
	\endfirsthead
	
	\caption[]{Summary of Challenges and Solutions in Emotional Understanding (Continued)} \\
	\toprule
	\textbf{Challenges} & \textbf{Sub-challenge} & \multicolumn{1}{c}{\textbf{Solutions}} \\
	\midrule
	\endhead
	
	\midrule
	\multicolumn{3}{r}{{Continued on next page}} \\
	\endfoot
	
	\bottomrule
	\endlastfoot
	
	\multirow{4}{2.5cm}{\textbf{Data-related}} & Small size of data & Semi-supervised graph contrastive learning \citep{ye2024semi}; GAN-based synthetic data generation \citep{schuller2024affective}; BERT-based augmentation techniques \citep{koufakou2023data} \\
	\cmidrule(lr){2-3}
	& Noise in data & Advanced data augmentation methods \citep{chowdary2023deep}; Uncertainty-aware multimodal fusion \citep{tellamekala2023cold}; Noise-robust CNN architectures \citep{oh2024noise} \\
	\cmidrule(lr){2-3}
	& Data imbalance & SMOTE and Tomek Links \citep{ghafourian2022facial}; GAN-based augmentation \citep{meng2024deep}; Cost-sensitive learning \citep{li2016real} \\
	\cmidrule(lr){2-3}
	& Low data quality & Multi-source transfer learning \citep{sarkar2023multi}; Probabilistic uncertainty modeling \citep{lo2023modeling}; Differential entropy features \citep{uyanik2022use} \\
	
	\midrule
	\multirow{4}{2.5cm}{\textbf{Learning-related}} & Low accuracy & Multi-modal integration \citep{umair2024emotion}; Ensemble learning methods \cite{younis2022evaluating}; Unsupervised representation learning \cite{ross2023unsupervised} \\
	\cmidrule(lr){2-3}
	& Interpretability & Attention mechanisms \cite{cortinas2023toward}; Grad-CAM visualization \cite{sharma2024emotion}; Neurosymbolic AI frameworks \cite{cambria2024senticnet}; Coarse-to-fine training \cite{lian2024affectgpt} \\
	\cmidrule(lr){2-3}
	& Standardized metrics & Zero-shot evaluation \cite{schuller2024affective}; Quantitative quality metrics \cite{langure2024improving} \\
	\cmidrule(lr){2-3}
	& Generalization & Subject-independent models \cite{younis2022evaluating}; Hybrid transformer approaches \cite{zanwar2022improving}; Emotion-specific pretraining \cite{aroyehun2023leia} \\
	
	\midrule
	\multirow{3}{2.5cm}{\textbf{Problem Nature}} & Variety of emotions & Graph contrastive learning \cite{ye2024semi}; Multi-modal gating mechanisms \cite{sharma2024emotion}; Sensory knowledge integration \cite{zhao2025leveraging} \\
	\cmidrule(lr){2-3}
	& Overlap of emotions & Uniform label annotation \cite{du2024unic}; Causal intervention \cite{yang2023context}; Graph attention networks \cite{li2023graphmft} \\
	\cmidrule(lr){2-3}
	& Experiment design & Real-world multimodal studies \cite{younis2022evaluating}; Cross-subject evaluation \cite{hu2021novel}; Continuous emotion assessment \cite{bota2019review} \\
	
	\midrule
	\multirow{4}{2.5cm}{\textbf{Multi-modal}} & Common representation & Rule-based conversion \cite{kumar2024multimodal}; Attention mechanisms \cite{geetha2024multimodal}; Semantic alignment networks \cite{ezzameli2023emotion} \\
	\cmidrule(lr){2-3}
	& Computational efficiency & Sparse cross-modal attention \cite{dai2021multimodal}; Hierarchical processing \cite{wei2022fv2es}; Multiplicative fusion \cite{mittal2020m3er} \\
	\cmidrule(lr){2-3}
	& Modality gap & Trimodal integration \cite{alsaadawi2024tac}; Cross-attention mechanisms \cite{rajan2022cross}; Contactless data fusion \cite{khan2024exploring} \\
	\cmidrule(lr){2-3}
	& Optimal fusion & Novel feature extractors \cite{middya2022deep}; Shifted Window Transformers \cite{kim2024enhancing}; Speaker-aware networks \cite{guo2024speaker} \\
	
	\midrule
	\multirow{3}{2.5cm}{\textbf{LLM/FM}} & Annotation cost & Synthetic data generation \cite{schuller2024affective}; LLM-based labeling \cite{li2024comparative}; Coarse-to-fine training \cite{lian2024affectgpt} \\
	\cmidrule(lr){2-3}
	& Contextual understanding & Enhanced prompting \cite{amin2024wide}; Multimodal datasets \cite{zhang2023dialoguellm}; Unified feature conversion \cite{kumar2024multimodal} \\
	\cmidrule(lr){2-3}
	& Hallucinations & Semantic entropy detection \cite{farquhar2024detecting}; Fact-checking methods \cite{sahoo2024comprehensive}; Reinforcement learning \cite{li2024dawn} \\
	
\end{longtable}
\subsubsection{Data-Related Challenges}
Challenges related to data availability and quality, including limited dataset sizes, noisy or low-quality samples, and class imbalance, significantly constrain the accuracy, robustness, and generalizability of emotion recognition systems. Addressing these challenges is critical for the development of reliable affective computing models.\\
\textbf{\textit{Small dataset size:}}
The limited size of datasets represents a major obstacle for emotion recognition models, as it restricts their ability to generalize across diverse emotional expressions and populations. Effective emotion recognition requires large-scale, diverse datasets to enable capturing the complexity and variability of human emotion. However, many existing datasets are small, domain-specific, and biased toward particular demographic groups, which undermines their practical applicability. To mitigate this issue, several studies have explored data-efficient learning and data augmentation strategies. For instance, \cite{ye2024semi} proposes a semi-supervised dual-stream self-attentive adversarial graph contrastive learning framework for EEG-based emotion recognition. By leveraging a small set of labeled samples alongside a larger pool of unlabeled data, this approach improves recognition performance under low-resource conditions. The dual-stream architecture effectively captures both structural and non-structural features, alleviating the limitations imposed by insufficient labeled data.\\
In a complementary direction, generative approaches have been investigated to expand dataset size and diversity. \cite{schuller2024affective} examines the use of generative techniques \cite{bond2021deep}, highlighting the role of adversarial learning and generative adversarial networks (GANs) \cite{hajarolasvadi2020generative} in synthesizing realistic affective data. By constructing large-scale synthetic datasets that approximate authentic emotional expressions, such methods can partially overcome the scarcity of annotated emotional data. In the context of textual emotion recognition, \cite{koufakou2023data} explores multiple data augmentation strategies to address both small dataset size and class imbalance. Their study applies contextual word replacement using BERT-based models \cite{devlin2019bert}, where words are substituted based on surrounding context to generate semantically coherent training samples. Additionally, synthetic data generation using pre-trained language models is employed to enhance dataset diversity and improve model generalization. Traditional augmentation techniques, such as Easy Data Augmentation (EDA) \cite{wei2019eda}, including synonym replacement, word insertion, deletion, and word shuffling, have also demonstrated effectiveness. By enriching datasets with balanced and diverse examples, these methods substantially improve the performance of deep learning models, like RoBERTa \cite{liu2019roberta}, especially in low-resource scenarios.\\
\textbf{\textit{Noise in data:}}
Noise or irrelevant information that disrupts the recognition process can affect emotion recognition systems, especially those relying on sensors or video recordings. Noise may arise from poor lighting, background interference in video data, or ambient noise in audio recordings. This noise complicates distinguishing relevant emotional cues from distracting information.  \cite{chowdary2023deep} addresses this issue by employing data augmentation techniques during the training of deep learning models. Data augmentation includes strategies such as translations, normalizations, cropping, adding noise, and scaling, which effectively increase the variety and quantity of training data. By incorporating these modifications, the model becomes more robust to variations and noise encountered in real-world data. This approach enhances the model's ability to generalize well across different scenarios, improving facial emotion recognition accuracy even in noise conditions. \cite{tellamekala2023cold} addresses noise in data, especially within the visual modality, by introducing an uncertainty-aware multimodal fusion approach that quantifies modality-wise aleatoric uncertainty. During the test phase, this paper simulates noise by applying face masking to half of the evaluation sequences. The proposed COLD fusion framework adapts to noise by dynamically estimating fusion weights. When the visual modality is obscured, COLD fusion increases reliance on the audio modality, effectively compensating for the loss of visual information. \cite{oh2024noise} presents a noise-robust deep learning model for emotion classification using facial expressions, addressing challenges posed by image noise, such as lighting variations, facial angles, and distortions. To make the model more intelligent and adaptable, it applies data augmentation tricks including image rotations, brightness shifts, and strategic cropping, mimicking real-world variations. State-of-the-art deep networks, such as CNNs \cite{o2015introduction}, ResNet \cite{he2016deep}, and VGG \cite{simonyan2014very}, enhanced with attention mechanisms \cite{vaswani2017attention}, facilitate the model to focus on the most informative facial regions, even under poor conditions. By integrating transfer learning with auto-encoder-based noise reduction, the model effectively learns from high-quality features while filtering out unwanted distortions. This boosts accuracy and enhances the reliability of emotion recognition in noisy environments.\\
\textbf{\textit{Data imbalance:}}
Another challenge in emotion recognition is data imbalance, where common emotions, such as happiness, are more prevalent than rare emotions like surprise. This imbalance can lead to model bias, resulting in better performance for recognizing common emotions while struggling with less frequent ones. Data imbalance can undermine the stability and fairness of emotion recognition systems. \cite{ghafourian2022facial} employs a combination of SMOTE (Synthetic Minority Over-sampling Technique)\cite{chawla2002smote} and Tomek Links. SMOTE generates synthetic samples for underrepresented emotions by interpolating between existing minority-class samples, ensuring more diverse and realistic training data. Meanwhile, Tomek Links removes borderline majority-class samples that may confuse the model, improving class separation. This dual approach balances the dataset without duplicating data, enabling deep learning models such as VGG-16 and ResNet-50 to learn more effectively from all emotion categories and improving overall classification accuracy. Data augmentation using GANs \cite{hajarolasvadi2020generative} produces new samples for less common emotional categories, which improves data distribution \cite{meng2024deep}. This paper employs sampling strategies to promote balanced learning and avoid incorrect conclusions that can result from imbalanced data. Another strategy is loss sensitivity, which involves designing a class-sensitive loss function that pays more attention to less prominent classes. Finally, feature integration is utilized, where a deep variational auto-encoder merges complementary semantic information from multiple modalities, such as text, audio, and images, to enhance emotion recognition. \cite{li2016real} reviews and compares several methods in imbalanced learning, which are categorized into algorithmic modification techniques and data distribution modification methods. Algorithmic approaches include Cost-Sensitive Learning (SVM-Weight)\cite{iranmehr2019cost}, which assigns higher misclassification penalties to minority emotions. This ensures that models pay more attention to these underrepresented categories. Data distribution approaches involve under-sampling to remove excess samples from majority classes and Over-Sampling (Random Over-Sampling and SMOTE) to generate new samples for minority classes synthetically. \cite{iranmehr2019cost} introduces virtual facial sample generation (VFSG), a novel technique that enhances data diversity by creating synthetic facial images with variations in lighting and angles, offering a more realistic and robust solution than standard over-sampling methods.\\\\
\textbf{\textit{Low data quality:}}
Low-quality data, such as blurry images, can harm the performance of emotion recognition systems. These algorithms rely on high-quality and precise data, so any decrease in quality can result in incorrect predictions and reduce the overall performance of the model. To tackle this issue, \cite{sarkar2023multi} incorporates multi-source transfer learning (MSTL) \cite{lee2019learning} combined with multivariate correlation analysis (MCA)\cite{abdi2003multivariate} for facial emotion recognition. This method focuses on extracting relevant and high-quality features from multiple pre-trained models that have been trained on diverse, high-quality datasets, rather than relying solely on noisy or low-quality target data. The MCA technique selects only the most correlated and valuable features, which helps minimize the negative effects of noisy or low-resolution images. The approach also reduces the risk of negative transfer, ensuring that only beneficial knowledge from source domains is transferred, while avoiding the inclusion of noisy or irrelevant patterns. This technique enhances model performance in low-data and few-shot scenarios by effectively integrating knowledge from multiple data sources. It enables offsetting the poor quality of the target dataset without needing explicit data denoising or super-resolution methods. \cite{lo2023modeling} proposes probabilistic data uncertainty learning, which models the uncertainty caused by poor resolution. They use the Emotion Wheel theory to handle label ambiguity by representing emotions in a continuous space, which enables the model to better distinguish between similar expressions. These methods contribute to improving recognition accuracy and uncertainty estimation in low-resolution settings. \cite{uyanik2022use} addresses the challenge of low-quality EEG data by leveraging Differential Entropy (DE)\cite{duan2013differential} as a robust feature extraction method, which is less sensitive to noise and variations in the signal. It also applies preprocessing techniques, such as band-pass filtering, to remove unwanted frequencies and reduce artifacts. To mitigate the impact of noisy data, the study employs machine learning models such as Support Vector Machine (SVM) \cite{hearst1998support} and Neural Networks, which can generalize and filter out unreliable patterns from the dataset. Combining these techniques enhances the accuracy of automated emotion recognition in virtual reality environments, even when dealing with low-quality EEG signals.
\subsubsection{Model-Related Challenges}
Learning models are crucial for emotion recognition, as they map the given input, such as facial expressions, vocal cues, or physiological measurements, into corresponding emotional states. However, developing robust models for this task involves several challenges. \\\\
\textbf{\textit{Low accuracy:}}
Low accuracy in emotion recognition models often stems from overfitting, where models that excel on training data struggle to generalize to unseen samples. This problem is especially common when training on small or imbalanced datasets, which often fail to capture the full complexity and diversity of human emotions. Relying on a single data modality (such as facial movements or text) limits a model’s ability to capture the multifaceted nature of emotions and can restrict accuracy. To combat this, researchers are developing innovative methods to boost the performance of emotion recognition systems, including multimodal fusion, more diverse datasets, robust learning strategies, and domain adaptation techniques. \cite{umair2024emotion} fuses multiple data streams, including facial expressions, voice, and text, to overcome the limitations of unimodal systems.  Ensemble learning methods have been utilized, as noted in \cite{younis2022evaluating}, which have significantly improved accuracy. These solutions highlight the importance of integrating diverse methods and ensemble techniques to boost performance. \cite{ross2023unsupervised} addresses the challenge of low accuracy by proposing an unsupervised multi-modal representation learning framework by utilizing stacked convolutional auto-encoders to learn latent representations from wearable bio-signals, specifically electrocardiogram (ECG) and electrodermal activity (EDA) data.\\\\
\textbf{\textit{Interpretability:}}
Understanding decision-making in complex deep learning models remains a significant challenge, particularly in sensitive domains such as healthcare and psychological analysis, where trust and transparency are critical.
Deep learning architectures, such as CNNs \cite{o2015introduction} and Recurrent Neural Networks (RNNs) \cite{schmidt2019recurrent}, are referred to as "black boxes" because their internal reasoning is difficult to interpret. While these models can produce acceptable results, explaining how or why a particular prediction was made remains difficult. This lack of transparency can be problematic, especially in applications where clarity is essential, such as medical or psychological contexts. To address this, \cite{cortinas2023toward} categorizes explainability approaches into three types: Pre-model (before training), In-model (during training), and Post-model (after training). Pre-model approaches focus on designing meaningful features and reducing data complexity. In contrast, in-model methods aim to develop architectures that are inherently interpretable, such as attention mechanisms\cite{vaswani2017attention} that highlight salient portions of the input. Post-model techniques, such as Grad-CAM\cite{selvaraju2017grad} or sensitivity analysis, enable visualization and analysis of model decisions after training. \cite{sharma2024emotion} employs a mechanism called Grad-CAM (Gradient-weighted Class Activation Mapping) to analyze and visualize the attention maps related to the predictions of the model. This method highlights the regions in meme images that the model references when making emotional classifications. This interpretative framework demonstrates how the model correlates visual elements of memes with their associated emotions. It provides insights into which aspects of a meme, such as facial expressions or contextual elements, contribute to the predicted emotions. \cite{cambria2024senticnet} introduces a neurosymbolic AI framework that combines commonsense knowledge representation with hierarchical attention networks to enhance transparency in affective computing. Unlike black-box deep learning models, this framework employs a three-step normalization process, including syntactic and pragmatic, to map input text to interpretable conceptual primitives. This approach ensures that the decision-making process is both understandable and explainable. By combining symbolic AI, which utilizes structured reasoning through commonsense knowledge graphs, with sub-symbolic AI that employs pattern recognition via deep learning, the model achieves traceable and trustworthy sentiment classification and personality prediction. \cite{lian2024affectgpt} tackles the challenge of interpretability in Explainable Multimodal Emotion Recognition (EMER) through a two-stage training framework. By integrating multimodal inputs, including audio, video, and text, the approach strengthens the link between the inputs and the predicted emotions, enabling more transparent reasoning about the outputs. In the first stage, the model learns a coarse mapping using a large-scale, coarsely labeled dataset (EMER-Coarse), which facilitates the identification of general emotional trends. In the second stage, the model utilizes a smaller, manually checked dataset (EMER-Fine) to refine these mappings, ensuring better alignment with reliable human-annotated labels. This systematic method not only improves emotion recognition accuracy but also provides a framework for tracing predictions to the supporting multimodal evidence, enhancing the model’s interpretability.\\\\
\textbf{\textit{Lack of standardized evaluation metrics:}}
The lack of standardized evaluation metrics for emotion recognition complicates the comparison of results across various studies and hinders the assessment of the actual effectiveness of proposed models. This inconsistency can lead to uneven reporting, hindering meaningful cross-study comparisons. \cite{schuller2024affective} discusses the challenges arising from the absence of standardized evaluation metrics in affective computing, particularly in assessing the capabilities of foundation models (FMs). This paper advocates for developing new methods and metrics that provide a rigorous scientific evaluation of these emerging models. It emphasizes the need for comparative benchmarks, such as the use of established datasets for emotion recognition tasks. Additionally, it highlights the importance of refined approaches, like zero-shot classification, to ensure that model performance can be assessed in a meaningful way. Similarly, \cite{langure2024improving} addresses the lack of standardized evaluation metrics in text emotion detection (TED). The study proposes a comprehensive framework for assessing dataset quality based on 14 quantitative metrics across four key dimensions: representativity, readability, structure, and part-of-speech (POS) tag distribution. By systematically measuring factors such as data balance, linguistic complexity, lexical diversity, and syntactic structure (POS tag distribution), the framework ensures that datasets are evaluated using consistent and reproducible criteria. It demonstrates that variations in these quality metrics impact model performance, reinforcing the need for standardization in TED research. By introducing these metrics and validating their influence through experiments with BiLSTM \cite{huang2015bidirectional} and Bert models \cite{devlin2019bert}, the paper establishes a foundation for more rigorous dataset assessment, enabling more reliable and comparable TED model evaluations.\\\\
\textbf{\textit{Generalization:}}
A critical challenge in emotion recognition systems lies in their limited ability to generalize across diverse domains, including cultural variations, environmental contexts, and different input modalities (e.g., speech versus facial expressions). This domain shift problem often leads to performance degradation when models trained on one dataset are applied to new scenarios. Recent work by \cite{younis2022evaluating} emphasizes the need for subject-independent predictive models to improve cross-domain robustness. These models do not rely on the unique characteristics of specific individuals, making them more versatile for real-world applications. They are designed to identify emotions across different populations with varying physiological and psychological characteristics, enhancing their utility in diverse settings. In \cite{zanwar2022improving}, a hybrid approach is proposed that combines transformer-based models, such as Bert \cite{devlin2019bert} and RoBERTa \cite{liu2019roberta}, with psycholinguistic features extracted using BiLSTM \cite{huang2015bidirectional} networks. This combination improves the model’s generalization in unseen datasets by leveraging state-of-the-art language representations and deeper contextual information. Similarly, in \cite{aroyehun2023leia}, a novel pre-training method called eMLM (emotion-specific Masked Language Model) is introduced. This method masks emotion-related words during training, encouraging the model to learn broader contextual relationships rather than relying solely on specific terms. Combining eMLM with fine-tuning allows the LEIA model to adapt effectively to domain shifts and perform well on new datasets. \cite{ross2023unsupervised} reduces the reliance on human-annotated labels, enabling the aggregation of multiple datasets into a more extensive and diverse training set. This approach enhances the model's generalization ability across different emotional contexts, improving classification performance. The results demonstrated that the proposed method achieved state-of-the-art accuracy across multiple datasets compared to baseline techniques, illustrating its effectiveness in overcoming the limitations associated with lower accuracy in existing models.\\
\subsubsection{Problem Nature Challenges}
The nature of the problem in emotion recognition poses significant challenges, as emotions are inherently subjective, overlapping, and context-dependent. Unlike fixed categorical tasks, emotions exist on a continuous spectrum, making it difficult to establish clear distinctions between them. Additionally, emotional expressions can vary widely between individuals and cultures, and environmental factors influence them, leading to inconsistencies in labeling. The presence of multiple modalities adds another layer of complexity, as different modalities can convey conflicting emotional cues. These factors make emotion recognition a complex task that requires adaptive models capable of handling ambiguity, variability, and contextual dependencies.\\\\
\textbf{\textit{Variety of emotions:}}
The variety of human emotions poses a significant challenge for emotion recognition systems, which need to address the complex range of emotional states that individuals experience. While basic emotions such as happiness, sadness, anger, and fear serve as foundational categories, human emotional experiences extend beyond these simple labels. Furthermore, emotions can manifest differently in each individual due to factors like personality, context, and socio-cultural norms. These complexities make it difficult to distinguish between emotions. Consequently, emotion recognition systems risk oversimplifying and misinterpreting the depth and context of human emotions.
\cite{ye2024semi} proposes a semi-supervised learning framework designed specifically for EEG-based emotion recognition. This framework employs graph contrastive learning \cite{NEURIPS2020_3fe23034} to enhance the model's ability to differentiate between various emotions. The framework leverages advanced learning strategies to tackle emotional diversity, enabling it to more accurately detect and represent emotions in EEG signals, even when labeled data are scarce. \cite{sharma2024emotion} proposes a multimodal neural framework designed to tackle the challenge of emotional diversity in understanding human emotions conveyed through memes. This framework specifically models enhanced visual cues related to emotions and utilizes a gating mechanism for effective integration of multiple modalities, ultimately improving emotion recognition in memes. \cite{zhao2025leveraging} presents a model that integrates sensory knowledge into the T5 framework \cite{xue2020mt5} to enhance emotion classification. This model embeds sensory information within the attention mechanism, which enhances contextual understanding and increases sensitivity to subtle emotional states. Notably, the model employs an adapter approach that facilitates the joint training of contextual and sensory information through a unified loss function.\\\\
\textbf{\textit{Overlap of emotions:}}
The overlap and coexistence of emotions complicate the accurate identification and classification of emotional states. For example, emotions such as fear and surprise share similar physiological responses and facial expressions, making it difficult to discern where one emotion ends and another begins. The subjective nature of emotional experiences, the dynamic shifts between different emotional states, and cultural differences in emotional expression heighten this ambiguity.
\cite{du2024unic} presents a dataset featuring uniformly independent labels across different modalities, which enables a more accurate understanding of emotions and tackles the challenges posed by mixed and overlapping emotions in emotion recognition. This approach addresses the contradictions frequently encountered in emotional expression. It implements robust annotation methods and evaluates inter-annotator agreement \cite{braylan2022measuring} to ensure the reliable labeling of mixed emotions. These improvements enhance the dataset's quality for training emotion recognition models, thereby increasing the technology's capability to interpret complex human emotions. \cite{yang2023context} offers a causal inference-based approach using the Contextual Causal Intervention Module (CCIM) to remove spurious correlations between context and emotion labels. Traditional emotion recognition models often misclassify emotions due to context bias, where certain emotions are disproportionately associated with specific backgrounds. The CCIM mitigates this by applying causal intervention to disentangle genuine emotional cues from misleading contextual dependencies. It constructs a confounded dictionary that clusters contextual features and re-weights their influence using a backdoor adjustment technique. This ensures emotions are classified based on intrinsic expressions rather than external biases. GraphMFT\cite{li2023graphmft} leverages a graph-based multi-modal fusion approach that effectively integrates intra-modal and inter-modal contextual information. GraphMFT develops three separate heterogeneous graphs: Visual-Acoustic, Visual-Textual, and Acoustic-Textual. This structure facilitates a more refined and adaptive fusion of different modalities. The model employs graph attention networks (GATs) to assign distinct importance weights to contextual and cross-modal relationships, allowing it to capture nuanced differences between emotions such as "frustration" and "anger" or "excitement" and "happiness." The improved GAT architecture also mitigates the over-smoothing problem often found in deep graph networks, preserving distinct emotional features. By incorporating speaker embeddings and learning dynamic interactions across different modalities, the model reduces the misclassification of overlapping emotions, thereby improving the overall accuracy of emotion detection.\\\\
\textbf{\textit{Proper design of experiments:}}
Proper experimental design significantly affects the reliability, generalizability, and accuracy of resulting models. It is essential to consider the diversity and complexity of human emotions, which can vary widely among individuals and cultures. Poorly designed experiments can lead to models that are inaccurate or biased due to cultural differences. 
Replicating real-world emotional dynamics in controlled settings is inherently challenging, as emotions are shaped by numerous complex and often unpredictable factors. Addressing these challenges necessitates careful attention to these elements to ensure that trained models accurately represent real-world scenarios and are applicable in practice. In this context, the article \cite{younis2022evaluating} conducts a real-world study that captures participants' natural emotional responses in their environment. It develops subject-independent models by integrating multi-modal data, combining physiological and environmental data to create a comprehensive dataset. 
Traditional emotion recognition models typically use predefined video clips to elicit emotions, so all participants watch the same videos.  This setup can introduce unwanted noise patterns in the data, leading models to focus on features related to the video content instead of capturing authentic emotional responses. To address this challenge, \cite{hu2021novel} introduces a novel experiment setup that mitigates the impact of stimulus materials on classification accuracy in cross-subject studies. The paper proposes an innovative approach to data partitioning, where the training and testing sets are sourced from different video-induced datasets, instead of using multiple participants who are watching the same videos. This method ensures that the model learns generalizable emotion patterns and reduces the risk of overfitting to specific stimuli. The study evaluates this approach using public emotion datasets and demonstrates that it enhances cross-subject emotion recognition. As a result, the findings are more reliable and applicable to real-world scenarios. \cite{bota2019review} addresses the challenges of designing experiments for emotion recognition, especially when comparing controlled laboratory settings to real-life situations. It identifies several issues, such as the difficulty in accurately annotating emotional responses in unconstrained environments, the degradation of signal quality due to noise from uncontrolled variables, and the significant impact of individual factors like mood and cultural background. To tackle these challenges, the authors suggest focusing on unconstrained scenarios that feature a diverse pool of participants and conducting continuous evaluations of emotional responses. This approach aims to enhance the validity, reliability, and generalizability of findings in emotion recognition research.\\
\subsubsection{Multi-Modal Challenges}
Fusing multiple modalities is a crucial step in developing robust emotion understanding models. These models integrate diverse data sources, such as audio, visual, and textual cues, each offering distinct insights into emotional states. Effective fusion requires precise alignment of modalities to preserve their individual contributions, a process that can be computationally costly, especially with large-scale datasets. Furthermore, the inherent differences in how each modality encodes emotional information can complicate the optimization of their interactions. To ensure these sources complement rather than conflict with one another, fusion strategies must be carefully designed to balance performance and efficiency. Addressing these challenges demands advanced architectures and substantial computational resources, ultimately enhancing the system’s ability to generalize across varied emotional expressions and contexts.\\\\
\textbf{\textit{Common representation space:}}
Achieving a common representation space in emotion recognition requires integrating diverse modalities, each capturing emotions in distinct ways. These modalities do not naturally align, creating gaps that make unifying them into a single representation challenging. Moreover, the variability of emotional expressions across individuals and cultures adds another layer of complexity, demanding advanced techniques to align these heterogeneous data sources effectively for accurate emotion recognition. \cite{kumar2024multimodal} developed a rule-based system that converts non-verbal cues into text to address this issue. These textual representations are combined into prompts and processed by a pre-trained LLM, allowing for more effective emotion recognition. This approach facilitates multimodal integration while maintaining flexibility to incorporate additional modalities in the future.   \cite{geetha2024multimodal} highlights the challenges associated with integrating various emotional cues, which often have different feature representations and temporal dynamics. It discusses advanced deep learning techniques, such as attention mechanisms and transformers, that can effectively focus on relevant features from multiple modalities to create a unified representation. Moreover, specialized architectures improve the interpretation of complex emotional states by incorporating spatiotemporal context. These approaches offer potential solutions for achieving a cohesive common representation in emotion recognition systems. \cite{ezzameli2023emotion} employs a fusion and alignment method to create a shared representation that effectively associates different modalities, which is essential for multi-modal learning. It reviews various fusion techniques, including feature-level, decision-level, and model-level fusion, highlighting the importance of identifying relationships between different modalities and integrating data to enhance recognition accuracy.
The Semantic Alignment Network (SAN)\cite{hou2023semantic} utilizes a Cross-Modal Alignment (CMA) module that projects heterogeneous features into a unified semantic embedding space, reducing semantic discrepancies and enabling effective fusion of emotionally relevant information. SAN further uses attention mechanisms to dynamically weight each modality’s contribution, improving alignment precision and the recognition of subtle emotional nuances.\\\\
\textbf{\textit{Computational resource and training cost:}}
Emotion recognition systems often process data from multiple sources, such as audio, video, and physiological signals, which requires substantial computational resources for feature extraction, data alignment, and model training. Training advanced algorithms or deep learning models on such large and complex datasets is both time- and resource-intensive, demanding high-performance hardware. These requirements significantly increase financial costs and development time, posing challenges to the scalability and widespread adoption of emotion recognition systems. \cite{dai2021multimodal} addresses this issue in multimodal emotion recognition by proposing a fully end-to-end model that integrates feature extraction and emotion recognition into a single pipeline, eliminating the need for hand-crafted feature engineering. 
To enhance efficiency, it introduces a sparse cross-modal attention mechanism, which selectively focuses on the most relevant features from different modalities, reducing redundant computations and unnecessary processing. Additionally, the model is optimized to learn directly from raw multi-modal data, removing the dependency on separately pre-processed datasets and reducing training overhead. This method effectively lowers computational costs while preserving high accuracy, making it well-suited for real-time applications and environments with limited resources. 
A fully end-to-end system for fast and efficient video-based emotion recognition is proposed in \cite{wei2022fv2es}. This paper incorporates a hierarchical attention mechanism that enhances the contribution of the audio modality without significantly increasing computational overhead,  thereby optimizing computational efficiency. Additionally, the authors introduce a single-branch inference module for visual processing, which replaces traditional multi-branch architectures with a simplified yet effective structure. This change reduces both computational complexity and storage requirements. By integrating data preprocessing and multi-modal learning into a unified system, this system eliminates redundant computations and speeds up inference. \cite{mittal2020m3er} presents a multiplicative fusion strategy that adaptively adjusts the influence of each modality based on its reliability, thereby reducing redundant computations and avoiding the processing of noisy or unreliable data. By suppressing less informative modalities and prioritizing the most reliable ones, the model maintains high accuracy while optimizing computational efficiency, leading to significant reductions in both training and inference costs.
\cite{Wu2025} adopts a prompt learning approach, which enables efficient fine-tuning with fewer labeled data and eliminates the need for extensive training. Task-specific prompts guide pre-trained models, substantially decreasing computation time and memory usage. Furthermore, their text–audio fusion strategy improves feature integration, reduces redundancy, and enhances inference efficiency.
\\\\
\textbf{\textit{Modality Gap:}}
The modality gap refers to the differences and mismatches between various data sources, such as audio, video, and physiological signals, each capturing emotional information in distinct ways. These variations arise from factors such as sensor limitations, environmental conditions, and the inherent characteristics of each modality. For instance, facial expressions may offer clear visual cues, while voice tone or heart rate might convey subtler or complementary aspects of emotion. This gap poses a challenge, as features extracted from different modalities may not align well or may even conflict, hindering effective fusion for accurate emotion recognition. Bridging the modality gap requires advanced techniques to align, integrate, and harmonize heterogeneous data sources into a unified and reliable representation.
\cite{alsaadawi2024tac} addresses this issue by advocating for a trimodal affective computing approach that integrates textual, vocal, and visual data to enhance the accuracy and understanding of emotions. It outlines various data fusion strategies for combining features from multiple modalities, bridging their differences. It demonstrates improvements in emotion classification performance by employing advanced algorithms designed to process these diverse cues simultaneously. \cite{khan2024exploring} investigates contactless data collection methods and proposes better solutions to integrate these diverse modalities. It focuses on bridging the modality gap by developing frameworks for effectively combining different sensing techniques, emphasizing the need for fusion strategies to handle the disparities in the type, quality, and granularity of data from various sources. The paper presents novel methods to align and integrate multimodal cues by leveraging the strengths and addressing the weaknesses of each modality, ensuring meaningful contributions from all modalities and enhancing the overall accuracy and robustness of emotion recognition systems. The performance of cross-attention and self-attention mechanisms in integrating multimodal data is compared in \cite{rajan2022cross}. This paper examines how attention mechanisms address discrepancies across modalities. While self-attention captures within-modal dependencies, cross-attention facilitates interactions between modalities, effectively aligning features from multiple sources. The paper demonstrates that cross-attention is more effective in overcoming modality gaps by enabling the model to learn cross-modal relationships.\\\\
\textbf{\textit{Optimal fusion:}}
Optimal multimodal fusion aims to integrate diverse data modalities to maximize emotion classification accuracy. Each modality provides unique emotional cues, but its reliability varies by context, and modalities may not always align. The goal is to identify the most effective stage, including early, intermediate, or later, in the processing pipeline for fusing these modalities to optimize system performance.
\cite{middya2022deep} proposes a model-level fusion approach that employs specialized feature extractor networks for audio and video data to derive relevant features before integration into a cohesive multimodal model. The study systematically evaluates various combinations of these extractors, analyzing their performance across diverse configurations to determine the optimal synergy between audio and video features. This methodical approach significantly improves emotion recognition accuracy. It highlights the significance of choosing the right combination of modalities and their features to improve overall classification results. \cite{kim2024enhancing} employs a shifted window transformer encoder combined with symmetric cross-attention mechanisms to model complex interactions among diverse data modalities. Physiological signals are transformed into 2D images, enabling effective extraction of spatial and temporal features. The model incorporates metadata, such as environmental conditions and personal traits, to enhance the emotional context. This integrative approach facilitates adaptation to individual variability, thereby improving classification accuracy. \cite{guo2024speaker} proposes a speaker-aware cognitive network with cross-modal attention to address optimal fusion in emotion recognition, effectively integrating multimodal data from text, audio, and video sources while incorporating speaker-specific information. The model employs a cross-modal attention fusion module to synergistically combine features across modalities, capturing their complementary information. Subsequently, a GRU-based cognitive network module simulates conversational dynamics and leverages speaker-specific data to extract richer emotional cues.
\subsubsection{Usage of LLMs/FMs}
The application of foundation models and LLMs to emotion recognition presents significant challenges due to the nuanced and diverse nature of emotional cues. Although these models excel at processing vast datasets and performing across multiple tasks, their generalization to diverse emotional expressions, particularly in complex social or cultural contexts, remains inconsistent. Unlike traditional emotion recognition models tailored to specific datasets or tasks, foundation models and LLMs are designed for broad applicability, which can compromise their precision in capturing subtle emotional nuances. Fine-tuning these models for emotion recognition is challenging, as it requires balancing general knowledge with context-specific adaptability to achieve high accuracy. To address this, techniques such as domain-specific pretraining, transfer learning with emotionally rich datasets, and the integration of multimodal data (e.g., text, audio, and visual cues) can enhance model performance. Additionally, incorporating cultural and social metadata into training pipelines may improve the models’ ability to interpret diverse emotional expressions, thereby advancing their effectiveness in real-world applications.\\\\
\textbf{\textit{Annotation cost}}
The high cost of data annotation poses a significant challenge in applying foundation models(FMs) and LLMs to emotion recognition. These models require extensive labeled datasets that capture the diversity of emotional expressions across varied contexts, cultures, and modalities. Annotating such datasets is labor-intensive and costly, as it involves human experts labeling nuanced emotional cues in multimodal data, such as text, audio, and video. Moreover, maintaining and updating these datasets to reflect evolving emotional expressions and emerging contexts further escalates costs. This challenge limits the accessibility and scalability of foundation models for emotion recognition, hindering their widespread adoption. \cite{schuller2024affective} uses foundation models in emotion recognition to reduce reliance on specialized, annotated affective data. Large multimodal FMs, pretrained on diverse, extensive datasets, achieve robust emotion recognition performance without requiring extensive labeled datasets. This approach significantly lowers the cost and effort associated with manually annotating emotional data for training affective computing models. Furthermore, FMs can generate synthetic emotional data, further decreasing dependency on human-annotated datasets. \cite{li2024comparative} investigates whether LLMs can serve as a cost-effective alternative to traditional human-based crowdsourcing for data annotation. The study compares the quality of individual annotations from human crowd workers and LLMs, evaluating various label aggregation methods to enhance overall annotation quality. Findings indicate that integrating LLM-generated labels with crowdsourced data improves the quality of aggregated annotations while reducing costs. This approach offers a scalable solution for producing high-quality labeled datasets for tasks such as emotion recognition. EMER-Coarse dataset \cite{lian2024affectgpt} consists of large-scale, coarsely labeled data for emotion recognition tasks. To tackle the high costs of manual annotation, the framework replaces costly closed-source models with open-source alternatives, streamlining the pre-labeling process and eliminating the need for manual verification. The AffectGPT framework leverages this dataset in a dual-stage training approach. In the first stage, it learns general alignments between multimodal inputs and emotion-related descriptions using the coarsely labeled data. In the second stage, the model is fine-tuned with a smaller, finely labeled dataset to enhance accuracy. This approach significantly reduces annotation effort and costs while maintaining high-quality performance in emotion recognition. \\\\
\textbf{\textit{Contextual understanding:}}
Foundation models and LLMs face challenges in recognizing emotions within nuanced contexts. Emotional expressions are strongly influenced by factors such as speaker background, social environment, and cultural norms, which these models often inadequately represent. Despite their large-scale data processing strengths, LLMs may misinterpret emotional cues, leading to inaccurate recognition in complex situations.
\cite{amin2024wide} examines ChatGPT models on multiple affective computing tasks, such as sentiment analysis and opinion extraction, focusing on their ability to understand contextual factors in emotional expressions. To enhance this capability, the study introduces a novel prompting framework and reformulates regression tasks as pair-wise ranking classification tasks. This structured methodology enables a robust assessment of ChatGPT’s performance in discerning subtle emotional cues in text, improving contextual understanding. \cite{lu2024gpt} demonstrates that GPT-4V struggles to accurately recognize certain facial expressions and emotions without sufficient contextual information, particularly in complex emotional states such as fear or subjective tasks. The lack of adequate contextual information often results in errors or omissions in emotion recognition, underscoring the importance of effective contextual modeling in affective computing. DialogueLLM\cite{zhang2023dialoguellm} is an emotion and context knowledge-enhanced Large Language Model (LLM) designed explicitly for emotion recognition in conversations (ERC). The model is fine-tuned on multimodal datasets, including text and video, to leverage contextual cues and emotional relationships within conversations. This approach enhances DialogueLLM’s ability to accurately detect emotions by integrating conversational context and visual cues, improving its performance across diverse scenarios. \cite{kumar2024multimodal} proposes a unified feature representation approach to enhance contextual understanding in LLMs for emotion recognition. By converting audio and visual features into a standardized text format using rule-based systems before inputting them into the LLM, the model effectively captures the full emotional context across modalities. This unification ensures consistent processing of diverse emotional cues, reducing misinterpretations and improving the LLM’s ability to recognize complex emotions in varied contexts. \\\\

\textbf{\textit{Hallucinations:}}
Hallucinations in LLMs pose a significant challenge in emotion recognition, as models may generate incorrect emotional labels or infer spurious emotional cues not present in the input data. For instance, an LLM might lose an emotion based on learned patterns that are irrelevant or contradictory to the actual emotional state. This issue stems from LLMs’ tendency to draw on unrelated or erroneous associations within their extensive training data, leading to inaccurate predictions. Such hallucinations undermine the reliability of emotion recognition systems, particularly in sensitive applications like mental health assessment, where precision is paramount. 
\cite{sahoo2024comprehensive} provides a comprehensive analysis of hallucinations in LLMs, categorizing them into contextual disconnection, semantic distortion, and factual inaccuracies. The study proposes detection and mitigation strategies, including self-checking mechanisms, fact-checking techniques, and data augmentation, to reduce hallucination risks. It also explores fine-tuning, prompt engineering, and consistency checking to enhance the factual reliability of LLMs. By establishing a structured taxonomy of hallucination types and corresponding detection methods, the paper lays a foundation for developing robust approaches to mitigate hallucinations across diverse domains, including affective computing. \cite{farquhar2024detecting} introduces a novel semantic entropy-based approach to detect confabulations, a specific type of hallucination in LLMs. The method measures semantic uncertainty by clustering model responses based on their semantic meaning, rather than lexical variations, and computing entropy across these clusters. High entropy reveals potential hallucinations in LLMs. Entropy indicates greater uncertainty, correlating with unreliable outputs, such as confabulations. This approach enhances the accuracy and reliability of LLMs in question-answering systems by identifying potentially erroneous outputs, and its task-agnostic nature makes it versatile across various domains, including affective computing.
\cite{li2024dawn} systematically studies the sources, detection, and mitigation of factuality hallucinations. It introduces a comprehensive approach using the HaluEval 2.0 benchmark, which includes 8,770 questions from various domains, including biomedicine, finance, science, education, and open domain. The paper proposes a practical framework for detecting hallucinations by extracting factual statements from model responses and then evaluating their correctness. It explores factors contributing to hallucinations, such as pre-training data, fine-tuning, prompt design, and inference methods. To mitigate hallucinations, the paper tests techniques like reinforcement learning from human feedback (RLHF), retrieval augmentation, and self-reflection, finding that RLHF significantly reduces hallucinations, especially in open-domain and biomedicine contexts.

\section{Affective Cognition }
\label{cognition}
The Affective Theory of Mind extends beyond emotion recognition, requiring reasoning about emotions and responding appropriately. Achieving this requires not only emotional recognition but also internal modeling of affective strategies and context-appropriate expressive behavior \cite{Raggioli2025}. Although current affective computing primarily addresses emotion recognition and expression synthesis, advancing artificial emotional intelligence depends on modeling both emotional elicitation and experience \cite{zall2022comparative}. Thus, integrating cognitive and affective theories of mind is essential to enable affective cognition. This process involves two central components: first, identification of events and cognitive states that elicit emotions, and second, recognition of behaviors and cognitive states shaped by those emotions. This paper examines affective cognition from varied perspectives, highlighting its challenges and discussing approaches for progressing artificial emotional intelligence.
\subsection{Approaches}
\begin{figure*}[t]
	\centering
	\resizebox{1\textwidth}{!}{
		\begin{tikzpicture}[
			node distance=10mm,
			inner/.style={
				rectangle, 
				draw=blue!70!black, 
				fill=blue!10, 
				rounded corners=3pt, 
				minimum width=3.7cm, 
				minimum height=1.1cm,
				align=center,
				font=\normalsize\bfseries,
				thick
			},
			outer/.style={
				rectangle, 
				draw=black!70, 
				fill=gray!10, 
				rounded corners=3pt, 
				minimum width=3.7cm, 
				minimum height=1.1cm,
				align=center,
				font=\normalsize\bfseries,
				thick
			},
			context/.style={
				rectangle, 
				draw=teal!70!black, 
				fill=teal!10, 
				rounded corners=3pt, 
				minimum width=3.7cm, 
				minimum height=1.1cm,
				align=center,
				font=\normalsize\bfseries,
				thick
			},
			arrow/.style={
				-Stealth,
				very thick,
				draw=black!80,
				shorten >=2pt,
				shorten <=2pt
			}
			]
			\node[context] (social) at (-11.2, 4) {Social Context};
			\node[inner] (appraisal) at (0, 2) {Cognitive Appraisal Computing};
			\node[inner, below left=of appraisal] (mapping) {Mapping Appraisal to\\Emotional States};
			\node[inner, below right=of appraisal] (experience) {Emotional Experiences};
			\node[outer] (perception) at (-4.5, 4) {Perception};
			\node[outer] (model) at (4, 4) {Cognitive Computational Model};
			\node[outer] (expression) at (0, -2) {Emotional Expression Synthesis};
			\draw[arrow] (appraisal) -- (mapping);
			\draw[arrow] (appraisal) -- (experience);
			\draw[arrow] (experience) -- (expression);
			\draw[arrow] (experience) -- (model);
			\draw[arrow] (perception) -- (appraisal);
			\draw[arrow] (model) -- (appraisal);
			\draw[arrow] (social) -- node[above, sloped, font=\small\itshape, text=black!70] {stimuli} (perception);
			\begin{scope}[on background layer]
				\node[draw=black!70, thick, dashed, rounded corners, inner sep=10mm, 
				fill=pink!20, fit=(perception) (model) (expression) (appraisal) (mapping) (experience)] (outerboundary) {};
				\node[draw=blue!80!black, thick, dashed, rounded corners, inner sep=4mm, 
				fill=white!20, fit=(appraisal) (mapping) (experience)] (innerboundary) {};
			\end{scope}
	\end{tikzpicture}}
	\caption{Block diagram of intelligent agent with affective cognition}
	\label{fig:block_elicitation}
\end{figure*}
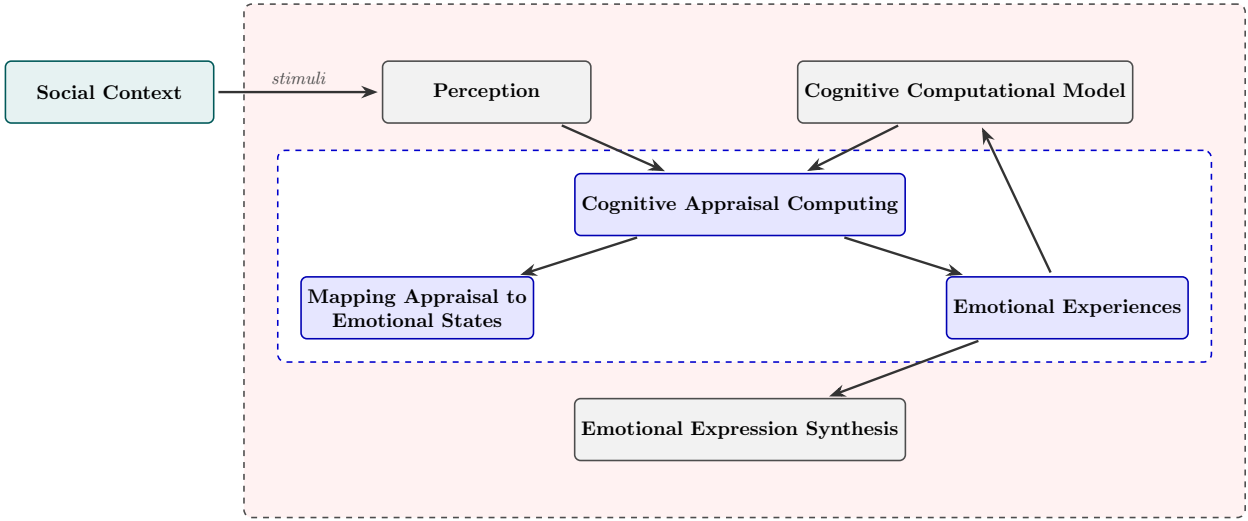

Overall, the proposed methods for modeling artificial emotional elicitation are categorized into two main groups: theory-driven \cite{zall2024towards} and data-driven approaches \cite{liu2024emotion}. Theory-driven approaches often focus on cognitive appraisal theory, which is one of the most well-established psychological theories of emotional elicitation in humans. This theory outlines that emotions arise in response to both internal and external events. To elicit the appropriate emotion, it is necessary to evaluate the experienced event through various cognitive appraisals and personal experiences. These theories emphasize cognitive appraisal variables that assess an event from multiple perspectives. Some theory-driven approaches are based on a cognitive computational framework \cite{jokinen2025introduction} that outlines essential cognitive information, such as concerns, goals, and needs, which are crucial for forming theory-based appraisals. The overall overview of these approaches is illustrated in Figure \ref{fig:block_elicitation}. As shown in this figure, the perceived stimuli are evaluated to calculate the cognitive appraisal variables. The mental states and experiences necessary for determining these appraisal variables are described within the cognitive computational model. Following this, the appraisal variables are mapped to specific emotional states. These resulting emotional states then influence emotional modulation, which in turn affects decision-making and emotional expression. Theory-based methods predict emotional states in future utterances by concentrating on cognitive processes. In contrast, data-driven methods analyze the conversation's history to identify specific events within the observed data and determine the emotions associated with those events for upcoming contributions.
\subsection{Challenges}
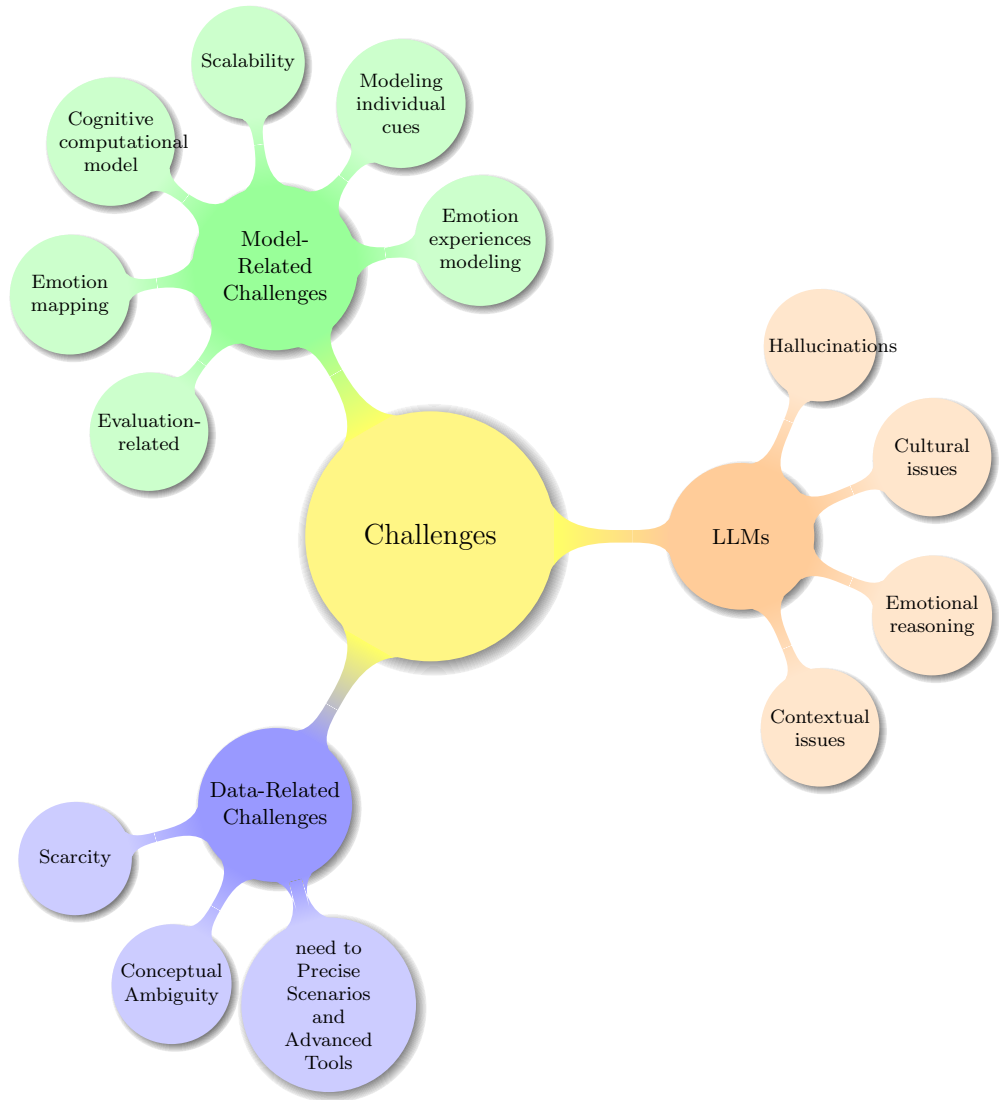
\begin{figure*}[t]
	\centering
	\resizebox{1\textwidth}{!}{
		\begin{tikzpicture}[mindmap, grow cyclic, every node/.style={concept, circular drop shadow, minimum size=1.5cm}, concept color=yellow!60, text=black,
			level 1/.append style={level distance=4.5cm,sibling angle=120},
			level 2/.append style={level distance=3cm,sibling angle=45}]
			\node[concept] {Challenges }
			child[concept color=blue!40] { node {Data-Related\\Challenges}
				child[concept color=blue!20!white] { node {Scarcity } }
				child[concept color=blue!20!white] { node {Conceptual Ambiguity} }
				child[concept color=blue!20!white] { node {need to Precise Scenarios and Advanced Tools  } }
			}
			child[concept color=orange!40] { node {LLMs}
				child[concept color=orange!20!white] { node {Contextual  \\ issues} }
				child[concept color=orange!20!white] { node {Emotional reasoning} }
				child[concept color=orange!20!white] { node {Cultural\\issues} }
				child[concept color=orange!20!white] { node {Hallucinations} }
			}
			child[concept color=green!40] { node {Model-Related\\Challenges}
				child[concept color=green!20!white] { node {Emotion experiences modeling} }
				child[concept color=green!20!white] { node {Modeling individual cues } }
				child[concept color=green!20!white] { node {Scalability} }
				child[concept color=green!20!white] { node {Cognitive \\computational model} }
				child[concept color=green!20!white] { node {Emotion mapping } }
				child[concept color=green!20!white] { node {Evaluation-related  } }
			};
		\end{tikzpicture}
	}
	\caption{Challenges in affective cognition}
	\label{fig:elicitation_challenges}
\end{figure*}

The modeling of emotional elicitation and experiences in intelligent agents is a crucial aspect of the affective theory of mind. A key component of this process involves developing a cognitive model that serves as a foundational framework for managing knowledge, information, experiences, and cognitive mental states related to emotions. 
One of the main challenges in this area is establishing reliable methods for computing cognitive appraisal variables based on underlying cognitive theories. Data-driven approaches also face significant obstacles, such as limited and ambiguous data, which impede the accurate capture and representation of the complex and multifaceted nature of human emotions. Additionally, technological and methodological barriers, including the need for robust models and the integration of LLMs, further complicate the task. Addressing these challenges is essential for advancing precise and effective models of emotional responses that are crucial for creating empathetic and socially intelligent systems. The primary challenges in this domain, along with their associated sub-challenges, are illustrated in Figure~\ref{fig:elicitation_challenges}. Furthermore, Table~\ref{tab:challenges_elicitation} provides an overview of various studies and the solutions they propose for tackling these issues.
In the following sections, we will explore these challenges in detail and review potential solutions to overcome them, to enhance the fidelity and applicability of emotion modeling in intelligent agents.
\subsubsection{Data-Related Challenges}
The field of emotion elicitation and experiences currently lacks a comprehensive and cohesive dataset. There is a clear need for data collections that explicitly clarify the relationship between various emotions and the appraisal variables and events that influence them. Integrating data-driven approaches with theoretical frameworks requires the availability of extensive and complete datasets. Although some efforts have been made in this area, a fully integrated and systematic dataset is still essential for advancing research. Developing such datasets would lead to a better understanding of emotional responses and their underlying mechanisms, ultimately contributing to the improvement of both theoretical models and practical applications in emotional analysis.
Emerging evidence indicates that induction mechanisms are essential for eliciting emotions by simulating the sensory experiences needed for experimental paradigms. Over the years, many reviews have predominantly focused on passive elicitation methods, where individuals act as observers, often neglecting the importance of self-relevance in emotional experiences. To address this gap in the literature, \cite{somarathna2022virtual} explores the potential of Virtual Reality (VR) as an active mechanism for emotion induction. Furthermore, to ensure the effectiveness and reliability of research outcomes, VR environments must incorporate well-selected stimuli to successfully evoke specific emotional responses. \cite{bayro2025systematic} presents experimental protocols for collecting datasets with virtual reality. \\
\cite{gandhi2024human} proposes an automatic dataset generation that includes diverse scenarios for benchmarking affective cognition, specifically focusing on understanding and reasoning about human emotions in foundation models. It presents a pipeline for generating diverse and naturalistic stimuli that can systematically and scalability evaluate affective reasoning. The generated scenarios explore relationships between appraisals, emotions, expressions, and outcomes.
\subsubsection{Model-related challenges}
In this section, we review the challenges are related to learning model. \\

\textbf{\textit{Cognitive computational modeling}}
Theoretical approaches to emotional elicitation modeling primarily depend on cognitive models to calculate appraisal variables. These models provide the foundational information required to determine emotional states. Most methods not only specify the emotion to be expressed but also examine how emotions influence decision-making and other cognitive processes, such as inference. A significant challenge in emotion elicitation modeling is the effective implementation of cognitive models. Several computational cognitive models are available for this purpose, including cognitive architectures \cite{anderson2004integrated, laird2019soar}, Belief-Desire-Intention (BDI) frameworks \cite{pereira2005towards, sanchez2019abc}, and other specific designs \cite{Hoorn2021, BeckerAsano2010}. These computational models exhibit substantial differences, with each group demonstrating unique structural and functional characteristics.
Cognitive architectures such as ACT-R \cite{anderson2004}, SOAR \cite{laird2019}, and LIDA \cite{franklin2007} provide structured frameworks for modeling the essential components required to produce intelligent behavior in artificial agents. By integrating diverse sources of knowledge, these architectures enable agents to address a wide range of complex problems. Typically, cognitive architectures incorporate specialized memory systems: procedural memory for action execution and planning, semantic memory for environmental knowledge, episodic memory for past experiences, and perceptual memory for object recognition and classification. In addition to memory systems, they include control mechanisms, processing modules, learning algorithms, structured data representations, and input/output interfaces. The approach described in \cite{juvina2018} utilizes the ACT-R cognitive architecture, assigning emotional values (positive or negative) to memory elements. The development of modern conversational agents requires the seamless integration of emotional and cognitive models to achieve human-like interactions.

\clearpage
\begin{longtable}{@{}p{2.5cm}p{2.8cm}p{9.5cm}@{}}
	\caption{Summary of challenges and solutions in affective cognition}
	\label{tab:challenges_elicitation} \\
	\toprule
	\textbf{Challenge} & \textbf{Sub-challenge} & \multicolumn{1}{c}{\textbf{Solution}} \\
	\midrule
	\endfirsthead
	
	\caption[]{Summary of Challenges and Solutions in Affective Cognition (Continued)} \\
	\toprule
	\textbf{Challenge} & \textbf{Sub-challenge} & \multicolumn{1}{c}{\textbf{Solution}} \\
	\midrule
	\endhead
	
	\midrule
	\multicolumn{3}{r}{{Continued on next page}} \\
	\endfoot
	
	\bottomrule
	\endlastfoot
	
	\multirow{3}{2.5cm}{\textbf{Data-Related}} & Scarcity & Automatic dataset generation with diverse scenarios for benchmarking affective cognition \citep{gandhi2024human} \\
	\cmidrule(lr){2-3}
	& Conceptual Ambiguity & Experimental protocols for VR dataset collection \citep{bayro2025systematic} \\
	\cmidrule(lr){2-3}
	& Precise Scenarios \& Tools & VR as active mechanism for emotion induction \citep{somarathna2022virtual} \\
	\midrule
	\multirow{5}{2.5cm}{\textbf{Model-Related}} & Cognitive computational modeling & ACT-R architecture (emotional memory valuation) \citep{juvina2018}; E-VOX SOAR + ALMA (real-time processing) \cite{perez2017}; EMA Automatic/deliberate appraisal \cite{gratch2004}; Emotional-BDI Resource management \cite{pereira2005}; Fuzzy-based BDI Cultural/linguistic modeling \cite{taverner2021}; ABC-EBDI BDI + ABC theory \cite{sanchez2019}; InFra Stimuli appraisal \cite{rodriguez2016}; Silicon Coppelia Ethics/aesthetics variables \cite{Hoorn2021}; EIAEC Appraisal theories + memory \cite{zall2024towards} \\
	\cmidrule(lr){2-3}
	& Scalability & Knowledge graphs + continuous learning \cite{zall2024towards} \\
	\cmidrule(lr){2-3}
	& Emotion mapping & EIAEC Data-driven approach \cite{zall2024towards}; EEGS Weighted mapping strategy \cite{ojha2020eegs}; CPM + VR data-driven \cite{somarathna2024}; Regret-based RL \cite{soman2024regret} \\
	\cmidrule(lr){2-3}
	& Emotion experiences modeling & SUSAN Inner speech modeling \cite{corvaia2025}; Neural network framework \cite{hernandez2024generic}; EIAEC Context-dependent values \cite{zall2024towards}; Memory valuations \cite{juvina2018}; Time-based annotations \cite{tsfasman2025emotion}; RL + appraisal theory \cite{zhang2024modeling}; Value prediction \cite{zhang2024simulating}; Weber-Fechner + Q-learning \cite{wu2025automated}; Emotional motivation \cite{Berto2025}; Reward modulation \cite{nikodemou2024deconstructing}; TD-based expressions \cite{nijeholt2023role} \\
	\cmidrule(lr){2-3}
	& Evaluation & Scenario simulation \& human comparison \cite{sanchez2019abc, zall2024towards}; Multi-agent performance assessment (EBDI) \cite{jiang2007}; Emotional reasoning in critical situations \cite{bourgais2016}; Human feedback evaluation (E-VOX) \cite{perez2017} \\
	\midrule
	\multirow{2}{2.5cm}{\textbf{LLM-based}} & Emotional reasoning & Distinguishes self-attribution vs perception of emotions using appraisal theory \cite{tak2024journal}; Combines cognitive/affective reasoning with intrinsic motivation \cite{Raggioli2025}; Explainable emotion alignment for Metaverse agents \cite{khan2025large} \\
	\cmidrule(lr){2-3}
	& Contextual misunderstandings & Assesses affective cognition in LLMs (GPT-4, Claude-3, Gemini-1.5-Pro) \cite{gandhi2024human} \\
	
\end{longtable}

The E-VOX \cite{perez2017} system exemplifies this by integrating emotional and cognitive models, combining a SOAR-based framework with an ALMA affective model \cite{flavian2011}. It enables real-time emotional processing, ongoing mood tracking, and stable personality representation, leading to improvements in knowledge retrieval, adaptive learning, and emotional intelligence. The Emotion and Adaptation Model (EMA) \cite{gratch2004}involves both fast (automatic) and slow (deliberate) appraisal processes. It systematically analyzes perceived events based on the virtual human’s goals and beliefs, generating appraisal frames that help interpret environmental features. These frames connect appraisal variables to specific emotions, which in turn determine the agent’s current emotional state.   \\
Cognitive frameworks explore how beliefs, desires, and intentions interact to demonstrate the reasoning processes in agent architectures, as seen in the belief-desire-intention (BDI) model. While these frameworks primarily focus on high-level decision-making structures, cognitive architectures offer a more comprehensive range of cognitive capabilities. This includes reinforcement learning mechanisms, specialized memory systems (procedural, semantic, episodic, and perceptual), advanced data representations, processing components, and control mechanisms—features that cognitive frameworks often lack or implement only partially. Emotional-BDI\cite{pereira2005} is a conceptual framework that merges emotional components with the traditional BDI cognitive architecture, focusing on the management of resources, capabilities, and emotional states. While it primarily emphasizes the emotion of fear, it also lays the groundwork for future research into emotion-driven behavior in agents. Fuzzy-based affective BDI \cite{taverner2021} is a computational model that incorporates cultural and linguistic differences into emotion modeling. It consists of two primary components: the event appraisal process, which evaluates stimuli using fuzzy appraisal rules to generate fuzzy emotions, and the affect-adaptation process, which converts these fuzzy emotions into PAD (Pleasure, Arousal, Dominance) dimensions through defuzzification. ABC-EBDI \cite{sanchez2019, sanchez2019b} is an emotion model designed for BDI agents that connects beliefs to emotional and behavioral outcomes through the ABC theory. It takes into account mood and personality to produce more realistic behavior. \\
Some methods develop cognitive components selectively, based on their specific requirements. The Integrative Framework (InFra) \cite{rodriguez2016} outlines the essential components required for generating emotions. It includes input and output interfaces that connect emotional processes with cognitive modules. InFra takes environmental information and processes it through perception, then conducts an emotional appraisal of stimuli. The resulting emotions subsequently influence behavior. Additionally, the framework considers the effects of personality and culture on the appraisal process, leading to more realistic emotional responses. Silicon Coppelia \cite{Hoorn2021} is designed to respond emotionally to users. It evaluates perceived features using appraisal variables such as ethics, aesthetics, epistemics, and affordances. Ethics guide the agent's moral behavior, while aesthetics relate to its social appearance. Epistemics assess the realism of perceptions, and affordances indicate possible actions in the current context. The agent determines the relevance and positivity or negativity of observations based on its beliefs. To handle multiple emotions occurring simultaneously, Silicon Coppelia employs fuzzy sets and operators. This approach allows it to manage ambiguous emotions and feelings, enabling it to make human-like decisions during interactions. The EIAEC\cite{zall2024towards} framework is designed to develop emotion-aware intelligent agents by utilizing appraisal theories to determine the agents' emotional states in different situations. It features an efficient episodic memory that stores events and their contexts, allowing the agent to recall relevant information for emotional expression and decision-making. The framework learns the affective values associated with events based on the agent's experiences in various contexts, extracting appraisal variables from memory metadata. It also employs a data-driven approach to map these variables to emotional states. Furthermore, a method is implemented to update action activation levels based on the agent's emotional states, effectively modeling how emotions influence decision-making. \\\\
\textbf{\textit{Scalability}}
Most existing emotion models are explicitly designed for specific scenarios or contexts \cite{hernandez2024generic}. For example, the MAMID model \cite{hudlicka2004} is exclusively tailored to treatment settings, relying on specialized if–then rules that render its action evaluation and selection highly domain-dependent. Similarly, WASABI \cite{becker2008} has been primarily validated in virtual character gaming environments, which limits its applicability to broader use cases. More generally, many computational emotion models concentrate on only one or two components of the emotional process, depending on the target application. Due to limited scalability and difficulties in seamless integration, extending these models with additional components remains challenging. In contrast, the EIAEC framework \cite{zall2024towards} supports scalability across diverse scenarios and applications by leveraging multiple memory types enriched with knowledge graphs. Rather than relying on predefined rules or fixed relationships, this framework enables the continuous learning and refinement of emotional relationships and experiences over time.\\
\textbf{\textit{Emotion mapping modeling}}
An emotion model implanted in appraisal theory aims to map event assessments (appraisals) into corresponding emotion intensity levels, as specified by the theoretical framework. However, most appraisal theories do not provide explicit, quantifiable rules for mapping these appraisals to emotion intensities. Consequently, many computational implementations adopt heuristic or ad-hoc methods to approximate this relationship, primarily to facilitate research and experimental validation \cite{ojha2021}.
For instance, EEGS \cite{ojha2020eegs} introduces a mathematical formula for calculating emotion intensities based on appraisal theory, employing a weighted mapping strategy that utilizes quantitative appraisal variables to derive implications for various emotions. Similarly, \cite{somarathna2024} leverages the Component Process Model (CPM), which encompasses appraisal, expression, motivation, physiology, and feeling components, to explore their interrelations with emotion elicitation. This study adopts a data-driven approach using interactive virtual reality (VR) games and multi-modal data collection, including self-reports, physiological signals, and facial expressions, to establish a dataset that delineates the relationship between appraisal variables and emotional states. Furthermore, \cite{zall2024towards} presents an advanced mapping methodology grounded in datasets from \cite{mohammadi2020} and \cite{somarathna2023}, employing a data-driven framework to refine the mapping between appraisal processes and emotional responses. \\
Path planning (PP) in autonomous systems can significantly benefit from reinforcement learning (RL) due to its flexible decision-making capabilities. \cite{soman2024regret} has introduced emotion-inspired mechanisms, particularly regret-based models, to enhance the performance of RL. These models incorporate two key concepts: experienced regret, which measures past suboptimal actions, and anticipated regret, which predicts future errors. By utilizing regret dynamics to adjust the epsilon decay in epsilon-greedy policies, agents can achieve a better balance between exploration and exploitation. This leads to improved learning efficiency. \\\\
\textbf{\textit{Emotion experiences modeling}}
Developing emotional intelligence in an intelligent agent involves more than merely modeling emotional elicitation or artificial emotion synthesis. A fundamental and critical challenge in this domain is modeling emotional experience, specifically understanding how emotions influence various cognitive processes, such as decision-making \cite{hernandez2024generic}. The invention of personalized experiences in intelligent agents improves communication and fosters authentic connections with humans. When an intelligent agent experiences emotions during the process of emotional elicitation, these emotions have impacts on various cognitive components, including inference, reasoning, and decision-making. Self-dialogue Utility in Simulating Artificial Emotions (SUSAN) \cite{corvaia2025} proposes a novel perspective for modeling emotional experiences. Specifically, it aims to emulate human-like emotional experience through the use of inner speech modeling. By leveraging inner speech, the model infers and simulates underlying contextual and cognitive processes that contribute to emotional experiences. The framework is grounded in Damasio’s theory, providing a structured basis for understanding and replicating emotional dynamics through internal cognitive mechanisms. Mature emotional responses that help regulate behavior offer a significant evolutionary advantage. However, a clear theoretical foundation in cognitive neuroscience is lacking, particularly regarding how emotions are elicited. \cite{hernandez2024generic} suggests that emotions correspond to specific temporal patterns in essential environmental variables and introduces a self-learning emotional framework for intelligent agents. An unsupervised neural network trained on unlabeled experiential data to identify eight basic emotional patterns that are contextually coherent and mimic natural emotional dynamics. EIAEC \cite{zall2024towards} learns context-dependent affective values by associating events with the agent's emotional experiences across different situations, storing these associations in episodic memory. Additionally, it dynamically updates activation values for actions in procedural memory based on the agent's emotional state, effectively modeling the influence of emotion on decision-making processes. The approach outlined in \cite{juvina2018} learns to associate emotional valuations and arousal levels with each declarative memory chunk, taking into account their frequency of use and reward signals derived from core affect dynamics. Ultimately, the model assesses how emotional value and arousal levels affect memory retrieval and decision-making processes. The relationship between emotional experiences and their memorability has long been acknowledged, with highly emotional events generally considered to be more memorable. This relationship suggests that emotional annotations could potentially serve as proxies for assessing memorability. However, most existing emotion recognition systems depend heavily on third-party annotations, which may not accurately reflect individuals’ firsthand emotional relevance and memory encoding. Recognizing this gap, \cite{tsfasman2025emotion} empirically explores the connection between perceived emotions and collective memorability within conversational interactions. It involves continuous, time-based annotations of both emotional states and memorability in dynamic, unstructured group settings, aiming to approximate real-world conversational AI environments. \\
Recent research has explored the integration of emotional and cognitive variables into reinforcement learning (RL) agents, frequently using these variables as reward to enhance decision-making processes \cite{moerland2018emotion, alkam2025reinforcement}. Building upon this foundation, \cite{zhang2024modeling} introduced a computational model of emotion that synthesizes RL with appraisal theory, establishing a formal link between reward processing, goal-directed learning, cognitive appraisal, and emotional experience. The model formalizes four appraisal variables, such as suddenness, goal relevance, goal conduciveness, and power, derived from the component process model (CPM), and operationalizes them via temporal difference learning updates. Remarkably, this framework is task-independent and applicable to any environment modeled as a Markov decision process (MDP), allowing the analysis and simulation of emotional responses within RL-based decision-making contexts. Complementing this approach, \cite{zhang2024simulating} proposed a computational cognitive model that conceptualizes emotion as a dynamic, continuous process rather than a static state, particularly during interactive decision episodes. Their framework integrates cognitive theories of emotion with principles of computational rationality, using RL for value prediction to capture the evolving nature of emotional experiences during ongoing interactions. Agent-based multi-issue bilateral automated negotiation is gaining increased attention from researchers. However, the impact of emotional deception and multi-attribute preferences on negotiation outcomes is often neglected. \cite{wu2025automated} develops a negotiation model based on agents and employs reinforcement learning. It quantifies emotions and generates emotional deception using the Weber-Fechner law in automated negotiation. Furthermore, emotions are used to adjust the reward function of Q-learning to update opponents' attribute preferences. A time-dependent issue updating model is proposed, which integrates emotional deception, attribute preferences, and a fairness function. \cite{Berto2025} presents a cognitive architecture that simulates human cognitive functions, including perception, motivation, and decision-making. The intelligent agent operates based on motivations derived from internal states, which are influenced by affective states.\\
Self-control, a fundamental aspect of human decision-making, is often understood as the internal conflict between higher-order executive functions and lower limbic systems, especially in situations involving the dilemma between smaller, immediate rewards and larger, delayed rewards. Previous computational models have employed game-theoretic frameworks, such as the Iterated Prisoner’s Dilemma, with learning mechanisms to explore decision dynamics related to self-control. However, these models generally do not explicitly account for the influence of emotional states. \cite{nikodemou2024deconstructing} incorporates emotion as a modulatory factor by adjusting reinforcement rewards, thereby simulating temporal fluctuations in positive and negative emotional intensities. This approach captures the effects of emotions on self-control by modeling their dynamic influence, rather than treating emotions as static or isolated states.\\
Transparency in behavior is crucial for robots that interact with humans. When robots need to adapt to different users and tasks, they must optimize their behavior, and Reinforcement Learning offers a promising method for achieving this. However, the behaviors generated by RL are not inherently transparent due to the exploration/exploitation tradeoff involved in optimizing policies. Emotions serve as a natural way for humans to communicate intent and evaluate situational relevance. Incorporating emotional expressions into robots has been suggested as a way to enhance the transparency of their learning processes. In this regard, some studies \cite{nijeholt2023role} have implemented emotional expressions based on Temporal Differences (TD) to make the robot's learning trajectory more interpretable to human observers and instructors.\\\\
\textbf{\textit{Evaluation-related}}
A significant challenge in developing agents with emotional intelligence is evaluating their computational models of emotion. Typically, assessments rely on subjective measures within specified scenarios. In such cases, if the agent's behavior aligns with expected responses, the underlying model is regarded as effective. When evaluating the emotional behavior of cognitive agents, two fundamental criteria can be considered: believability and social acceptability\cite{ojha2020}. Believability assesses how natural and contextually appropriate the agent's actions appear, contributing to a sense of lifelikeness. While believable emotions are essential for an agent's acceptance, they alone may not guarantee social approval. An agent can appear lifelike but still fail to adhere to social norms. Therefore, social acceptability assesses whether the emotions expressed by the agent align with societal expectations and standards, ensuring that the agent's behavior is not only believable but also socially appropriate \cite{zall2022comparative}.\\
Each of the reviewed approaches has extended the cognitive components necessary for emotional elicitation from various perspectives. Consequently, a cognitive component based on a similar conceptual framework is not consistently evident across these models. This presents significant challenges for the development of a novel approach, particularly given that detailed implementation specifics of these architectures are often not readily accessible.\\
To assess the effectiveness of computational models of emotion, some methods involve simulating an emotional scenario and comparing the reactions of the virtual agent to those of a human in a similar situation \cite{sanchez2019abc, zall2024towards}. EBDI \cite{jiang2007} models a multi-agent environment capable of evaluating the performance of various agent architectures across different contexts by adjusting multiple parameters. It focuses on decision-making in agents with diverse emotional capabilities to analyze the influence of emotions on making appropriate choices. Similarly, \cite{bourgais2016} simulates the behavior of multiple agents that incorporate emotions into their reasoning processes during critical situations, highlighting the role of emotion in adaptive decision-making. E-VOX \cite{perez2017} is evaluated based on human feedback, where users interacting with the E-VOX-based agent assess its believability and acceptability by completing several questionnaires.\\
\subsubsection{LLMs-Related Challenges}
The emergence of LLMs has paved the way for developing emotionally intelligent agents that leverage these technologies. In response, the development of LLMs is revolutionizing agents, transforming them from mere instrumental tools into autonomous "Metaverse Citizens" equipped with decision-making and interaction abilities. The recent study \cite{khan2025large} demonstrates how intelligent agents can integrate emotions into their decision-making processes, resulting in behaviors that are both acceptable and satisfying to users. This research examines the role of LLMs in creating intelligent agents within the Metaverse service ecosystem, where digital avatars, digital twins, and digital natives play essential roles. To improve the reliability and social realism of these agents, the paper proposes an explainable emotion alignment framework that incorporates factual data into their decision-making processes. The paper \cite{Raggioli2025} introduces an integrated architecture that combines cognitive and affective reasoning using LLMs. In this framework, LLMs generate beliefs from natural language inputs. It employs an intrinsically motivated approach that relies on an internal emotional state to influence internal rewards.\\
\cite{tak2024journal} builds on recent studies of the emotional reasoning capabilities of LLMs. While prior work has assessed LLMs' general understanding of emotions, it has not distinguished between their predictions of self-attribution versus perception of others’ emotions, based on appraisal theory, which reflects the fundamental mechanisms involved in human emotion elicitation. It reveals that GPT-4 excels in reasoning about stimuli designed to evoke inferred emotional attributions, closely mirroring human judgments, especially in idiosyncratic scenarios. Conversely, GPT-4's interpretations align more with perceptions of others’ emotions than with self-assessments, supporting the idea that LLMs adopt an observer-like perspective. The paper \cite{gandhi2024human} assesses affective cognition in foundation models to specify the abilities of LLMs such as GPT-4, Claude-3, and Gemini-1.5-Pro in determining emotion elicitation in various scenarios. It manifests that these models constantly align with human judgments, occasionally exceeding average human accuracy, particularly with chain-of-thought reasoning. Recent studies have moved beyond surface-level emotion recognition to examine the cognitive mechanisms underlying emotional reasoning in large language models (LLMs). In particular, \cite{bhattacharyya2025machines} introduces CoRE, a large-scale benchmark grounded in cognitive appraisal theory, to evaluate whether LLMs exhibit coherent and interpretable cognitive reasoning when processing emotionally charged stimuli. This paper indicates that while many models demonstrate broadly human-like appraisal structures, they struggle with nuanced emotions and exhibit inconsistencies across abstract cognitive dimensions, revealing implicit biases in emotion representation and limitations of current training paradigms.
\section{Emotional Text Synthesis}
\label{text}
ETS in NLP focuses on producing emotionally resonant text through advanced techniques such as cross-modal emotion generation and emotion-controlled language models. The process involves creating written or spoken content that conveys specific emotional states, enabling machines to produce emotionally resonant and contextually appropriate responses. It employs techniques such as style transfer, conditional generation, and fine-tuning of LLMs. This capability is a cornerstone of affective computing, which seeks to enhance HCI by imbuing systems with emotional intelligence \cite{schuller2018age1}. Emotion generation plays a pivotal role in applications such as virtual assistants \cite{xue2024chat16}, mental health support systems \cite{feng2024bridging17}, and personalized content creation \cite{abilbekov2024kazemotts21}, where the ability to express and evoke emotions significantly impacts user engagement and satisfaction \cite{becker2024text2}. For instance, virtual assistants equipped with emotion generation capabilities can provide more empathetic and context-aware responses, fostering a more natural and human-like interaction \cite{zheng2021semantic3}.
In NLP, techniques like sentiment analysis and emotion detection lay the groundwork for generating emotionally charged text by identifying and interpreting emotional cues in existing data \cite{truong2024textual5}. Advances in AI, particularly in generative models and LLMs, have further propelled the field by enabling fine-grained control over emotional expression in generated text \cite{becker2024text2}. Here, we first briefly introduce approaches to generate emotional text. Then, we proceed with the challenges within this area and how different research studies have tried to overcome them.
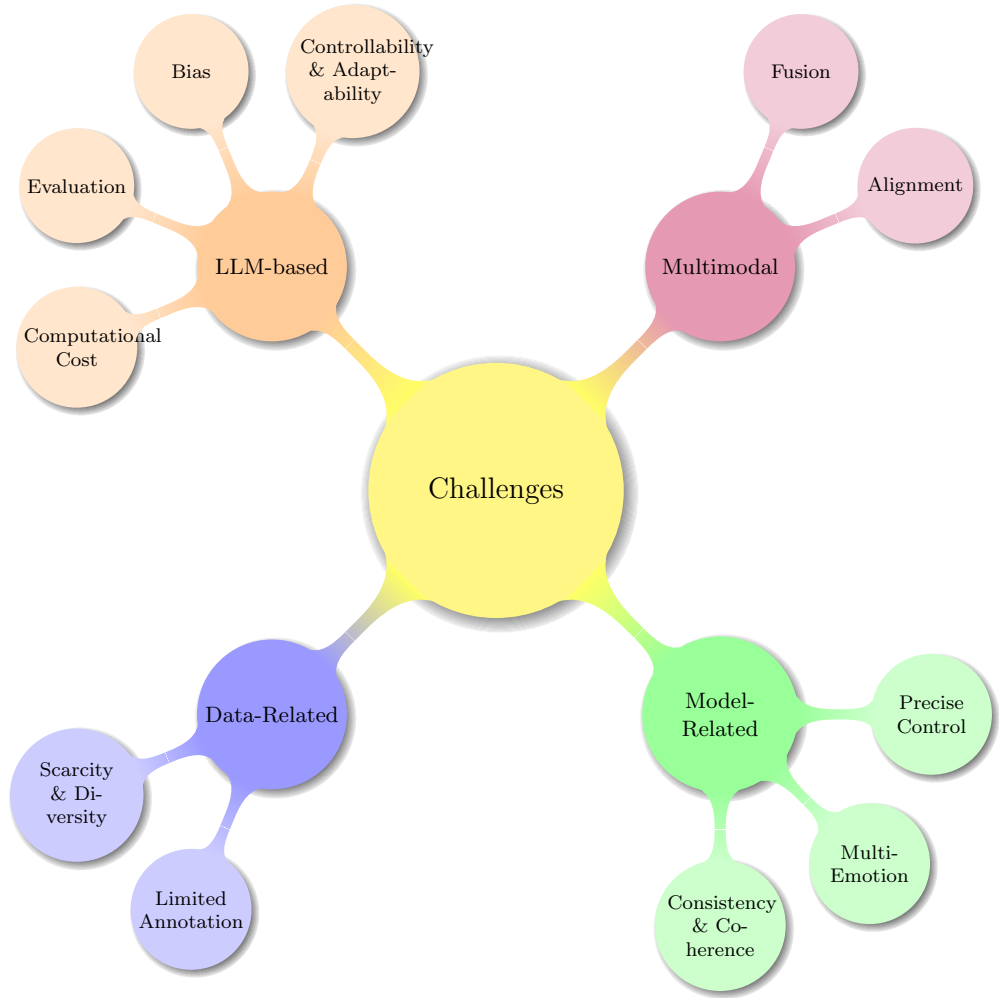
\begin{figure*}[t]
	\centering
	\resizebox{0.8\textwidth}{!}{
		\begin{tikzpicture}[mindmap, grow cyclic, every node/.style={concept, circular drop shadow, minimum size=1.5cm}, concept color=yellow!60, text=black,
			level 1/.append style={level distance=4.5cm,sibling angle=90},
			level 2/.append style={level distance=3cm,sibling angle=45}]
			\node[concept] {Challenges}
			child[concept color=blue!40] { node {Data-Related} 
				child[concept color=blue!20!white] { node {Scarcity \\ \& Diversity} }
				child[concept color=blue!20!white] { node {Limited \\ Annotation} }
			}
			child[concept color=green!40] { node {Model-Related} 
				child[concept color=green!20!white] { node {Consistency \\ \& Coherence} }
				child[concept color=green!20!white] { node {Multi-Emotion} }
				child[concept color=green!20!white] { node {Precise Control} }
			}
			child[concept color=purple!40] { node {Multimodal} 
				child[concept color=purple!20!white] { node {Alignment} }
				child[concept color=purple!20!white] { node {Fusion} }
			}
			child[concept color=orange!40] { node {LLM-based} 
				child[concept color=orange!20!white] { node {Controllability \\ \& Adaptability} }
				child[concept color=orange!20!white] { node {Bias} }
				child[concept color=orange!20!white] { node {Evaluation} }
				child[concept color=orange!20!white] { node {Computational \\ Cost} }
			};
		\end{tikzpicture}
	}
	\caption{Challenges in ETS}
	\label{fig:challenges_new}
\end{figure*}
\subsection{Approaches}
\textbf{\textit{Style Transfer}:}
Style transfer alters the emotional tone of a text while maintaining its meaning, which is beneficial for personalized content creation. However, it often struggles with maintaining semantic coherence and emotional consistency in more complex texts. Innovations such as the lexicon-based attention mechanism and methods separating content from style \cite{fu2018style27} have been developed to improve emotional nuance and address the lack of parallel corpora.\\
\textbf{\textit{Conditional Generation}:}
Conditional generation controls emotional content through specific prompts, enhancing applications like virtual assistants. Models such as the co-attention neural network \cite{li2018co36} and Emotional Tacotron \cite{lee2017emotional37} exemplify this approach, though they require high-quality datasets and struggle with multi-emotion scenarios. Affective Chatbot \cite{jiang2022affective38} and emotion prediction in TTS systems \cite{yoon2022language39} highlight its versatility.\\
\textbf{\textit{Fine-tuning LLMs}:}
Leveraging models like GPTs and BERT enables high-quality emotional text generation with minimal training \cite{singh2020adapting20}. This method requires significant computational resources and careful bias management.

\begin{table*}[!t]
	\centering
	\caption{Summary of challenges and solutions in emotional text synthesis}
	\label{tab:challenges_summary}
	\renewcommand{\arraystretch}{1.2}
	\begin{tabularx}{\textwidth}{@{}p{2.5cm}p{2.8cm}X@{}}
		\toprule
		\textbf{Challenge} & \textbf{Sub-challenge} & \multicolumn{1}{c}{\textbf{Solutions}} \\
		\midrule
		\multirow{4}{2.5cm}{\textbf{Data-Related}} & Scarcity & Data augmentation \citep{firdaus2023multi65}; Transfer learning \citep{li2018generative44}; SentiGAN \citep{wang2018sentigan52}; CS-GAN \citep{li2018generative44}; LeakGAN \citep{guo2018long53} \\
		\cmidrule(lr){2-3}
		& Limited Annotation & Desire annotation \citep{jia2022beyond57} \\
		\midrule
		\multirow{8}{2.5cm}{\textbf{Model-Related}} & Emotional Consistency \& Coherence & Appraisal theories \citep{xia2019emotion26, singh2020adapting20, resendiz2023affective55}; Emotion embeddings \citep{tan2023survey66}; MSEG \citep{firdaus2023multi65}; SEPRG \citep{li2022seprg70}; RCVAE \citep{zhou2017mojitalk64}; DecoupledESC \citep{zhang2025decoupledesc}; CARE \citep{care2025}; COMPEER \citep{compeer2025} \\
		\cmidrule(lr){2-3}
		& Handling Multi-Emotion & Emotion blending \citep{ghosh2017affect33, resendiz2023affective55}; ECM \citep{zhou2018emotional15} \\
		\cmidrule(lr){2-3}
		& Precise Emotion Control & FUDGE \citep{yang2021fudge51}; MOAEP \citep{resendiz2024mopo43}; Prefix-tuning \citep{qian2022controllable41}; DEXPERTS \citep{mahmood2023dexperts80} \\
		\midrule
		\multirow{4}{2.5cm}{\textbf{Cross-Modal}} & Alignment & Hierarchical attention \citep{zhang2017multimodal}; Word-level alignment \citep{zhang2017multimodal} \\
		\cmidrule(lr){2-3}
		& Fusion & Time-dependent fusion \citep{zhang2017multimodal}; FaceChat \citep{alnuhait2023facechat18}; Omni-perception Pre-Trainer \citep{wu2023large79}; MAGIC \citep{su2022language46}; emoTTS \citep{luo2024emotion12}; FIRES \citep{fires2025} \\
		\midrule
		\multirow{8}{2.5cm}{\textbf{LLM-based}} & Controllability \& Adaptability & Coda \citep{evuru2024coda}; MOPO \citep{resendiz2024mopo43}; Emotion Vectors \citep{emotionvectors2025} \\
		\cmidrule(lr){2-3}
		& Bias & Debiasing techniques \citep{sheng2021societal} \\
		\cmidrule(lr){2-3}
		& Computational Cost & Affective prompt-tuning \citep{gu2024affective42} \\
		\cmidrule(lr){2-3}
		& Evaluation & EmoBench \citep{sabour2024emobench}; Kardia-R1 \citep{kardia2025} \\
		\bottomrule
	\end{tabularx}
\end{table*}

\subsection{Challenges}
In affective text generation, defining and representing emotions presents significant challenges, crucial for accurately conveying emotional states and their contextual expressions. Accurately capturing and generating emotions requires a deep understanding of emotional states and their contexts. Foundational models like Ekman’s six basic emotions \cite{ekman1992} and Plutchik’s wheel of eight emotions \cite{plutchik1980general} provide structured frameworks for categorizing emotions, highlighting their interconnected nature \cite{zhang2024evaluation59}. This section explores these complexities, emphasizing their foundational role in emotion generation and how different studies have tried to overcome these challenges. Main challenges in this area and their related sub-challenges are depicted in Figure~\ref{fig:challenges_new}. Moreover, Table~\ref{tab:challenges_summary} highlights different studies and their solutions to these issues.
\subsubsection{Data-Related Challenges}
The performance of affective text generation models is heavily dependent on the availability of large, high-quality, and culturally diverse datasets. The scarcity of annotated datasets limits a model's ability to replicate nuanced emotional states, which in turn affects the accuracy of the generated text \cite{li2023emotion73}. This is particularly evident in the lack of annotations for emotions like desire \cite{jia2022beyond57}. To mitigate this, strategies such as data augmentation and transfer learning are used to enhance the diversity of training data and improve model generalization \cite{li2018generative44, firdaus2023multi65}. GANs have been used to address data scarcity. SentiGAN \cite{wang2018sentigan52} and CS-GAN \cite{li2018generative44}, for example, enhance emotional diversity in generated text, and LeakGAN \cite{guo2018long53} tackles challenges related to discrete outputs and provides continuous guidance.\\
\subsubsection{Model-Related Challenges}
Generating text that captures multiple emotions, emotional transitions, and maintains coherence is a complex task. Emotion blending and sequential emotion modeling are key approaches to integrate multiple emotions into a coherent expression \cite{ghosh2017affect33}. Despite progress, achieving fine-grained control over emotional expressions remains a significant challenge \cite{zhang2023survey50}. This is particularly difficult for non-native speakers who often struggle with expressing mixed emotions, highlighting the need for models that can replicate these nuanced expressions \cite{feng2024bridging17}. The Emotional Chatting Machine \cite{zhou2018emotional15} addresses coherence with specified emotions through the use of emotion embeddings and attention mechanisms. For precise emotion control, techniques like FUDGE \cite{yang2021fudge51} have been developed to offer fine-grained control over emotional intensity and category. Advanced model architectures like prefix-tuning \cite{qian2022controllable41} and DEXPERTS \cite{mahmood2023dexperts80} have been proposed to address challenges like mode collapse. The SemEval\cite{mohammad2025semeval} shared task on bridging the gap in text-based emotion detection highlights the importance of developing culturally aware and context-sensitive models that can accurately interpret and generate emotions across different languages and cultures. 
Modifying generative models to incorporate new emotional attributes without compromising the integrity of the content is a major obstacle, exacerbated by the non-differentiable nature of discrete text \cite{wang2018sentigan52}. Furthermore, cultural and linguistic diversity introduces additional complexity, necessitating advanced frameworks that merge linguistic and psychological theories with machine learning techniques \cite{jain2017extraction63}. The absence of non-verbal cues in text poses another layer of difficulty, often leading to misinterpretations, especially among non-native speakers.
Diffusion and flow-based models are also at the forefront of generating nuanced emotional text. Diffusion models, such as Diffusion-LM \cite{li2021diffusion78}, iteratively transform noise into structured data, allowing for precise emotional control. Flow-based models use invertible transformations for high-dimensional data generation, capturing emotional nuances while maintaining linguistic quality. These models can integrate affective parameters, enhancing the expressiveness of conversational language \cite{ghosh2017affect33}. DVAE-interVA \cite{chen2023decoupled} is proposed to solve the emotion-semantic entanglement problem in continuous sentiment text generation via adversarial sentiment decoupling, continuous sentiment embeddings, denoising training, and interactive attention to maintain emotional consistency throughout generation.
Another challenge is maintaining emotional consistency and coherence throughout a piece of text or a conversation. This requires models to sustain a consistent emotional tone and narrative. Integrating nuanced understanding, such as appraisal theories, is one approach to enable contextually relevant emotional expression \cite{truong2024textual5, xia2019emotion26, resendiz2023affective55}. However, the variability of language and the informal nature of social media continue to make emotional consistency a complex problem \cite{tan2023survey66}. Context understanding is critical, as many chatbots struggle to maintain coherent conversations and lose track of the emotional context \cite{shum2018eliza68}. To address this, methods like MSEG \cite{firdaus2023multi65} have been developed to generate responses that align with identified emotions. The SEPRG model \cite{li2022seprg70}, for instance, aims to maintain emotional connections while adhering to a consistent persona. Disentangled representations are vital for isolating emotion-specific features, enabling precise emotional expression while maintaining semantic coherence. This is particularly crucial for applications like virtual assistants and mental health support systems \cite{truong2024textual5}. The integration of these representations with LLMs, like the MOPO method, improves emotion control and prediction accuracy \cite{resendiz2024mopo43}. The RCVAE model demonstrates the effectiveness of disentangled representations in generating emotionally rich responses from tweets and emojis \cite{zhou2017mojitalk64}.
A key challenge in emotional support conversation is that supervised fine-tuning of large language models produces psychologically inconsistent responses, and applying direct preference optimization is hindered by the entanglement of psychological strategies and response content in existing emotional support conversation data, leading to ambiguous optimization objectives. DecoupledESC \cite{zhang2025decoupledesc} addresses this by decomposing the emotional support conversation task into two sequential subtasks, strategy planning and empathic response generation, inspired by Gross's Extended Process Model of Emotion Regulation. An Inferential Preference Mining method constructs high-quality preference pairs for each subtask, and each is independently trained via supervised fine-tuning followed by direct preference optimization. This decoupled approach reduces preference bias and improves both strategic soundness and emotional appropriateness of generated responses.
Current emotional support conversation approaches often overlook the deeper cognitive reasoning processes that underpin effective emotional support, focusing instead on data augmentation and synthetic corpus construction. CARE \cite{care2025} proposes a framework that strengthens cognitive reasoning in emotional support conversation without relying on large-scale synthetic data. It leverages the original emotional support conversation training set to guide models in generating logically coherent and supportive responses, then employs reinforcement learning to refine and reinforce the reasoning process. Experimental results demonstrate significant improvements in both the logical soundness and supportive quality of generated responses.
Generating emotionally supportive text that is both empathetically grounded and non-repetitive remains challenging, as current models lack deep empathetic reasoning rooted in psychological principles, and reinforcement learning-based training often suffers from entropy collapse leading to repetitive outputs. COMPEER \cite{compeer2025} proposes controllable empathetic reasoning that combines natural language reasoning with structured psychological steps. It constructs a fine-grained dataset annotated with reasoning correctness and response preferences, and employs reinforcement learning with a unified process-outcome reward model for precise feedback. To mitigate repetitiveness, COMPEER introduces personality-based dialogue rewriting and a redundancy-aware reward reweighting strategy.\\
\subsubsection{Multimodal-Related Challenges}
Cross-modal learning enhances emotional accuracy by integrating text with other modalities like speech and facial expressions. Since this type of challenge is associated with the following sections in emotional content synthesis, here some challenges are underscored briefly and meticulous information will be delivered there. A key challenge in this area is alignment, which involves synchronizing information from different sources, such as aligning visual cues with corresponding text segments. This is particularly difficult when the relationship between modalities is not explicit, as is often the case with text that only describes a video or audio track in general terms \cite{zhang2017multimodal}. Another significant challenge is fusion, which refers to the method of combining features from heterogeneous data sources. Many early models fused modalities only at an abstract level, which fails to capture the time-dependent interactions between them \cite{zhang2017multimodal}. Addressing these challenges is crucial for creating seamless and contextually aware affective experiences. Systems like FaceChat \cite{alnuhait2023facechat18} have demonstrated real-time improvements in emotional accuracy by combining speech recognition, NLP, and facial analysis. The Omni-perception Pre-Trainer \cite{wu2023large79} integrates emotional cues from multiple modalities to enhance the quality of generated text. The MAGIC model highlights the potential of cross-modal learning to improve emotional accuracy and adapt text to diverse user needs \cite{su2022language46}. The emoTTS model \cite{luo2024emotion12} is another example of a cross-modal application that enhances emotion control and generation accuracy.
Chain-of-thought based emotional support conversation methods typically employ rigid, text-only reasoning, which limits adaptability in dynamic multimodal interactions and introduces reasoning noise that degrades support quality. FIRES \cite{fires2025} introduces ``Flexible Thinking'' for multimodal emotional support conversation, enabling models to adaptively select contextually relevant reasoning aspects, including visual scene, emotion, situation, and response strategy. The framework integrates supervised fine-tuning for initial learning with reinforcement learning for refinement, directly linking thinking processes to response quality via tailored rewards. Experiments on the MESC and EMOTyDA datasets demonstrate improved quality and generalizability of emotional support responses through this adaptive multimodal reasoning approach.\\\\
\textbf{\textit{LLM-Based Challenges}:}
Recent research has also focused on leveraging LLMs and developing novel techniques to enhance emotional expression while maintaining semantic coherence. The high computational cost associated with fine-tuning large models for emotional text generation is of importance. APT-LM \cite{gu2024affective42}, a parameter-efficient solution, overcomes computational inefficiency in emotional text generation by freezing pre-trained language models and using minimal affective parameters through prompt-tuning, enhanced by affective decoding that systematically strengthens emotional expression at multiple linguistic levels while preserving fluency. Another key challenge is the evaluation of emotional intelligence in LLMs. While models are becoming more sophisticated, their ability to understand and apply emotional intelligence is still limited. To this end, benchmarks like EmoBench \cite{sabour2024emobench} have been developed to assess the emotional intelligence of LLMs through a series of hand-crafted multiple-choice questions that cover both emotional understanding and application. This benchmark has revealed that even the most advanced LLMs still have a considerable gap to bridge to reach human-level emotional intelligence. 
Multi-Objective Prompt Optimization (MOPO) \cite{resendiz2024mopo43}, a method that optimizes prompts for LLMs to generate text meeting multiple objectives simultaneously, specifically transforms the desired emotion and fits the stylistic requirements of different domains. MOPO employs a three-layer optimization process: Layer-1 generates initial text prompts, Layer-2 refines these by paraphrasing or combining them using genetic operations, and Layer-3 provides fixed prompts to guide the optimization and improves performance by up to 2\% for individual objectives while providing balanced, flexible solutions, enhancing the adaptability and effectiveness of affective text generation across diverse applications.
Generating affective text that conveys specific emotions with controlled intensity and topic relevance, while preserving grammatical accuracy, has been addressed by enhancing the GPT-2 model and integrating emotional priors and applying a gradient descent-based perturbation technique \cite{singh2020adapting20}.
CoDa \cite{evuru2024coda} introduces a training-free data augmentation framework for low-resource NLP that extracts simple heuristic-based constraints from limited training data and verbalizes them to prompt off-the-shelf instruction-tuned LLMs, achieving controlled generation that balances diversity and consistency while outperforming existing methods across 11 datasets.
While large language models exhibit strong reasoning capabilities, they struggle to express emotions in a consistent, controllable, and contextually appropriate manner, and existing prompt-based or fine-tuning-based methods lack flexibility or require costly retraining. The Emotion Vector framework \cite{emotionvectors2025} addresses this by extracting latent representations from internal activation shifts between neutral and emotion-conditioned responses. By injecting these vectors into the hidden states of pretrained large language models during inference, the method enables fine-grained, continuous modulation of emotional tone and intensity without any additional training or architectural modification, while preserving semantic fidelity and linguistic fluency.
Existing emotional support conversation systems rely on situation-centric datasets that lack persistent user identity, which limits the capture of personalized affective nuances. Moreover, opaque and coarse reward signals hinder the development of verifiable empathetic reasoning in large language models. Kardia-R1 \cite{kardia2025} introduces KardiaBench, a large-scale user-grounded benchmark comprising 178,080 question-answer pairs anchored to 671 real-world user profiles, and proposes Rubric-as-Judge Empathetic Reinforcement Learning. This group relative policy optimization-based method uses explainable, human-aligned rubric rewards that tightly couple user understanding, emotional inference, and supportive response generation, enabling interpretable stepwise empathetic cognition with consistent improvements in emotion accuracy, empathy, and persona consistency across multiple large language model backbones.
Finally, another hurdle is modifying generative models to incorporate new emotional attributes without compromising the integrity of the content, a problem exacerbated by the non-differentiable nature of discrete text \cite{wang2018sentigan52}. The community is also addressing the ethical implications of affective computing, particularly the issue of bias. As models become more capable of generating emotional text, there is a growing need to ensure that they are used responsibly and do not perpetuate biases present in the training data \cite{sheng2021societal}. Integration of disentangled representations with LLMs improves emotion control and prediction accuracy.
\section{Emotional Speech Synthesis}
\label{speech}
Emotional speech generation is a key area in the field of AI and HCI, aimed at enabling machines to produce emotionally natural, meaningful speech. This process involves transferring the intended emotions in the speech generation process. In recent years, advancements in machine learning, NLP, and signal processing have propelled speech generation technologies forward, enabling virtual assistants, chatbots, and social robots to engage with users in more natural and fluid ways.
Emotional speech generation can be categorized into two primary subfields: Emotional voice Conversion and ESS. These areas, while related, serve different purposes and are implemented through distinct technologies and techniques.
Even though TTS technology has achieved mature and reliable outcomes, ESS still faces significant challenges. Developing robust EVC systems capable of performing any-to-any emotion conversion could address some of the key issues related to ESS in a two-stage manner and pave the way for substantial progress in this field. Inspired by \cite{Triantafyllopoulos2023}, we present a typical ESS workflow that leverages EVC. Initially, based on the stimuli and the given context, an appropriate response is generated, along with the recognition of the desired emotion to be imbued. A TTS system then generates the spectrum of the target speech in a neutral format. Finally, the EVC component processes the input speech and converts it to the desired emotional output. During the generation and utterance of emotional speech, capturing feedback and stimuli from the environment is crucial. This is because the state of the situation may change, requiring adjustments to both the emotion being expressed and the context of the speech.
In the following, the first approaches for emotional voice generation are discussed. Then, we dive into the challenges in the realm and propose solutions for them.
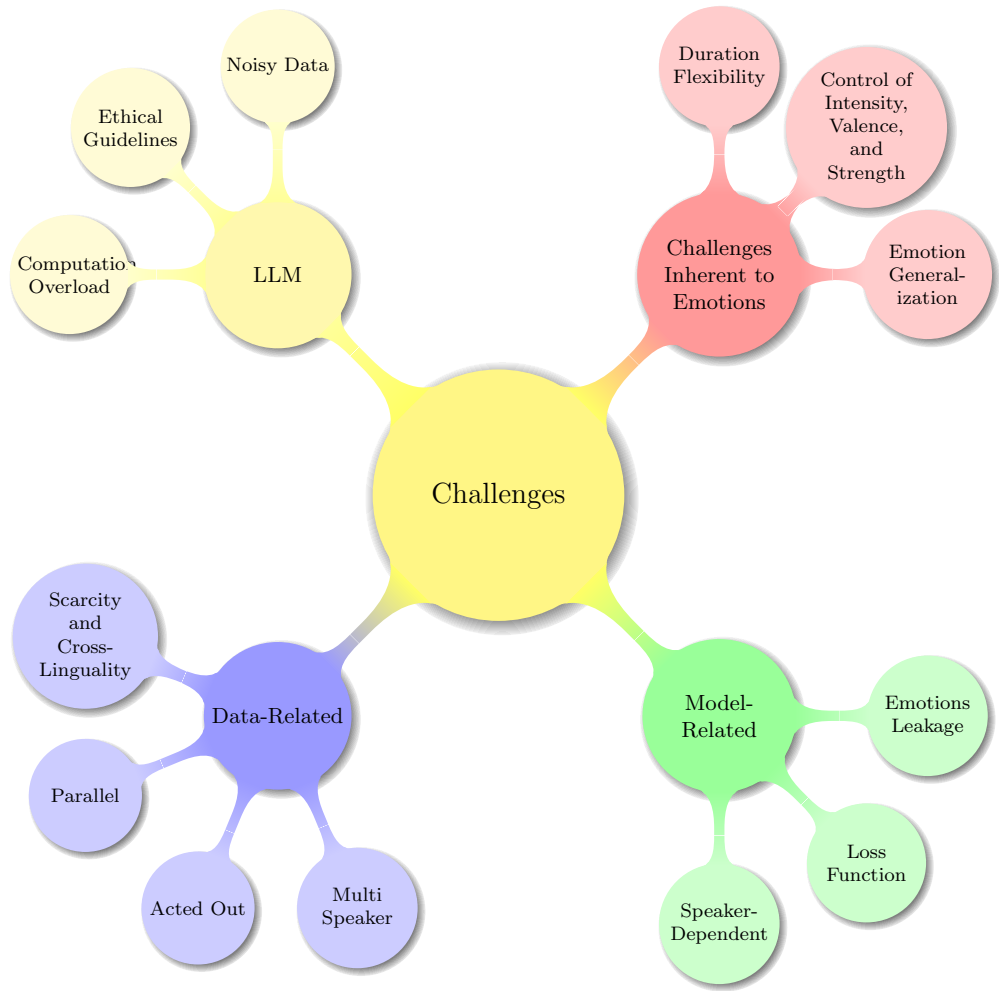
\begin{figure*}[t]
	\centering
	\resizebox{0.8\textwidth}{!}{
		\begin{tikzpicture}[mindmap, grow cyclic, every node/.style={concept, circular drop shadow, minimum size=1.5cm}, concept color=yellow!60, text=black,
			level 1/.append style={level distance=4.5cm,sibling angle=90},
			level 2/.append style={level distance=3cm,sibling angle=45}]
			\node[concept] {Challenges}
			child[concept color=blue!40] { node {Data-Related}
				child[concept color=blue!20!white] { node {Scarcity\\and\\Cross-Linguality} }
				child[concept color=blue!20!white] { node {Parallel} }
				child[concept color=blue!20!white] { node {Acted Out} }
				child[concept color=blue!20!white] { node {Multi Speaker} }
			}
			child[concept color=green!40] { node {Model-Related}
				child[concept color=green!20!white] { node {Speaker-Dependent} }
				child[concept color=green!20!white] { node {Loss Function} }
				child[concept color=green!20!white] { node {Emotions Leakage} }
			}
			child[concept color=red!40] { node {Challenges\\Inherent to\\Emotions}
				child[concept color=red!20!white] { node {Emotion Generalization} }
				child[concept color=red!20!white] { node {Control of Intensity, Valence,\\and Strength} }
				child[concept color=red!20!white] { node {Duration Flexibility} }
			}
			child[concept color=yellow!40] { node {LLM}
				child[concept color=yellow!20!white] { node {Noisy Data} }
				child[concept color=yellow!20!white] { node {Ethical Guidelines} }
				child[concept color=yellow!20!white] { node {Computation Overload} }
			};
		\end{tikzpicture}
	}
	\caption{Challenges in ESS}
	\label{fig:challenges_speech}
\end{figure*}
\subsection{Approaches}
\textbf{\textit{ESS}:	}
ESS or Affective Speech Synthesis focuses on generating speech imbued with emotional expressivity, enabling machines to produce human-like, emotionally rich communication\cite{Triantafyllopoulos2023}. ESS primarily relies on text-to-speech (TTS) technology and plays a crucial role in enhancing HCI and enriching audio broadcast scenarios\cite{Lei2022}. Conversational Emotional Speech Synthesis (CESS) builds upon ESS by incorporating the ability to maintain context from previous interactions. While ESS focuses on generating speech with emotional expressiveness, CESS enhances this by enabling the system to remember and reference past conversations. This memory capability distinguishes CESS from ESS, which typically lacks such continuity. \\\\
\textbf{\textit{Emotional Voice Conversion}:}
Emotional voice conversion (EVC) involves transforming the emotional expression in speech from a source emotion to a target emotion while preserving both the linguistic content and the speaker identity \cite{zhou2022}. In contrast, voice conversion (VC) focuses on modifying one’s voice to sound like another’s without changing the linguistic content \cite{Sisman2020}. EVC involves altering specific acoustic features, such as pitch, timbre, and formant frequencies, to reflect the target emotion while keeping the original speech intact. This process can be used in various applications, including virtual assistants\cite{elgaar2020}, Human-Robot Interaction\cite{crumpton2016}, and HCI\cite{pittermann2010}. Early approaches to EVC relied on statistical models such as Gaussian Mixture Models (GMMs)\cite{aihara2012}, which modeled the relationship between source and target features, and techniques like frequency warping\cite{sheikhan2012}, which directly modified spectral features to reflect emotional changes. While effective for their time, these methods were limited by their reliance on parallel data and their inability to capture complex, non-linear transformations. With advancements in machine learning, modern approaches have shifted towards deep learning techniques\cite{Walczyna2023}, which enable more flexible and robust emotional conversions. Encoder-decoder frameworks have been widely adopted to disentangle speaker identity, linguistic content, and emotional features, facilitating more effective style transfer\cite{zhou2021}.
\subsection{Challenges}
In this section, we investigate the challenges associated with emotional speech generation from diverse standpoints. Consequently, we review the studies on these challenges and their proposed solutions. As illustrated in Figure~\ref{fig:challenges_speech}, we categorize these challenges into four main domains: data-related challenges, model-related challenges, challenges inherent to emotions, and challenges related to LLMs. Table~\ref{tab:emotion_speech} illustrates these categories, detailing their associated sub-issues and highlighting studies attempting to mitigate these challenges. Following, we will discuss challenges and methods proposed to address them.\\
\subsubsection{Data-Related Challenges}
One of the main issues in developing emotional speech generation methods is related to speech datasets. Data-related challenges encompass issues such as the reliance on parallel data\cite{cycleGANemotionalVC, zhou2020, prabhu2024}, data scarcity\cite{schnell2021, zhou2022}, monolingual datasets\cite{ zhou2022}, and the need for realistic (not acted), multi-speaker datasets\cite{lotfian2019}. \\
There are two major types of EVC according to data: 1)Parallel and 2)Non-Parallel. In the former, methods utilize pairs of utterances that contain the same content from the same speaker but are expressed with different emotions. During training, the conversion model learns to map features from the source to the target emotion using these paired feature vectors. Most EVC systems are implemented in this manner \ cite {zhou2022}. In contrast, non-parallel EVC involves learning to map source speech features (e.g., spectral, prosodic) to target emotional features without relying on paired utterances \cite{gao2018}. Parallel data requires both input and output audio recordings for each utterance, which is resource-intensive and time-consuming to collect. Consequently, there is a growing preference for non-parallel data to alleviate the burdens associated with parallel datasets. \\
Adversarial Generative Network (GAN) models, such as CycleGAN\cite{zhou2020transforming} and StarGAN\cite{rizos2020}, have proven particularly powerful in non-parallel EVC scenarios by learning mappings between source and target emotional styles without requiring paired data. CycleTransGAN\cite{cycletransgan}, a CycleGAN-based model enhanced with transformers for non-parallel EVC, uses transformers to capture temporal intra-relations over wider receptive fields. Also, it adopts curriculum learning, gradually increasing the frame length during training to improve feature comprehension, and incorporates fine-grained discriminators for detailed emotional mapping. In \cite{meftah2023}, the use of the StarGANv2-VC framework \cite{li2021starganv2vc} for EVC in English has been explored. Current systems often struggle with the ability to handle multiple speakers and emotions, especially when limited data is available. This research aims to address these gaps by evaluating the StarGANv2-VC model across various configurations, including speaker-dependent, gender-dependent, and gender-independent scenarios.
CycleGAN \cite{cycleGANemotionalVC} is a non-parallel EVC model that incorporates two discriminators to differentiate between natural and converted speech, along with a classifier to identify the underlying emotion from both types of speech. Recently, diffusion models have emerged as a promising approach in emotional speech generation \cite{ma2024}, leveraging their ability to model complex data distributions and generate high-quality emotional transformations. EmoConv-Diff \cite{prabhu2024} presents a diffusion-based model utilizing non-parallel and in-the-wild data. It effectively disentangles lexical content, speaker identity, and emotional information through a diffusion-based decoder, enabling precise emotion transformation via reverse stochastic differential equations.	\\
Current emotional TTS systems struggle to authentically capture human emotions due to reliance on oversimplified emotional labels and single-modality inputs. To solve this, the authors propose UMETTS \cite{li2025umetts}, featuring (1) EP-Align that uses contrastive learning to align emotional features across text, audio, and visual modalities, and (2) EMI-TTS that integrates these aligned embeddings with state-of-the-art TTS models, resulting in significantly improved emotion accuracy and speech naturalness. EmoCat \cite{schnell2021}, a language-agnostic EVC model designed to convert neutral speech to emotional speech, incorporates gradient inverter \cite{ganin2015} blocks to suppress emotional leakage and uses a VAE-based encoder-decoder structure inspired by the CopyCat \cite{karlapati2020}. EmoCat effectively reduces the emotional training data required in the target language by leveraging emotional data from another language. As a result, it achieves high-quality emotional conversion in German, using only 45 minutes of German emotional data while being supported by extensive emotional datasets in US English. EmoSphere++ \cite{cho2025emosphere++} introduces an emotion-adaptive spherical vector (EASV) that represents emotional style through angular position and intensity through radial distance in a spherical coordinate system, enabling fine-grained emotion control without predefined labels. The framework employs a joint attribute style encoder with orthogonality loss to effectively disentangle speaker identity and emotional features, achieving high-quality zero-shot emotional speech synthesis for unseen speakers while eliminating the need for additional discriminators.\\
\begin{table*}[!t]
	\centering
	\caption{Summary of challenges and solutions in emotional speech synthesis}
	\label{tab:emotion_speech}
	\renewcommand{\arraystretch}{1.2}
	\begin{tabularx}{\textwidth}{@{}p{2.2cm}p{3cm}X@{}}
		\toprule
		\textbf{Challenge} & \textbf{Sub-challenge} & \multicolumn{1}{c}{\textbf{Solutions}} \\
		\midrule
		\multirow{8}{2.2cm}{\textbf{Data-Related}} & Parallel Data & Non-parallel data with Adversarial Networks \citep{zhou2020transforming, cycleGANemotionalVC, rizos2020}; Transformer \citep{cycletransgan}; MSP-Podcast dataset \citep{lotfian2019} \\
		\cmidrule(lr){2-3}
		& Scarcity \& Cross-linguality & Emotional datasets in other languages \citep{schnell2021}; ESD dataset \citep{zhou2022}; EmoSphere++ EASV \citep{cho2025emosphere++}; UMETTS contrastive learning \citep{li2025umetts} \\
		\cmidrule(lr){2-3}
		& Mono Speaker & ESD dataset \citep{zhou2022} \\
		\cmidrule(lr){2-3}
		& Acted vs. Natural & MSP-Podcast dataset \citep{lotfian2019} \\
		\midrule
		\multirow{10}{2.2cm}{\textbf{Model-Related}} & Loss Function & Disentangled loss \citep{chou2024}; Perceptual losses \citep{Zhou2022d}; Mutual information minimization \citep{yang2025emotional} \\
		\cmidrule(lr){2-3}
		& Speaker Dependency & VAW-GAN encoder-decoder \citep{zhou2020}; Seq2seq with attention \citep{choi2021}; Self-supervised distillation \citep{cho2025diemo} \\
		\cmidrule(lr){2-3}
		& Emotion Leakage & Gradient reversal layer \citep{schnell2021} \\
		\cmidrule(lr){2-3}
		& Long Sequence Processing & Long-sequence processing \citep{cycletransgan}; Emotion alignment \citep{cycletransgan} \\
		\midrule
		\multirow{6}{2.2cm}{\textbf{Emotion-Related}} & Emotion Generalization & VAW-GAN \citep{zhou20210}; Two-stage training \citep{chen2023}; Ranking-based SVM \citep{zhou2022mixed}; Dual-granularity diffusion \citep{su2025diffemotionvc} \\
		\cmidrule(lr){2-3}
		& Intensity \& Control & Continuous arousal scales \citep{prabhu2024}; Two-stage training \citep{chen2023}; Expressive guidance \citep{chou2024}; Seq2seq attention \citep{choi2021}; Activation steering \citep{xie2025emosteer} \\
		\midrule
		\multirow{8}{2.2cm}{\textbf{LLM-Related}} & Duration Flexibility & Expressive guidance for enhanced emotional diffusion \citep{oh2024durflex} \\
		\cmidrule(lr){2-3}
		& Noisy Output & Emotion/intensity encoders \citep{zhang2025proemo}; Prosody adjustment \citep{zhang2025proemo} \\
		\cmidrule(lr){2-3}
		& Human Expectations & ECoT prompting \citep{cai2024} \\
		\cmidrule(lr){2-3}
		& Computational Overhead & Efficient model design \citep{Mishra2023} \\
		\bottomrule
	\end{tabularx}
\end{table*}
\subsubsection{Model-Related Challenges}
Challenges associated with model structures include difficulties in adapting frameworks for emotional speech generation, defining effective loss functions for various components \cite{cycleGANemotionalVC, chou2024}, preventing emotion leakage \cite{schnell2021}, and overcoming speaker dependency \cite{zhou2020, choi2021} and achieving disentangled emotion and speaker representations. Moreover, models must be capable of processing long sequences of data and accurately aligning each segment with its corresponding emotional expression \cite{cycletransgan}. Although these challenges are significant, they are generally considered less critical than those related to data.\\	
Using minimum generation loss as the objective function in speech generation can constrain the learning process, often leading to over-smoothing of the generated speech parameters. In contrast, employing a more sophisticated loss function that incorporates perceptual aspects, diversity, and emotional qualities can enhance the emotional quality of speech generation, resulting in outputs that resonate more authentically with human listeners. \cite{cycleGANemotionalVC} utilizes various losses, including adversarial, cycle consistency, and emotion classification, to effectively learn parameters in the training process. The adversarial loss measures how distinguishable the generated speech is from the true speech. Cycle consistency loss deposits that the input can keep its original form after passing through the two generators. To capture emotional aspects, this method employs an emotion classification loss. 
\cite{chou2024} uses a disentangled loss in the diffusion EVC model, which reduces the correlation between different speech representations, particularly emotion information and speaker identity. Additionally, an expressive guidance mechanism enhances emotional expressiveness throughout the reverse diffusion process. \cite{Zhou2022d} leverages perceptual losses in the training process to enhance the intelligibility of the generated emotional speech. It uses a contrastive loss to ensure that the text and audio embeddings are similar, effectively disentangling linguistic and emotional elements. Additionally, by using pre-trained Speech Emotion Recognition (SER), this method predicts the emotion category of the generated speech and calculates the emotion classification loss at the utterance level. \\
Emotion leakage occurs when emotional intensity is weak during conversion, leading to a significant decline in signal quality and intelligibility. EmoCat \cite{schnell2021} addresses this issue by incorporating a gradient reversal block before the emotion classifier during training \cite{ganin2015}. This approach reverses the gradients during backpropagation to eliminate any input activation that could aid the emotion classifier, thereby reducing emotional leakage.
\\
(ProEmo) \cite{zhang2025proemo} generates synthesized speech with refined emotional expressiveness and controlled intensity, which conventional TTS systems often lack. It extends the FastSpeech 2 framework \cite{ren2020fastspeech2} by incorporating an emotion encoder and an intensity encoder. The emotion encoder employs a fine-tuned HuBERT transformer \cite{hsu2021hubert} with a classification head to derive robust emotion embeddings from the waveform, while the intensity encoder, also based on HuBERT and equipped with a regression head, estimates continuous emotion intensity through a speaker- and emotion-specific ranking function applied to acoustic features. During inference, GPT-4 \cite{openai2023gpt4} produces global and local scaling factors that modulate the FS2-predicted duration, energy, and pitch via multiplicative and additive adjustments, using a quadratic mapping function to yield speech with human-like expressiveness while preserving linguistic content. \\
There is a need to develop efficient emotional TTS systems that capture nuanced distinctions between emotions while generating expressive speech. Emo-DPO \cite{gao2024emo} combines a text encoder, an emotion-aware decoder, and a flow-matching model with a vocoder, enabling the synthesis of nuanced, controllable emotional speech with improved fidelity and expressiveness. The proposed Emo-DPO framework leverages Direct Preference Optimization (DPO) to refine emotional outputs by optimizing for preferred emotions. Integrating an LLM-based TTS architecture (LLM-TTS), Emo-DPO combines instruction tuning and DPO fine-tuning with Jensen-Shannon regularization to enhance emotional control. Leveraging emotional representation from various perspectives and levels is critical in developing TTS systems that disentangle various emotions. MsEmoTTS \cite{Lei2022}, a multi-scale framework, models emotions at global, utterance, and local levels. It employs a pre-trained BERT-based classifier for global emotion prediction, models intonation variations at the utterance level, and controls syllable-level strengths using a ranking-based method. Built on a Tacotron2 backbone with GMM-based attention, it unifies emotion transfer, prediction, and manual control, achieving superior performance in expressive speech synthesis.\\
Capturing the disentangled emotion and speaker representations in developing emotional speech generation is critical. 
Zhang, G. at \cite{Zhang2023} proposes iEmoTTS, a TTS system for cross-speaker emotion transfer that disentangles prosody from timbre. This approach allows robust emotion transfer even when the speaker and emotion features are entangled. iEmoTTS also supports zero-shot emotion transfer to unseen speakers through a timbre encoder with an information bottleneck mechanism, which retains only speaker-specific timbre features while excluding prosodic information. The system is trained end-to-end in a semi-supervised framework, reducing reliance on extensive labeled data. Additionally, it incorporates pre-trained models for style encoding and SER to reduce training data dependency. It introduces perceptual loss functions to enhance emotion intelligibility and discrimination in converted speech. \\
According to independent emotion expression from the speaker style, \cite{zhou2020} proposed a VAW-GAN-based encoder-decoder structure to learn the spectrum and prosody mapping in a speaker-independent manner. DurFlex-EVC\cite{oh2024durflex} model uses de-stylize and stylize transformers, which separate the source style from input features and apply the target emotion style.\\
Cross-speaker emotion transfer in text-to-speech synthesis requires extracting speaker-independent emotion embeddings, yet existing timbre compression methods such as gradient reversal layers and vector quantization fail to fully separate speaker and emotion characteristics, causing speaker leakage and degraded synthesis quality. DiEmo-TTS \cite{cho2025diemo} addresses this through a self-supervised distillation approach that minimizes emotional information loss while preserving speaker identity. The method introduces cluster-driven sampling and information perturbation to retain emotional content while removing speaker-related factors, and proposes an emotion clustering and matching strategy using emotional attribute prediction and speaker embeddings to generalize to unlabeled data. A dual conditioning transformer integrates style features more effectively, and experimental results on the Emotional Speech Dataset confirm that the approach achieves state-of-the-art performance in learning speaker-irrelevant emotion embeddings, excelling in expressiveness, naturalness, and speaker identity preservation.\\
Current emotional text-to-speech and style transfer methods rely on reference encoders that compress the reference speech into a single global style or emotion vector, which fails to capture nuanced phoneme-level acoustic details and leaves timbre and emotion features entangled in the same representation. To overcome this limitation, \cite{yang2025emotional} proposes a novel emotional text-to-speech method built on the FastSpeech 2 backbone that predicts fine-grained phoneme-level emotion embeddings while disentangling them from global timbre information. The architecture employs two parallel feature extractors within a dedicated style encoder, a global timbre extractor and a phoneme-aware emotion extractor that aligns reference acoustics with target phonemes via multi-head cross-attention. Mutual Information Neural Estimation explicitly minimizes the mutual information between the two representations, ensuring that the timbre embedding retains only speaker-specific information while the emotion embeddings capture prosodic nuance. Experimental results demonstrate that this combination of phoneme-level emotion modeling with principled feature disentanglement outperforms strong baselines in both naturalness and style similarity, producing well-separated emotion clusters and enabling more expressive and controllable emotional speech synthesis.\\
Emotional voice conversion faces persistent challenges due to the complexity of emotion features, which are deeply entangled with speaker identity and linguistic content characteristics, making it difficult to achieve high-quality any-to-any emotion conversion. DiffEmotionVC \cite{su2025diffemotionvc} proposes a diffusion-based framework for any-to-any emotional voice conversion that integrates a dual-granularity emotion encoder capturing both utterance-level emotional context and frame-level acoustic details. The framework employs an orthogonality-constrained condition encoder that disentangles emotion features through gated cross-attention while preserving feature independence with an orthogonal loss. Additionally, multi-objective diffusion training enhances both reconstruction fidelity and emotion discriminability via contrastive learning. Experimental results demonstrate the effectiveness of the framework in maintaining speech quality while optimizing emotional expression.\\
\subsubsection{Challenges Inherent to Emotions}
The third category pertains to the intrinsic nature of emotions. Emotions are not merely discrete states; rather, they are highly dynamic \cite{prabhu2024, zhou20210, chen2023, zhou2022mixed}, varying in duration, intensity, valence, and strength \cite{prabhu2024, chen2023, chou2024, choi2021, oh2024durflex}. This variability introduces complexities that make it challenging to model and reproduce emotional expressions accurately. Among the three categories, this challenge ranks second in importance, underscoring its substantial impact on achieving human-like emotional speech generation.\\
EmoConv-Diff \cite{prabhu2024} uses conditioning on continuous arousal dimensions, allowing for effective control over emotional intensity. 
\cite{chou2024} utilized the generative power of diffusion models to tackle significant issues in previous deep learning approaches that use GANs and Autoencoders (AE), specifically concerning quality degradation and limited control over emotions. This method employs a stochastic differential equation-based diffusion process to progressively transform speech features, enabling precise emotion transformation while preserving speaker identity and linguistic content. A disentangled loss distinguishes speaker and emotion representations, while an expressive guidance mechanism enhances emotional expressiveness during reverse diffusion. \cite{zhou20210}  employs a variational auto-encoding Wasserstein generative adversarial network (VAW-GAN) to transfer seen and unseen emotional style during training and run-time inference. Attention-based Interactive Disentangling Network (AINN)\cite{chen2023} uses a two-stage training process to transfer emotional attributes, such as emotional strength and category, from a reference speech to a source speech while preserving the source's content. \cite{li2022cross} addresses the challenges of speaker leakage, emotion strength control, and effective disentanglement of speaker and emotion features in cross-speaker emotion transfer for TTS. It proposes a modified Tacotron2-based framework incorporating an Emotion Disentangling Module (EDM), which uses emotion and speaker encoders with orthogonality constraints to ensure speaker-irrelevant and emotion-discriminative embeddings. Additionally, a scalar value is introduced to control emotion strength in synthetic speech, enabling flexible adjustments without dependency on manually labeled data. The challenges of mono-scale emotion modeling, limited flexibility in emotion transfer and prediction, and the lack of fine-grained emotion control in emotional speech synthesis are explored in \cite{Lei2022}. To address the challenge of intensity variation, Emovox \cite{Zhou2022d}, a Seq2Seq EVC framework, was developed. Emovox controls emotion intensity using relative attributes to capture fine-grained variations. Zhang, G. at \cite{Zhang2023} proposes iEmoTTS, which uses a probability-based method for emotion intensity control, enabling a nuanced generation of emotional speech with varying strengths. Although prior frameworks mark a significant advancement in the quality and versatility of EVC systems \cite{Sisman2020}, enabling applications like personalized speech synthesis and cross-lingual conversion, they process speech on a frame-by-frame basis, limiting their ability to modify speech duration and also intensity. The first issue was addressed in DurFlex-EVC \cite{oh2024durflex}. The DurFlex-EVC \cite{oh2024durflex} model incorporates a style AE to disentangle emotional style from linguistic content. This is achieved using de-stylize and stylize transformers, which separate the source style from input features and apply the target emotion style. The unit aligner further compresses the features to unit-level representations and predicts durations, creating an efficient and context-aware framework for emotional style transformation.\\
Most existing text-to-speech systems offer only coarse and rigid emotion control, typically relying on discrete emotion labels or carefully crafted emotional text prompts, which makes fine-grained emotion manipulation either inaccessible or unstable, and these models require extensive high-quality datasets for training. EmoSteer-TTS \cite{xie2025emosteer} proposes a training-free approach to achieve fine-grained speech emotion control, including emotion conversion, interpolation, and erasure, through activation steering. The method builds on the empirical observation that modifying a subset of internal activations within a flow matching-based text-to-speech model can effectively alter the emotional tone of synthesized speech. It develops an efficient algorithm comprising activation extraction, emotional token searching, and inference-time steering that can be seamlessly integrated into a wide range of pretrained models. Extensive experiments demonstrate that this approach enables fine-grained, interpretable, and continuous control over speech emotion, outperforming the state of the art as the first method to achieve training-free and continuous fine-grained emotion control in text-to-speech synthesis.\\
Transferring a mix of primary emotions is an essential task in speech synthesis. Most existing methods focus on imitating a single emotion; however, to facilitate natural and engaging interactions between humans and agents, it is essential to incorporate mixed emotions into speech synthesis. Unfortunately, the development of such models is hindered by the lack of extensive multi-speaker corpora that contain mixed emotion labels \cite{kreibig2017}. To tackle this issue, \cite{zhou2022mixed} uses a ranking-based SVM to model emotional styles as attributes reflecting the relevance of different emotions. This approach enables the system to quantify relationships between emotion pairs and synthesize new emotional mixtures by manually defining these attributes during conversion. Its architecture uses a seq2seq emotional voice conversion framework that integrates these attributes for mixed emotion synthesis. The architecture integrates a text encoder, an emotion encoder for embeddings, and a decoder with bidirectional LSTMs and attention mechanisms to generate natural and expressive emotional speech. Moreover, \cite{zhou2022speech} uses a pre-trained model on massive speech corpora without emotional annotations, which is then fine-tuned using emotional speech data. It similarly uses a ranking function to determine the level of primary emotion according to variations between pairs of emotional speech samples.
\cite{bott2024controlling} deals with controllable emotional prosody in TTS systems using natural language prompts. A FastSpeech 2-based architecture integrates prompt embeddings, derived from DistilRoBERTa, with speaker embeddings through a squeeze-and-excitation mechanism for accurate prosodic control. The system produces high-quality, emotionally expressive speech while preserving the speaker's identity by training with curriculum learning. Period VITS \cite{Shirahata2023} focuses on the challenge of unstable pitch contours and artifacts in end-to-end emotional TTS systems caused by prosodic diversity. The proposed Period VITS integrates a periodicity generator for explicit pitch modeling, producing sample-level sinusoidal sources to enhance pitch stability and waveform quality. A frame pitch predictor within the prior encoder estimates frame-level prosodic features while normalizing flows augment prior distributions for richer acoustic variation. The HiFi-GAN-based decoder aligns pitch signals with latent acoustic features through down-sampling layers. The model is optimized end-to-end using variational inference with combined loss functions to ensure stability and expressiveness. The challenge of incorporating fine-grained intonation control, particularly questioning intonation, into emotional speech synthesis was considered in \cite{Tang2023}. It claims existing TTS models can transfer emotions but struggle to model nuanced prosody like "angry question" versus "angry statement." The proposed QI-TTS builds on FastSpeech 2 and introduces a multi-style extractor to capture emotion at the sentence level and intonation at the final syllable level. By using relative attributes, it models intonation intensity in an unsupervised manner, enabling fine-grained control. A gradient reversal layer ensures content and style disentanglement to prevent interference. \\
Humans can experience roughly 34,000 distinct types of emotions. This includes eight basic emotions, along with secondary emotions that arise from combinations of these basics, as explained in the Theory of the Emotion Wheel \cite{plutchik2001nature}. While secondary emotions are critical in social human interaction, synthesizing these emotions is often overlooked. EmoMix \cite{Tang2023a} addresses the challenge of synthesizing emotional speech with controllable intensity and the ability to express mixed emotions, a significant limitation of current text-to-speech (TTS) systems. EmoMix overcomes this issue using a diffusion probabilistic model conditioned on emotion embeddings from a pre-trained SER model. It achieves flexible emotion control by blending predicted noise for different emotions during the sampling process and mixing neutral noise with the target emotion's noise. The architecture of EmoMix incorporates several components: GradTTS, a U-Net model with linear attention, the SER model, the HiFi-GAN vocoder, and a style reconstruction loss that ensures emotional consistency and naturalness in the synthesized speech. \\
\subsubsection{LLM Challenges}
The application of LLMs has shown great potential in regulating emotional expression in synthesized speech, particularly through prompt-based techniques \cite{Guo2023, Sigurgeirsson2024}. This approach enhances the expressiveness and naturalness of generated speech while preserving clarity and quality \cite{zhang2025proemo}. While LLMs excel at producing diverse and contextually rich text, their output can sometimes be noisy and inconsistent when directly applied to emotional modifications in speech synthesis. In (ProEmo) \cite{zhang2025proemo}, researchers noted that relying solely on LLM outputs for emotion control could compromise expressiveness due to output noise. They addressed this by integrating specific emotion and intensity encoders to guide prosody adjustment in systems like FastSpeech2. Moreover, approaches that modify backbone architectures (e.g., FastSpeech2) by adding emotion and intensity encoders have been explored to bridge the gap of effective integration with TTS systems. Such systems leverage the linguistic expressiveness of LLMs while fine-tuning acoustic outputs to convey emotion, although this remains an active area of research with room for improvement. Emotional expression is inherently subjective, and LLMs may generate outputs that do not align with human emotional expectations or ethical guidelines. \cite{Li2024} have proposed methods like the Emotional Chain-of-Thought (ECoT) prompting technique. This plug-and-play method guides LLMs through multiple reasoning steps to generate emotionally appropriate content, thereby improving human preference alignment in generated outputs. \\
Computational overhead and latency present significant challenges, particularly for real-time applications such as human-robot interaction. \cite{Mishra2023} explored the real-time use of LLMs for tasks like emotion prediction in dialogue systems. The study indicated that efficient model design and careful engineering are crucial for the practical deployment of these technologies.

\begin{figure*}[t]
	\centering
	\resizebox{0.8\textwidth}{!}{
		\begin{tikzpicture}[mindmap, grow cyclic, every node/.style={concept, circular drop shadow, minimum size=2.0cm}, concept color=yellow!60, text=black, level 1/.append style={level distance=4.6cm,sibling angle=72}]
			\node[concept] {Challenges}
			child[concept color=blue!40] { node {Data-Related\\Challenges}
				child[concept color=blue!20!white] { node {Scarcity \\ \& Diversity} }
				child[concept color=blue!20!white] { node {Occlusions} }
				child[concept color=blue!20!white] { node {Limited Training Data} }
			}
			child[concept color=green!40] { node {Model-Related\\Challenges}
				child[concept color=green!20!white] { node {Realism \\\& \\ Identity Preservation} }
				child[concept color=green!20!white] { node {Scalability \\\& \\ Adaptability} }
				child[concept color=green!20!white] { node {Generalization} }
				child[concept color=green!20!white] { node {Mode collapse \\\& Training instability} }
			}
			child[concept color=red!40] { node {Emotion Synthesis}
				child[concept color=red!20!white] { node {Nuanced Expressions} }
				child[concept color=red!20!white] { node {Localized Control \\ over facial muscles} }
				child[concept color=red!20!white] { node {Micro-expressions} }
			}
			child[concept color=purple!40] { node {Multimodal\\Integration}
				child[concept color=purple!20!white] { node {Cross-Modal Alignment} }
				child[concept color=purple!20!white] { node {Contextual Relevance \\ of Expressions} }
			}
			child[concept color=orange!40] { node {Computational\\Aspects}
				child[concept color=orange!20!white] { node {Efficiency \\ \& Cost} }
				child[concept color=orange!20!white] { node {Real-time Performance} }
			};
		\end{tikzpicture}
	}
	\caption{Challenges in emotional face synthesis}
	\label{fig:face_challenges}
\end{figure*}
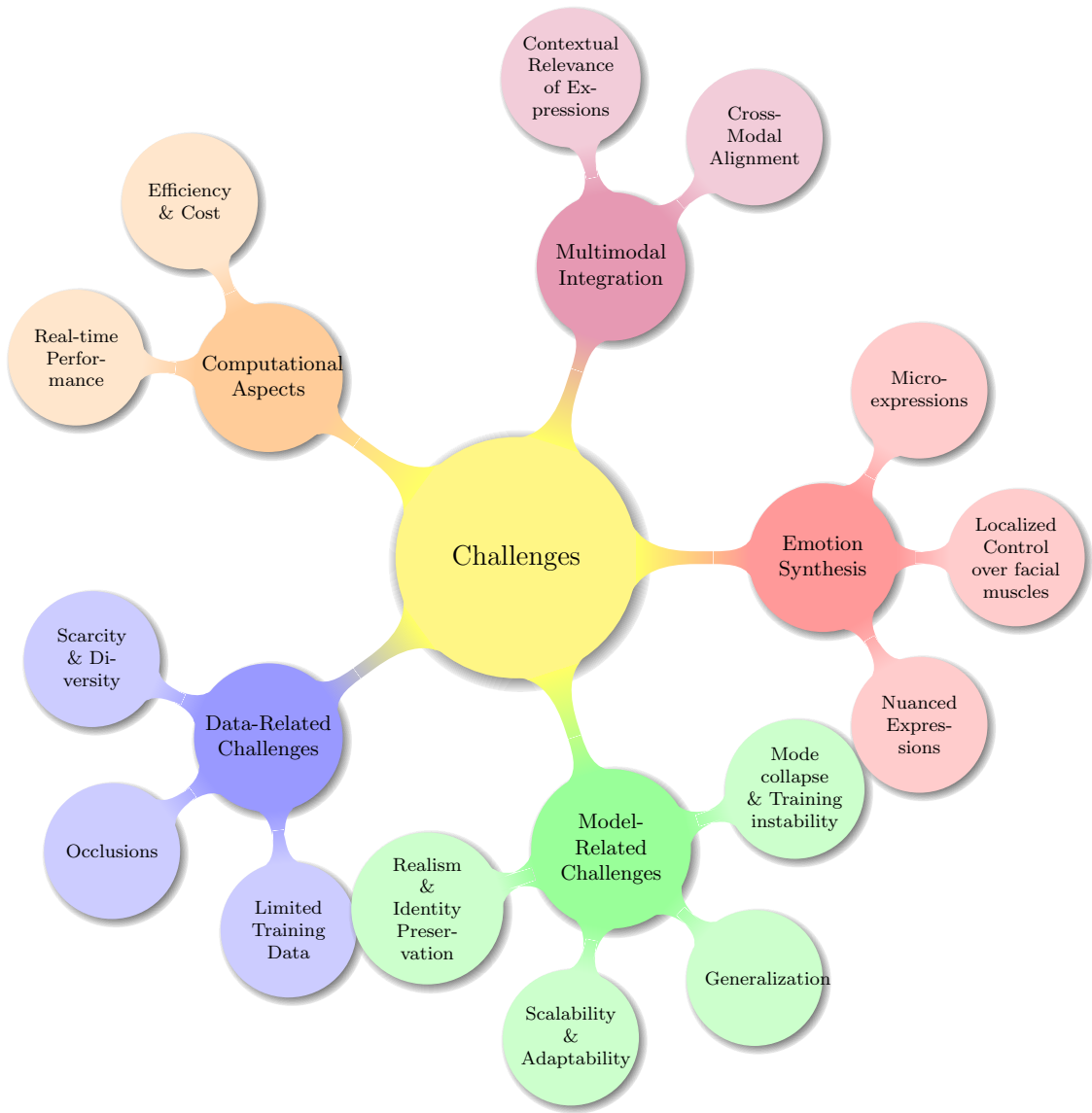
\section{Emotional Face Synthesis}
\label{face}
Emotional facial expressions are fundamental nonverbal signals that enable accurate assessment of internal states in psychiatric applications and bolster the integration of machine learning in mental-health diagnostics \cite{coda2023inducing}. In therapeutic contexts, these cues guide clinicians in recognizing and addressing patient emotions \cite{iftikhar2024therapy}, while in everyday social exchanges, their contextual interpretation is critical for effective communication and rapport building \cite{han2024knowledge}. Beyond human-to-human interaction, facial expressions drive dynamic user engagement across interactive systems, enhancing empathetic response generation in dialogue agents \cite{wang2025sibyl} and underpinning social intelligence in conversational AIs \cite{chen2024socialbench}. The breadth of these roles underscores the centrality of facial affect in advancing emotion research and human-machine interaction \cite{chen2024using}.
Emotional face synthesis builds on this foundation by enabling virtual avatars, social robots, and conversational agents to exhibit realistic emotional expressions, highlighting its transformative potential in interactive applications \cite{tan2024edtalk}. By integrating multimodal cues—visual, auditory, and textual—the emotional authenticity and contextual sensitivity of AI systems are greatly enhanced, which is critical for education, therapy, and collaborative AI \cite{zhan2022multimodal}. Moreover, precise lip synchronization combined with coordinated nonverbal signals (facial expressions, gaze, head pose) yields seamless, engaging animations for film, gaming, and HCI \cite{liang2024emotional,ma2024,wang2024emotivetalk}. These advancements not only meet ethical and equitable AI standards but also continue to drive innovation across virtual/augmented reality and teleconferencing environments. 
Here, we briefly outline three core approaches for generating realistic talking faces from audio inputs, static images, and their integration. We then analyze the primary challenges in this field and evaluate the various methods researchers have developed to overcome them. We highlight diverse application domains—virtual avatars, social robotics, sentiment analysis, and HCI—where advanced emotional synthesis drives deeper engagement and improved user experience \cite{pandey2024target, saunders2023readavatars, xu2024textto3d, pataranutaporn2021aicharacters}, and discuss ethical considerations such as bias mitigation and responsible deployment of facial manipulation technologies. 
\subsection{Approaches}
\textbf{\textit{Photo-realistic facial animation:}}
Photo-realistic facial animation techniques are advanced computational methods designed to generate digital facial animations that closely resemble real human expressions and movements in terms of both visual detail and emotional subtlety. These techniques, such as the Warp-Guided GANs introduced by Geng et al. \cite{geng2018warpgan}, advance visual fidelity and emotional nuance, significantly enhancing the realism of animated characters. \\
\textbf{\textit{Audio-driven synthesis:}}
Audio-driven synthesis systems, exemplified by EDTalk \cite{tan2024edtalk}, integrate auditory and visual inputs to create interactions imbued with emotional depth and realism. Unlike photo-realistic facial animation techniques, which prioritize high visual fidelity to achieve lifelike faces, these methods do not require a highly realistic facial appearance. Instead, they focus on synchronizing audio and visual cues to produce expressive and engaging facial animations that enhance the quality of virtual interactions.\\
\textbf{\textit{Multi-modal synthesis and editing:}}
Multi-modal synthesis and editing techniques, encompassing GAN-based, autoregressive, diffusion, and NeRF approaches, enhance emotional authenticity and contextual awareness in facial animations \cite{zhan2022multimodal}.
\subsection{Challenges}
Emotional facial generation faces several critical challenges, including challenges related to data, models, emotions, multi-modal integration, and computational inefficiencies. Addressing these issues is essential for creating inclusive, robust systems capable of accurately representing the full spectrum of human emotions. These insights provide a foundation for advancing the field and overcoming current limitations. Figure~\ref{fig:face_challenges} classifies these challenges into five major domains: data, learning models, emotion synthesis, multimodal, and computational issues. Table~\ref{tab:face_synthesis} summarizes these categories, outlining their respective sub-issues and highlighting studies that address these challenges. Next, we will discuss the challenges and the methods proposed to overcome them.\\
\begin{table*}[!t]
	\centering
	\caption{Summary of challenges and solutions in emotional face synthesis}
	\label{tab:face_synthesis}
	\renewcommand{\arraystretch}{1.2}
	\begin{tabularx}{\textwidth}{@{}m{2.8cm}p{2.7cm}X@{}}
		\toprule
		\textbf{Challenge} & \textbf{Sub-challenge} & \multicolumn{1}{c}{\textbf{Solutions}} \\
		\midrule
		\multirow{6}{2.8cm}{\textbf{Data-Related}} & Scarcity \& Diversity & Promoting diversity in datasets \citep{xu2024id3}; Enhanced annotation for expressions \citep{chen2018vgan} \\
		\cmidrule(lr){2-3}
		& Limited Training Data & Synthetic data generation \citep{xu2024vasa1}; StyleGAN-generated data \citep{akamatsu2025comface}; EmoVOCA dataset combination \citep{nocentini2025emovoca} \\
		\cmidrule(lr){2-3}
		& Occlusions & Robust model design \citep{geng2018warpgan}; Specialized training data \citep{xu2024vasa1} \\
		\midrule
		\multirow{12}{2.8cm}{\textbf{Model-Related}} & Realism \& Identity & Advanced GANs (StyleGAN) \citep{jiang2023styleipsb}; Diffusion Models \citep{yin2022styleheat}; Identity-emotion disentanglement \citep{tan2025disentangle}; Instruction-driven 3D generation \citep{vo2025instruction} \\
		\cmidrule(lr){2-3}
		& Generalization & Robust architectures \citep{xu2024vasa1}; Domain adaptation \citep{geng2018warpgan}; Long-range autoregressive diffusion \citep{zhang2025x} \\
		\cmidrule(lr){2-3}
		& Scalability & Unified frameworks (UniPortrait) \citep{he2024uniportrait}; Motion modeling (MotionGAN) \citep{otberdout2022sparse}; Temporal layers \citep{xu2024magicanimate}; Sparse landmark methods \citep{tu2024stableanimator} \\
		\cmidrule(lr){2-3}
		& Training Stability & Improved GAN architectures \citep{jiang2023styleipsb}; Regularization techniques \citep{yin2022styleheat}; Alternative loss functions \citep{jiang2023styleipsb} \\
		\midrule
		\multirow{6}{2.8cm}{\textbf{Emotion Synthesis}} & Nuanced Expressions & Fine-grained control (Action Units) \citep{yin2022styleheat}; Continuous modeling (GC-GANs) \citep{qiao2018geometry}; Subtle cue analysis \citep{retsinas2024analysis}; Emotion-audio spatial attention \citep{ma2026esgaussianface}; Continuous valence-arousal conditioning \citep{cha2025emotalkinggaussian}; FLAME-guided diffusion \citep{zhang2024emodiffhead}; Semantic expression parameters \citep{shen2025emohead} \\
		\cmidrule(lr){2-3}
		& Micro-expressions & High-res temporal modeling \citep{xu2024magicanimate} \\
		\cmidrule(lr){2-3}
		& Localized Control & Advanced facial modeling \citep{retsinas2024analysis}; Anatomical models \citep{varanka2024localized} \\
		\midrule
		\multirow{6}{2.8cm}{\textbf{Multi-Modal}} & Feature Alignment & Collaborative diffusion \citep{huang2023collaborative}; Self-supervised learning \citep{pham2017speech}; Audio-visual integration (C3D-DBN) \citep{nguyen2017deep}; Text-to-expression synthesis \citep{cheng2024emotionllama}; Multi-modal emotion embedding \citep{yee2025synchrorama} \\
		\cmidrule(lr){2-3}
		& Contextual Relevance & Textual data integration \citep{cheng2024emotionllama}; Sentiment analysis (VECTN) \citep{pandey2024target}; Cross-reconstruction disentanglement \citep{liu2025medtalk} \\
		\midrule
		\multirow{6}{2.8cm}{\centering \textbf{Computational}} & Efficiency & Hybrid models (GANs+Diffusion) \citep{xu2024magicanimate}; Compact representations \citep{he2024uniportrait}; Efficient alignment \citep{liu2018efficient}; Model quantization/pruning \citep{liu2018efficient}; Knowledge distillation \citep{he2024uniportrait}; Hardware acceleration \citep{he2024uniportrait} \\
		\bottomrule
	\end{tabularx}
\end{table*}
\subsubsection{Data-Related Challenges}
Dataset bias significantly impacts the generalization capability of affective face generation models, particularly in representing diverse demographics, emotional expressions, and cultural nuances. Limited training data diversity further restricts the generalizability of models to diverse emotional states \cite{siddiqui2022acgan}. Biased datasets often fail to capture the complexity of human emotions, leading to inequities in model performance and reduced inclusivity \cite{washington2021}. This issue is compounded by the difficulty of achieving localized control over facial muscle movements and generating nuanced expressions \cite{varanka2024localized}. Moreover, existing benchmarks inadequately capture the fluidity and diversity of emotional expressions, limiting the generalization capabilities of models in real-world scenarios \cite{xu2024textto3d}. Current methods relying on detailed face modeling encounter challenges in unconstrained environments with variations in pose, lighting, and expression \cite{wiles2018x2face}. The need for comprehensive annotation of facial attributes further restricts the manipulation of identity and attributes, impeding the development of robust models \cite{bao2018openset}. 
Multi-modal datasets also struggle with aligning features across modalities, resulting in limited high-resolution, contextually relevant outputs \cite{zhan2022multimodal}. The lack of diversity within and across identities in datasets exacerbates these issues, underscoring the importance of inclusive data collection practices \cite{xu2024id3}. ComFace \cite{akamatsu2025comface} is introduced to address the challenge of capturing intra-personal facial changes for health and emotion monitoring, hindered by the scarcity of temporally varying real-world face images. Its novel representation learning approach leverages StyleGAN-generated synthetic data to simultaneously learn both inter-personal facial differences and intra-personal facial changes within individuals. Remarkably, their method trained exclusively on synthetic data achieves comparable or superior performance to state-of-the-art approaches trained on real images across multiple facial change estimation tasks.
Many approaches reconstructed 3D faces from 2D videos using parametric models (3DMMs), but these lacked precision for accurate lip-syncing. The authors solve this with EmoVOCA \cite{nocentini2025emovoca}, a data-driven framework that combines VOCAset (neutral 3D talking heads with speech) and Florence4D (expressive 3D faces without speech) using a Double Encoder/Shared Decoder architecture, where separate encoders learn speech and expression features while a shared decoder combines them to create synthetic emotional 3D talking heads that preserve both accurate lip movements and convincing expressions.
Privacy concerns intersect with dataset bias, as existing methods often compromise either privacy or the quality of expression recognition \cite{chen2018vgan}. Ethical implications arise from biased datasets, which hinder fairness and generalization across diverse populations \cite{stahl2023ethics}. Addressing these concerns requires promoting diversity in datasets while maintaining identity consistency, as well as enhancing annotation processes to improve the representation of nuanced and localized expressions \cite{xu2024id3, varanka2024localized}. These efforts are critical for fostering fairness, equity, and trust in affective face generation technologies.
\subsubsection{Model-Related Challenges}
Three primary sub-challenges reside in model-related challenges: preserving identity, ensuring the realism of the model's outputs, and achieving generalization to real-world scenarios. Advanced generative methodologies for generating emotionally expressive facial animations, such as GANs and diffusion models, could address these issues, and these frameworks have significantly improved the quality of affective facial generation, enabling applications in human-computer interaction, virtual avatars, and sentiment analysis.
GANs have been instrumental in advancing affective face generation, producing high-quality animations that capture nuanced emotional states. Methods like EDTalk \cite{tan2024edtalk} enhance realism by disentangling facial features such as mouth shape, head pose, and emotional expression. Similarly, techniques integrating driving videos with single-image inputs, as demonstrated in \cite{averbuch2017portraits}, expand the versatility of GANs in synthesizing dynamic animations.
Despite the greatness of GANs, challenges like generalization to real-world faces persist \cite{xu2024vasa1}. Diffusion models provide a robust alternative to GANs, leveraging iterative noise refinement to generate diverse emotional attributes while maintaining identity consistency. Conditional diffusion frameworks such as ID3 excel at preserving intra-class identity and generating emotionally coherent expressions \cite{wang2023score}. Innovations like Stable Animator address temporal stability, ensuring smooth transitions in dynamic animations \cite{tu2024stableanimator}. These models also bridge textual input and emotional synthesis, as demonstrated by frameworks like ContinuousText-to-Expression Generator (CTEG) and Globally-informed Gaussian Avatar (GiGA), which produce nuanced 3D avatars \cite{xu2024textto3d}.
These models excel at capturing subtle emotional cues and offer advantages in emotional synthesis \cite{zhan2022multimodal}. Innovations like READ Avatars, Emo3D \cite{xu2024textto3d} generation, and UniPortrait \cite{he2024uniportrait} leverage adversarial loss, autoregressive models, and unified frameworks to enable applications in hyper-realistic avatars, empathetic human-computer interaction, and nuanced sentiment analysis \cite{saunders2023readavatars,pataranutaporn2021aicharacters}.
Hybrid approaches combining GANs and diffusion models aim to leverage the strengths of both techniques. For instance, UniPortrait \cite{he2024uniportrait} employs a dual-module architecture to enhance identity preservation and adaptability, balancing the speed of GANs with the nuanced output of diffusion models. Additionally, frameworks utilizing Action Units for fine-grained expression control demonstrate the potential for bridging the gap between realism and expressiveness in affective face synthesis \cite{yin2022styleheat}. By integrating the strengths of GANs and diffusion models, researchers are advancing the synthesis of nuanced, contextually appropriate facial expressions. These models enhance realism and emotional dynamics, enabling applications in virtual avatars, human-computer interaction, and personalized content creation \cite{pumarola2018ganimation, bouzid2022fevgan, siddiqui2022acgan, huang2017dyadgan}. \\
Existing diffusion-based talking head generation methods struggle to produce emotionally expressive portraits while preserving speaker identity, due to insufficient utilization of audio's inherent emotional cues, identity leakage in emotion representations, and isolated learning of emotion correlations. DICE-Talk \cite{tan2025disentangle} addresses these limitations through a framework that disentangles identity from emotion and then cooperates emotions with similar characteristics. It develops a disentangled emotion embedder that jointly models audio-visual emotional cues through cross-modal attention, representing emotions as identity-agnostic Gaussian distributions. A correlation-enhanced emotion conditioning module with learnable Emotion Banks explicitly captures inter-emotion relationships through vector quantization and attention-based feature aggregation, while an emotion discrimination objective enforces affective consistency during the diffusion process through latent-space classification. Experiments on the MEAD and HDTF datasets demonstrate superiority in emotion accuracy while maintaining competitive lip synchronization performance.
Audio-driven portrait animation methods typically emphasize lip synchronization and short-range visual fidelity in constrained speaking scenarios, but fail to capture nuanced, dynamically evolving emotions that flow coherently with the rhythm and content of speech over long temporal contexts. X-Actor \cite{zhang2025x} presents an audio-driven portrait animation framework that generates lifelike, emotionally expressive talking head videos from a single reference image and an input audio clip, enabling actor-quality long-form portrait performance. Central to the approach is a two-stage decoupled generation pipeline: an audio-conditioned autoregressive diffusion model predicts expressive yet identity-agnostic facial motion latent tokens within a long temporal context window, followed by a diffusion-based video synthesis module that translates these motions into high-fidelity video animations. By operating in a compact facial motion latent space decoupled from visual and identity cues, the autoregressive diffusion model captures long-range correlations between audio and facial dynamics through a diffusion-forcing training paradigm, enabling infinite-length emotionally rich motion prediction without error accumulation.
Generating 3D facial expressions from natural language instructions remains challenging because most existing methods rely on discrete emotion labels or predefined expression categories, which cannot capture the richness and specificity of textual descriptions for both static expressions and dynamic expression transitions. \cite{vo2025instruction} proposes an instruction-driven approach for 3D facial expression generation and transition that takes text instructions as input and produces corresponding 3D facial animations. The method leverages a language model to interpret free-form textual descriptions and maps them to parametric 3D face model expression parameters, enabling both the generation of target expressions and smooth transitions between emotional states. This text-based control paradigm offers greater flexibility and user accessibility compared to conventional label-driven or reference-driven approaches, broadening the applicability of 3D facial animation in interactive systems and content creation.
\subsubsection{Multi-modal Integration Challenges}
Multi-modal integration is pivotal for generating emotionally authentic facial expressions. By combining visual, auditory, and textual cues, multi-modal frameworks align generated expressions with intended emotional states, ensuring contextual relevance. Frameworks like EmotiveTalk \cite{wang2024emotivetalk} and Emotion-LLaMA \cite{cheng2024emotionllama} use audio decoupling and self-supervised learning to align speech, lip movements, and expressions, producing realistic talking-head videos \cite{liang2024emotional, zhan2022multimodal}. Collaborative Diffusion \cite{huang2023collaborative} exemplifies this by integrating pre-trained uni-modal diffusion models, enhancing emotional synthesis through modality synergy. EDTalk \cite{tan2024edtalk} exemplifies the importance of synthesizing multiple modalities for lifelike outputs.
Audio-visual integration remains central to multi-modal systems; methods like C3D-DBN \cite{nguyen2017deep} align auditory cues (e.g., tone, pitch) with visual signals for coherent emotional outputs. Emotion-LLaMA further incorporates textual data, enabling contextually appropriate expression synthesis \cite{cheng2024emotionllama}. Techniques for compact multi-modal representations, like \cite{liu2018efficient} improve real-time processing and scalability, making these systems suitable for virtual avatars and conversational AI. Multi-modal frameworks like VECTN integrate textual data for sentiment analysis, aligning emotional synthesis with contextual information and capturing complex states like sarcasm \cite{pandey2024target}. By synthesizing facial expressions aligned with multi-modal cues, these systems enhance user engagement and emotional resonance in AI-driven interactions \cite{zhan2022multimodal}.
Most existing emotion-aware talking face generation methods rely on a single modality, either audio or image, for emotion embedding, which limits their ability to capture nuanced affective cues, and conditioning on a single reference image restricts the representation of dynamic changes in actions or attributes across time. SynchroRaMa \cite{yee2025synchrorama} introduces a framework that integrates a multi-modal emotion embedding by combining emotional signals from text via sentiment analysis and audio via speech-based emotion recognition and audio-derived valence-arousal features, enabling the generation of talking face videos with richer and more authentic emotional expressiveness. To ensure natural head motion and accurate lip synchronization, the framework includes an audio-to-motion module that generates motion frames aligned with the input audio. Additionally, scene descriptions generated by a large language model serve as additional textual input, capturing dynamic actions and high-level semantic attributes that enhance temporal consistency and visual realism.
Audio-driven emotional 3D facial animation typically relies on static and predefined emotion labels, which limits the diversity and naturalness of generated expressions and prevents fine-grained dynamic emotional control. MEDTalk \cite{liu2025medtalk} proposes a framework for fine-grained and dynamic emotional talking head generation that first disentangles content and emotion embedding spaces from motion sequences using a carefully designed cross-reconstruction process, enabling independent control over lip movements and facial expressions. Beyond conventional audio-driven lip synchronization, the method integrates audio and speech text to predict frame-wise intensity variations and dynamically adjust static emotion features for realistic emotional expressions. Furthermore, multimodal inputs including text descriptions and reference expression images guide the generation of user-specified facial expressions, and the generated results are compatible with MetaHuman for integration into industrial production pipelines.
\subsubsection{Computational Challenges}
These systems often face computational inefficiencies and scalability challenges, particularly in synthesizing realistic facial expressions under diverse conditions. 
Advanced generative models, such as GANs and diffusion models, require substantial computational resources for training and inference, limiting their applicability in resource-constrained environments \cite{zhan2022multimodal}. Slow inference speeds and extensive training data requirements further hinder real-time applications and large-scale deployments. Existing methods, such as Warp-Guided GANs, struggle with generalizing to unseen data or scenarios with significant occlusions, highlighting the limitations of current frameworks \cite{geng2018warpgan}. Additionally, techniques that assume a neutral face as a starting point or rely on simplified representations restrict adaptability to dynamic environments \cite{averbuch2017portraits}. Identity blending and the need for extensive fine-tuning also pose challenges, as frameworks like UniPortrait demonstrate the resource-intensive nature of achieving identity preservation and adaptability \cite{he2024uniportrait}. The inability to effectively model dynamic and contextually adaptive facial animations further limits the realism of generated outputs. 
Many methods fail to capture intricate temporal dynamics, such as micro-expressions and head movements, which are critical for producing emotionally resonant animations \cite{netsHilbert,xu2024vasa1}. Multi-modal frameworks introduce additional complexity during training and inference, requiring significant resources to align features and synthesize coherent outputs \cite{zhan2022multimodal}. Efforts to address these challenges include optimizing algorithms and architectures to reduce resource consumption while maintaining high-quality synthesis. Addressing computational constraints and improving multi-modal integration will enhance the robustness and accessibility of affective face generation systems, facilitating their application in diverse domains \cite{liu2018efficient, zhan2022multimodal}.\\
\subsubsection{Emotion Synthesis Challenges}
Temporal consistency is a cornerstone of dynamic facial animation, preventing abrupt transitions that undermine emotional impact and is critical for video synthesis, ensuring smooth transitions and coherent animations \cite{zhan2022multimodal, tan2024edtalk}. Techniques like those in \cite{tan2024edtalk} disentangle temporal dynamics from other facial attributes, while methods such as driving-video-based animation ensure smooth transitions \cite{averbuch2017portraits}.
Techniques like MotionGAN model expression transitions on a hypersphere, minimizing motion artifacts and enhancing temporal fidelity \cite{otberdout2022sparse}. Sparse landmark-based methods leverage anatomical priors for coherent facial deformations with reduced computational complexity \cite{xu2024magicanimate}. Disentangling temporal dynamics from attributes like expression and pose further enhances coherence. Frameworks integrating temporal layers mitigate interference with spatial priors, ensuring stable animations \cite{averbuch2017portraits, tu2024stableanimator,xu2024magicanimate}. 
Multi-modal cues, such as audio and textual signals, align facial dynamics with emotional intent, as seen in EmotiveTalk, which synchronizes lip movements and expressions with emotional audio cues \cite{pham2017speech, liang2024emotional, wang2024emotivetalk}. Innovations like Takin-ADA address challenges like expression leakage, while Action Unit-based techniques provide anatomically accurate representations of expressions. These advancements improve dynamic animations, broadening their applicability in teleconferencing, virtual reality, and digital media \cite{vougioukas2020realistic} \cite{pumarola2018ganimation} \cite{lin2024takin}.  
Techniques such as SMIRK \cite{retsinas2024analysis} and Auxiliary Classifier GANs \cite{siddiqui2022acgan} aim to address complex emotional nuances by enhancing the quality and diversity of generated expressions, yet challenges persist in capturing them. Occlusions often lead to assumptions of exaggerated expressions, compromising the fidelity of nuanced emotional synthesis \cite{retsinas2024analysis}. 
Temporal modeling is critical for generating continuous and smooth facial expression videos, yet existing approaches often fail to simulate intricate transitions and micro-expressions effectively \cite{netsHilbert,xu2024vasa1}. Discrete methods for facial expression generation also cannot capture the continuity of emotional transitions, though advancements like Geometry Contrastive GANs (GC-GANs) demonstrate the potential for high-fidelity continuous expressions \cite{qiao2018geometry}. 
Interactive generative adversarial networks (iGANs) and temporal behavioral biometrics illustrate the importance of incorporating temporal consistency and additional facial behaviors, such as head pose, to enhance emotional authenticity \cite{nojavanasghari2018interactive,agarwal2020detecting}. To advance nuanced emotional synthesis, researchers must focus on robust temporal modeling, effective handling of occlusions, and seamless multimodal integration. Innovations such as GC-GANs and advanced frameworks for continuous, contextually adaptive expression generation are pivotal for achieving emotionally intelligent systems. These advancements are essential for applications requiring high levels of emotional realism and human-like interaction capabilities, driving progress in affective face generation.
While some studies have addressed the generation of facial videos driven by emotional audio, efficiently generating high-quality talking head videos that integrate both emotional expressions and style features remains a significant challenge, as most current audio-driven facial animation research primarily focuses on generating videos with neutral emotions. ESGaussianFace \cite{ma2026esgaussianface} proposes a framework for emotional and stylized audio-driven facial animation that leverages 3D Gaussian Splatting to reconstruct 3D scenes and render videos, ensuring efficient generation of 3D consistent results. The method introduces an emotion-audio-guided spatial attention mechanism that effectively integrates emotion features with audio content features, enabling more accurate reconstruction of facial details across different emotional states. Two 3D Gaussian deformation predictors achieve emotional and stylized deformations of the Gaussian points through emotion and style features, and a multi-stage training strategy enables step-by-step learning of the character's lip movements, emotional variations, and style features.
3D Gaussian splatting-based talking head synthesis has gained attention for its ability to render high-fidelity images with real-time inference speed, but since it is typically trained on only a short video that lacks diversity in facial emotions, the resultant talking heads struggle to represent a wide range of emotions. EmoTalkingGaussian \cite{cha2025emotalkinggaussian} addresses this by proposing a lip-aligned emotional face generator that trains a 3D Gaussian splatting model capable of manipulating facial emotions conditioned on continuous emotion values, specifically valence and arousal, while retaining synchronization of lip movements with input audio. To achieve accurate lip synchronization for in-the-wild audio, the method introduces a self-supervised learning approach that leverages a text-to-speech network and a visual-audio synchronization network. Experiments on publicly available videos demonstrate improvements over existing methods in image quality, emotion expression accuracy, and lip synchronization.
While many existing approaches to audio-driven portrait animation focus on lip synchronization and video quality, few tackle the challenge of generating emotion-driven talking head videos with fine-grained control over both emotion categories and intensities. EMOdiffhead \cite{zhang2024emodiffhead} proposes a method for emotional talking head video generation that enables fine-grained control of emotion categories and intensities while supporting one-shot generation. Given the linearity of the FLAME 3D model in expression modeling, the method extracts expression vectors using the DECA approach and combines them with audio to guide a diffusion model in generating videos with precise lip synchronization and rich emotional expressiveness. This approach enables learning rich facial information from emotion-irrelevant data while facilitating the generation of emotional videos, effectively overcoming the limitations of emotional data such as the lack of diversity in facial and background information.
Generating emotion-specific talking head videos from audio input is a complex challenge because emotion is a highly abstract concept with ambiguous boundaries, necessitating disentangled expression parameters to produce emotionally expressive results. EmoHead \cite{shen2025emohead} presents a method to synthesize talking head videos via semantic expression parameters. An audio-expression module that can be specified by an emotion tag predicts expression parameters for arbitrary audio input, enhancing the correlation from audio input across various emotions. The method leverages a pre-trained hyperplane to refine facial movements by probing along the vertical direction, and the refined expression parameters regularize neural radiance fields to facilitate emotion-consistent generation of talking head videos. Experimental results demonstrate that semantic expression parameters lead to improved reconstruction quality and controllability.

\section{Discussion and Future Prospects}
\label{future}
The rapid progress in affective computing highlights both remarkable achievements and unresolved challenges. To move closer to emotionally intelligent agents, future research must address key gaps in emotion understanding, affective cognition, and expression. This section outlines promising research directions, emphasizing the need for robust datasets, interpretable models, multimodal integration, and ethical frameworks to ensure trustworthy and human-centered development.
\subsection{Future Prospects in Emotion Understanding}
Despite significant advances in emotion recognition, the field continues to encounter persistent challenges that hinder practical deployment and broad generalization of affective computing systems. Current solutions often depend on limited or biased datasets\cite{thakur2026emotion}, and models may struggle to interpret emotions that are complex, overlapping, or deeply shaped by cultural nuances. Although recent progress in advanced multimodal fusion strategies has led to improved performance, the optimal integration of diverse modalities, particularly in noisy, real-world environments, remains an open problem \cite{Nandini2025}.
Additional barriers, such as model interpretability, the lack of unified evaluation metrics, and limited cross-domain generalizability, require sustained research attention. As highlighted by \cite{DILUZIO2025107177}, deep learning models, while highly accurate, frequently operate as “black boxes,” making it difficult to justify or explain their predictions in sensitive applications. This fundamental limitation underscores the need for new interpretability frameworks in emotion recognition \cite{DILUZIO2025107177}. To address the lack of explainability in existing multimodal aspect-based sentiment analysis methods, \cite{wang2026explainable} reformulates this as a generative task using multimodal large language models, enabling joint aspect-level sentiment prediction and natural language explanation generation. The proposed framework incorporates dependency-syntax-guided sentiment cues to enhance aspect-oriented reasoning and improve the faithfulness of generated explanations. \cite{Alharbi2024} demonstrated that employing explainable feature selection methods in virtual reality environments not only increases user trust but also facilitates clinical adoption, though effectively communicating technical insights to end-users remains challenging.
With the emergence of LLMs and foundation models, the landscape of affective computing is rapidly evolving, offering new opportunities such as zero-shot and few-shot emotion recognition. Nonetheless, these models introduce critical risks, including hallucinated or spurious emotional attributions, difficulties in context understanding, and high annotation costs \cite{Wognum2024}.
Future research should prioritize the creation of large-scale, diverse, and high-quality datasets that capture real-world variability across individuals and cultures. Further efforts are needed to develop robust and explainable emotion recognition models, along with standardized evaluation metrics tailored to the complexities of multimodal affective computing. Finally, realizing the full potential of LLMs and foundation models in this domain will require systematic solutions for mitigating hallucinations, improving context awareness, and reducing labeling costs—ensuring that advances in hybrid affective systems translate to trustworthy and generalizable real-world applications.

\subsection{Future Prospects in Affective Cognition}
Emotional intelligence has been extensively studied from psychological, neuroscientific, and technological perspectives, with prior work reviewing its theoretical foundations, neurobiological underpinnings, and measurement methods, as well as discussing its integration into emotion-aware and human-centered AI systems and related ethical considerations\cite{espinosa2026emotional}. One of the critical tasks for advancing human-centered AI systems is the integration of emotion prediction and elicitation within complex interactive scenarios \cite{zhang2024simulating}. This involves developing models that can accurately forecast emotional responses and trigger appropriate emotional states in dynamic, real-world human-computer interaction (HCI) settings, thereby enhancing the adaptability and effectiveness of such systems. This could involve controlled experiments to elicit targeted emotions or the collection of large-scale naturalistic datasets capturing diverse human emotions. Additionally, replacing the simple moving window average with advanced time-dependent models could better capture the dynamic persistence of emotions. The implications of models include its potential integration into adaptive interactive systems that anticipate user emotional states, providing designers with insights into how task progression and user goals influence emotional outcomes. Future developments should focus on personalizing the model to individual user objectives and proficiency, thereby improving its alignment with cognitive states and enhancing prediction accuracy in affective computing applications.\\
While language models play a significant role in the advancement of modern affective computing, their contributions to emotional elicitation and emotional experiences are still limited. These models are mainly used for understanding emotions and generating emotional expressions. Most existing research has concentrated on assessing their ability to model emotional elicitation \cite{khan2025large}. However, the development of intelligent agents based on language models, particularly those with specific cognitive skills such as decision-making in social contexts, has been relatively neglected. To create an intelligent agent with emotional intelligence using language models, it is essential to combine these models with a specific cognitive computational framework for the agent. This integration enables the knowledge and experiences to influence agent decision-making processes and emotional responses, ultimately enhancing its emotional intelligence and adaptability.
\subsection{Future Prospects in ETS}
The future of affective text synthesis is poised for significant advancements, driven by the integration of more sophisticated AI models and a deeper understanding of human emotion \cite{picard1997affective}. As models become more adept at capturing the nuances of emotional expression, we can expect to see more personalized and empathetic human-computer interactions \cite{singh2020adapting20}. The development of large-scale, high-quality emotional datasets will be crucial in training models that can generate truly authentic and contextually appropriate affective text. Furthermore, the exploration of cross-modal emotion generation, where text is synthesized in conjunction with other modalities like speech and facial expressions, will open up new frontiers in creating immersive and emotionally resonant experiences \cite{mcdarby2003affective}. Ethical considerations will be a major focus, particularly in preventing the misuse of emotion generation for manipulation and ensuring fairness in algorithmic emotional expression. Methods like emotion embeddings and reinforcement learning can be used to improve emotional consistency through reward mechanisms. However, achieving a balance between emotional consistency and coherence, particularly for subtle emotions, remains a significant challenge \cite{lu2022makes69}.
\subsection{Future Prospects in ESS}
As previously discussed, a primary challenge in speech generation is acquiring high-quality audio data. An optimal dataset for TTS should encompass diverse emotions and duration, multi-speaker, and multi-gender recordings, featuring authentic, unacted voices in parallel formats. Sources such as podcasts, films, and theater performances are valuable for collecting varied vocal expressions. Additionally, preserving the emotional nuances present in performed poetry can enhance the expressiveness of TTS outputs. A dataset similar to Emilia \cite{He2024}, but tailored for emotional speech generation (ESG), would be ideal.
Models must be capable of converting and synthesizing various emotions with different durations, ensuring speaker independence and preventing emotion leakage. GANs and diffusion models have emerged as promising approaches to address these challenges. GANs, through their adversarial training mechanism, effectively generate complex, high-dimensional data, making them suitable for producing nuanced emotional speech \cite{ma2024}. Diffusion models, on the other hand, have been utilized to synthesize speech with mixed emotions or varying intensities, offering fine-grained control over emotional expression \cite{Tang2023a}. Reinforcement learning (RL) also presents a valuable framework, particularly for conversational speech synthesis. In this context, an agent interacts with the environment, receiving feedback that guides the learning process. This interactive paradigm enables the model to learn and adapt to various emotional expressions, even from a limited set of basic emotions, by refining its performance through continuous interaction \cite{Liu2021}.
LLMs have demonstrated proficiency in generating TTS systems. Advancements in LLMs, like rising Deepseek \cite{Guo2025}, will lead to more efficient and less computationally intensive emotional speech generation systems.
In conclusion, ESG can be improved by synthesizing diverse emotions with accurate control and preventing emotion leakage. Techniques like GANs, diffusion models, and reinforcement learning show promise in achieving this. Advancements in efficient LLMs further enhance ESG systems, and the rise of speech-language models \cite{Cui2024} is a great opportunity to have efficient ESG systems. These innovations will lead to more natural and expressive emotional speech synthesis, enriching human-computer interaction.
\subsection{Future Prospects in Emotional Face Synthesis}
The future of emotional face synthesis presents exciting opportunities at the intersection of diffusion models, multimodal integration, and ethical AI development. Recent advancements in diffusion model-based approaches have demonstrated superior performance in generating realistic and emotionally expressive faces, with works like \cite{xu2024id3} achieving significant improvements in emotion similarity metrics. The field is rapidly moving toward more nuanced expression control, with emerging research focusing on micro-expression modeling and anatomically informed facial muscle simulation \cite{retsinas2024analysis, varanka2024localized}. Multimodal integration represents another promising direction, as researchers develop collaborative diffusion techniques that synchronize facial expressions with speech and text inputs \cite{huang2023collaborative, pham2017speech}. Additionally, the development of efficient architectures and model compression techniques addresses computational challenges, making real-time emotional face synthesis increasingly feasible \cite{he2024uniportrait, liu2018efficient}.
However, recent research has highlighted significant ethical considerations that must be addressed as these technologies mature. Studies from 2024 have challenged the universality hypothesis of facial expressions, emphasizing that emotions are expressed and perceived differently across cultures and contexts \cite{katirai2024ethical}. This cultural variability raises concerns about potential biases in emotional face synthesis systems. Furthermore, as the global market for emotion recognition and synthesis technologies expands, issues of privacy, consent, and potential misuse in sensitive domains like employment, healthcare, and surveillance demand careful consideration \cite{wu2024social}. Future development will require robust ethical frameworks that address fairness, non-discrimination, and a defined scope of use, particularly as these technologies become more integrated into human-computer interaction systems \cite{ballesteros2024facial}. The responsible advancement of emotional face synthesis will depend on balancing technological innovation with ethical guidelines that protect individual privacy and prevent emotional manipulation, ensuring these powerful generative systems serve beneficial purposes.
\section{Conclusion}
This study has provided a comprehensive exploration of the integration of emotional intelligence into intelligent agents, highlighting the critical roles of emotion understanding, affective cognition, and emotion expression in fostering naturalistic and empathetic human-computer interactions. By systematically analyzing the challenges, ranging from dataset limitations and model interpretability in emotion understanding to contextual and cognitive complexities in affective cognition and multimodal synchronization in emotion expression, we have underscored the multifaceted hurdles that impede progress in affective computing. We outline and explore recent solutions for revealing promising pathways to overcome these obstacles. This work not only synthesizes current advancements but also proposes a roadmap for future research, paving the way for the development of emotionally intelligent agents capable of truly adaptive, empathetic, and human-like interactions. Our findings underscore the necessity of continued innovation in data collection, model design, and evaluation frameworks to overcome existing barriers and pave the way for emotionally intelligent agents that can foster trust, empathy, and more effective human-computer interaction.\\

\bibliographystyle{abbrv}

\end{document}